\documentclass[fleqn,usenatbib]{mnras}

\usepackage{newtxtext,newtxmath}

\usepackage[T1]{fontenc}

\DeclareRobustCommand{\VAN}[3]{#2}
\let\VANthebibliography\thebibliography
\def\thebibliography{\DeclareRobustCommand{\VAN}[3]{##3}\VANthebibliography}

\usepackage{graphicx}	
\usepackage{amsmath}	
\usepackage{subcaption}
\usepackage{scalerel}

\newcommand{\be}{\begin{equation}}
\newcommand{\ee}{\end{equation}}
\def\Lsun{\, L_{\odot}}
\def\Msun{\, M_{\odot}}
\def\Mh{M_{\rm h}}
\def\Mv{M_{\rm V}}
\def\muV{\mu_{\rm V}}

\def\Ms{M_{\star}}
\def\HH{\rm H_2}

\def\chid{\chi_{\rm d}}
\def\Rcrit{R_{\rm 200c}}
\def\Rd{R_{\rm d}}
\def\zextract{z_{\rm extract}}
\def\dhelio{d_{\rm helio}}

\def\rhalf{r_{\rm 1/2}}
\def\Mpeak{M_{\rm peak}}
\def\Mvir{M_{\rm vir}}
\def\Mcrit{M_{\rm 200c}}
\def\zrei{z_{\rm rei}}

\def\rvir{r_{\rm vir}}
\def\menc{M(\rm < 100 kpc)}
\def\pho{p_{\scaleto{\rm HO}{4pt}}}

\title[GRUMPY treatment of the Milky Way satellites]{Forward-modelling the Luminosity, Distance, and Size distributions of the Milky Way Satellites}

\author[V. Manwadkar \&  A.V. Kravtsov ]{
Viraj Manwadkar$^1\thanks{E-mail: virajmanwadkar@gmail.com}$ and Andrey V. Kravtsov$^{1,2,3}\thanks{E-mail: kravtsov@uchicago.edu}$
\\
$^{1}$Department of Astronomy  \& Astrophysics, The University of Chicago, Chicago, IL 60637 USA\\
$^{2}$Kavli Institute for Cosmological Physics, The University of Chicago, Chicago, IL 60637 USA\\
$^{3}$Enrico Fermi Institute, The University of Chicago, Chicago, IL 60637 USA
}

\date{Accepted XXX. Received YYY; in original form ZZZ}

\pubyear{2021}

\begin{document}
\label{firstpage}
\pagerange{\pageref{firstpage}--\pageref{lastpage}}
\maketitle

\begin{abstract}
We use \texttt{GRUMPY}, a simple regulator-type model for dwarf galaxy formation and evolution, to forward model the dwarf galaxy satellite population of the Milky Way (MW) using the Caterpillar zoom-in simulation suite. We show that luminosity and distance distributions of the model satellites are consistent with the distributions measured in the DES, PS1 and SDSS surveys, even without including a model for the orphan galaxies. We also show that our model for dwarf galaxy sizes can simultaneously reproduce the observed {\it distribution} of stellar half-mass radii, $\rhalf$, of the MW satellites and the overall $\rhalf-\Ms$ relation exhibited by observed dwarf galaxies. The model predicts that some of the observed faint stellar systems with $\rhalf<10$ pc are ultra-faint dwarf galaxies. Scaling of the stellar mass $\Ms$ and peak halo mass $\Mpeak$ for the model satellites is not described by a power law, but has a clear flattening of $\Ms-\Mpeak$ scaling at $\Mpeak<10^8\,\Msun$ imprinted by reionization. As a result, the fraction of low mass haloes ($\Mpeak < 10^8 \Msun$) hosting galaxies with $M_V<0$ is predicted to be 50\% at $\Mpeak \sim 3.6 \times 10^7\,\Msun$. We find that such high fraction at that halo mass helps to reproduce the number of dwarf galaxies discovered recently in the HSC-SSP survey. Using the model we forecast that there should be the total of $440^{+201}_{-147}$ (68\% confidence interval) MW satellites with $\Mv < 0$ and $\rhalf > 10$ pc within 300 kpc and make specific predictions for the HSC-SSP, DELVE-WIDE and LSST surveys.
\end{abstract}




\section{Introduction}
\label{sec:intro}

Most galaxies in the Universe are {\it dwarf} galaxies with luminosities of $L \lesssim 10^{9} \Lsun$. Due to their shallow potential wells, dwarf galaxies are particularly sensitive to feedback processes \citep[e.g.,][]{Dekel.Silk.1986,Efstathiou.2000,Pontzen.Governato.2012}, such as heating of intergalactic gas during reionization \citep[e.g.,][]{Efstathiou.1992,Thoul.Weinberg.1996,Barkana.etal.1999,Bullock.etal.2000,Gnedin.2000,Somerville.2002,Benson.etal.2002a} and galactic outflows \citep[e.g.,][]{Hopkins.etal.2014,Muratov.etal.2015,Pandya.etal.21}. Dwarf galaxies are thus a key laboratory for probing and understanding these feedback processes.   


In addition, the faintest of dwarf galaxies---the \textit{ultra-faint} dwarf galaxies \citep[UFDs, $\Mv \gtrsim -8$ or $L \lesssim 10^5\Lsun$, see][for a review]{Simon.2019}--predicted by theoretical models \citep{Ricotti.Gnedin.2005,Gnedin.Kravtsov.2006,Bovill.Ricotti.2009,Salvadori.Ferrara.2009}, showed that such galaxies are expected to be the \textit{fossils} of the earliest stages of galaxy formation. These predictions were born out by observations so far, which showed that UFDs have formed most of their stars very early  \citep[][]{Tolstoy.etal.2009,Brown.etal.2012,Brown.etal.2014,Weisz.etal.2014,Sacchi.etal.2021} and their stellar populations are very metal poor  \citep[see][for a review]{Simon.2019}. Due to the broad cosmological significance of dwarf galaxies, observational searches for faintest and lowest surface brightness galaxies have been the subject of much observational effort over the past two decades \citep[e.g.,][]{Belokurov.etal.2010,Martin.etal.2015,Longeard.etal.2018,Mau.etal.2020,Cerny.21.eriIV,MutluPakdil.etal.2021.NGC253}.

Prior to 2000, discoveries of dwarf galaxies were mainly based on searches using photographic plates \citep[e.g.,][cf. Fig. 1 in \citealt{Simon.2019}]{Harrington.etal.1950,Wilson.1955,Cannon.etal.1977,Irwin.etal.1990,Ibata.etal.1994}. However, in the last two decades wide area digital sky surveys, such as the Sloan Digital Sky Survey (SDSS), Dark Enery Survey (DES), PanSTARRS (PS1), Hyper
Suprime-Cam Strategic Survey Program (HSC-SSP), ATLAS, MagLiteS, etc. have revolutionized studies of the faintest dwarf galaxies. The number of known dwarf galaxies in the Local Group volume was tripled and dozens of the UFDs were discovered \citep[e.g.,][see \citealt{Simon.2019} and \citealt{DrlicaWagner.etal.2020} for recent summaries]{Willman.etal.2005,Belokurov.etal.2007,Koposov.etal.2015,DrlicaWagner.etal.2015}.  The advent of the Vera C. Rubin Observatory \citep{Ivezic.etal.2019} is expected to greatly expand the discovery space of dwarf galaxies  again \citep[e.g.,][]{MutluPakdil.etal.2021}.

The low surface brightnesses and faint luminosities  of UFDs make them  challenging to detect even in the most advanced modern surveys and so far observations only probed the tip of the iceberg \citep[e.g., Figure 5,6 in][]{DrlicaWagner.etal.2020} of the UFD galaxy population expected in the Local Group volume \citep[e.g.,][]{Gnedin.Kravtsov.2006}. Although ongoing and future surveys are expected to discover many more UFDs, most of the expected UFD discoveries, especially $\Mv \gtrsim -4$, are expected to be confined to within $\sim$300 kpc from the MW \citep[e.g.,][]{MutluPakdil.etal.2021}. Thus, for the foreseeable future, the faintest satellites of the MW will continue to be the window into formation of the faintest galaxies \citep[e.g.,][]{Koposov.etal.2009,Jethwa.etal.2018,Newton.etal.2018,Nadler.etal.2019,Nadler.etal.2020}. 

The luminosity function (LF) is a fundamental property of galaxy population in general, with the faint-end of LF in the dwarf regime of particular importance as it is expected to be shaped by feedback processes accompanying galaxy formation \citep[e.g.,][]{Bullock.etal.2000,Benson.etal.2002a,Somerville.2002,Busha.etal.2010,Maccio.etal.2010,Font.etal.2011,Bose.etal.2018,Rey.etal.2019,Katz.etal.2020}. The UFD tail of the LF, which is expected to be particularly sensitive to both outflows and suppression of gas accretion due to UV heating after reionization, is particularly interesting. In the next decade, the luminosity function of the MW satellites will remain the most powerful observational probe of these processes in the UFD regime.

Likewise, distribution of the MW satellites around Milky Way can be used to probe processes shaping evolution of dwarf satellite galaxies around $\sim L_\star$ galaxies, such as disruption of subhaloes by the central disk \citep[e.g.,][]{Nadler.etal.2018,Kelley.et.al.2019,Richings.etal.2020,Webb.Bovy.2020,Green.etal.2021}, merger history of their host halo \citep[e.g.,][]{Bose.etal.2020} or details of  reionization \citep[e.g.,][]{Ocvirk.etal.2011}. In fact, so far models have struggled to reproduce the concentration of the radial distribution of the luminous satellites of the Milky Way and a number of other nearby galaxies \citep[e.g.,][although see \citealt{Mao.etal.2021}]{Carlsten.etal.2020.radial}, leading to conclusions that a significant fraction of subhaloes is missing in simulations for numerical reasons and that special treatment of the ``orphan'' galaxies of these missing subhaloes is needed \citep{Bose.etal.2020}. Indications that a surprisingly large fraction of UFDs are close to the pericentre of their orbit \citep[e.g.,][]{Simon.2018,Fritz.etal.2018,Li.etal.2021} is another puzzle to be solved. 

Cosmological simulations that include all relevant physical and environmental processes should ultimately provide the basis for our understanding of dwarf galaxy formation  \citep[e.g., see][for recent examples]{Wheeler.etal.2015,Wheeler.etal.2019,Rey.etal.2019,Rey.etal.2020,Agertz.etal.2020,Katz.etal.2020,Applebaum.etal.2021,Jeon.etal.2021,Rey.etal.2021.edge.hi}. However, current simulations face  many challenges, such as the dynamic range challenge in resolving relevant small-scale structure both in dark matter and baryon distribution and associated computational expense. At the same time, key physical processes affecting galaxy evolution, such as radiative and stellar feedback or the process of star formation in low-metallicity diffuse environments of dwarf galaxies, are not yet well understood. All of this demands convergence studies and parameter exploration to understand the full range of scenarios, however, the complexity and computational expense of simulations make this difficult in practice. 

Semi-analytic models have thus been used as an alternative modelling approach for dwarf satellite galaxies \citep[e.g.,][]{Kauffmann.etal.1993,Somerville.2002,Benson.etal.2002a,Benson.etal.2002b}. Such approaches had good success in modelling the observed dwarf satellites \citep[e.g.,][]{Salvadori.Ferrara.2009,Starkenburg.etal.2013,Bose.etal.2018}. During the past decade, approaches based on simple phenomenological ansatzes for how galaxy properties are related to the properties of their dark matter haloes have become widely used  \citep[e.g.,][]{Kravtsov.2010.review,Jethwa.etal.2018,Kim.etal.2018,Nadler.etal.2020,Jiang.etal.2021}. However, the  simplicity of this approach has limitation as it does not include modelling of the full range of observed dwarf galaxy properties, which include sizes, gas masses and metallicities. 

``Regulator'' type models have recently been used successfully to model a variety of galaxy statistics and scaling relations across wide range of epochs \citep[see, e.g., \S 4.1.2 of ][and references therein]{Somverville.Dave.2015}. These models are simpler than classical semi-analytic models, as they aim to model only the most crucial aspects of galaxy formation. At the same time, they retain sufficient complexity to predict evolution of key observable properties of galaxies, such as their gas mass, stellar and gas phase metallicities, etc. The relative simplicity and flexibility of such models allows straightforward exploration of the effect different parts of the model have on galaxy properties. At the same time, the regulator models proved to be quite capable in reproducing a wide range of observations. 

Indeed, in \citet{Kravtsov.Manwadkar.2022} we presented a regulator-type model focusing on modelling galaxies in the dwarf regime and demonstrated that the model can {\it simultaneously} match a wide range of properties of observed nearby dwarf galaxies and their observed correlations, including star formation histories, stellar and gas phase metallicities, gas masses, sizes, star formation rates. In this paper, we use results of this successful model along with realistic survey detection probabilities to forward model  observed satellite population. Such forward modelling is particularly important for modelling UFD satellite population and was used in a number of previous theoretical studies of the MW satellite population based on phenomenological models of galaxy-halo connection \citep[e.g.,][]{Jethwa.etal.2018,Nadler.etal.2020}. 

In this study we focus on forward-modelling the satellite luminosity function down to the faintest luminosities, distribution of dwarf satellits in their radial distance to the Milky Way, and dwarf satellite sizes probed in various current and upcoming observational surveys. 
This paper is organized as follows. In Section~\ref{sec:model}, we describe the main features of the galaxy formation model and the forward modelling steps to produce predicted satellite populations. In Section~\ref{sec:comp_to_obs}, we compare model predictions to the luminosity function, radial distribution, and size distribution of the observed population of MW satellites. In Section~\ref{sec:discuss}, we discuss implications of our results for the census of satellite dwarf galaxies in the MW environment and their sizes and surface brightness and then summarize our results in Section~\ref{sec:summary}. In the Appendices, we discuss the impact of simulation resolution on predicted satellite statistics (Appendix~\ref{app:reso_effect}), parametrization of relations between peak halo mass, stellar mass and V-band luminosity (Appendix~\ref{app:shmrapprox}), sensitivity of model predictions to the current uncertainty in the MW halo mass (Appendix~\ref{app:mw_mass_unc}), and list the relevant properties of observed dwarf galaxies and ultra faint star clusters used in this study (Appendix~\ref{app:mw_sats}). 

\section{The Model}
\label{sec:model}

To model the satellite population of the Milky Way (MW) we use the \texttt{GRUMPY} dwarf galaxy model presented in \citet{Kravtsov.Manwadkar.2022}. As a baseline for modelling different galaxy components it uses halo mass accretion histories extracted from simulations. In this section we describe halo catalogs used to make model runs and a general overview of the model and details particularly relevant to this study.

\begin{figure}
    \centering
    \includegraphics[width = \columnwidth]{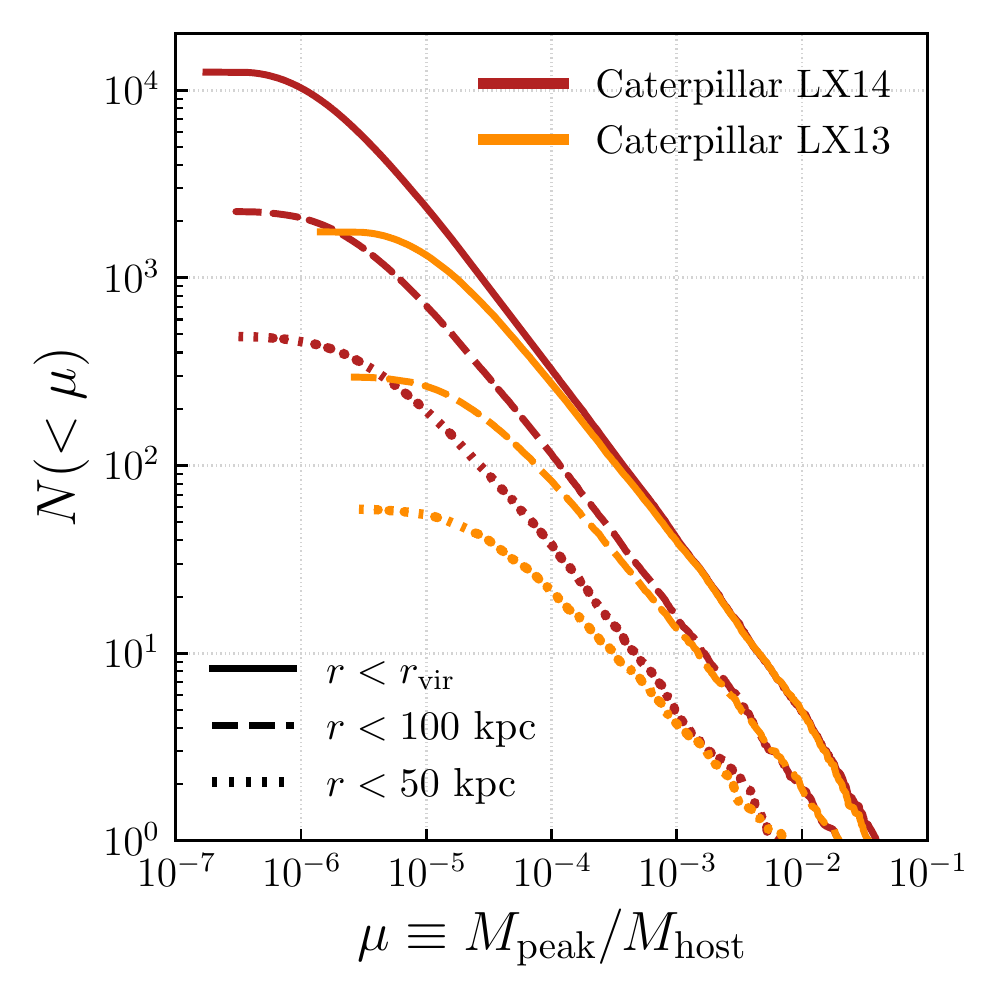}
    \caption{The average subhalo mass functions of subhaloes within i) 50 kpc (dotted lines), ii) 100 kpc (dashed lines) and iii) virial radius (solid lines) for the Caterpillar suites at the LX13 and LX14 resolution level. }
    \label{fig:cat_res}
\end{figure}

\subsection{Halo catalogs}
\label{ssec:dmo_sims}
 
 \begin{figure*}
    \centering
    \includegraphics[width = \textwidth]{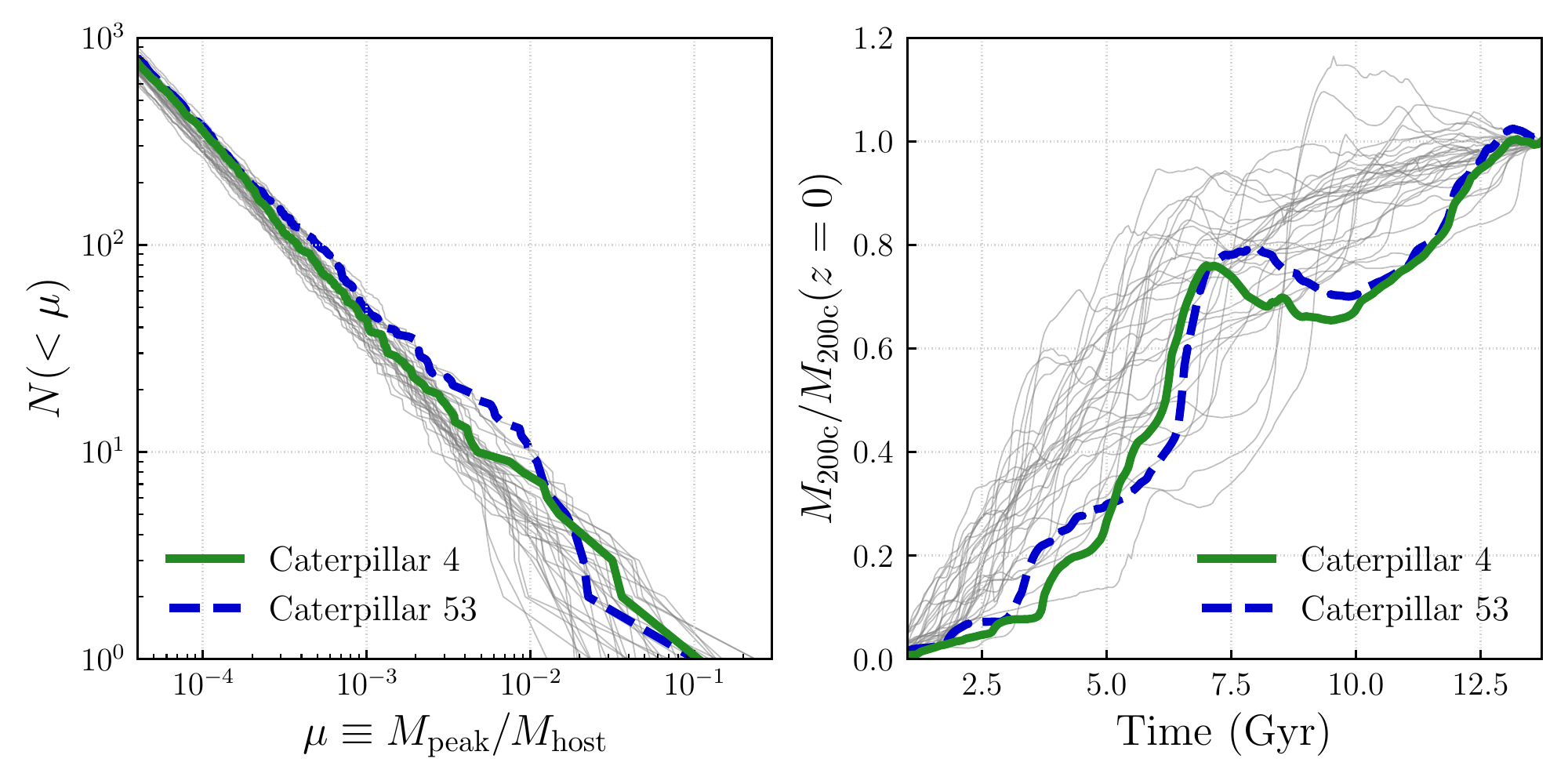}
    \caption{(\textit{Left}) The subhalo mass functions within virial radius for the two MW-sized hosts from Caterpillar LX14 suites (denoted by colored lines) that contain a LMC analog satellite according to criterion in Section~\ref{ssec:lmc_select}. For comparison, we also show the subhalo mass functions from other hosts in Caterpillar LX14 that do not have an LMC analog (thin grey lines). (\textit{Right}) The mass assembly histories (MAHs) of the two MW-sized hosts from Caterpillar LX14 suites (denoted by colored lines) that contain a LMC analog satellite according to criterion in Section~\ref{ssec:lmc_select}. For comparison, we also show the MAHs from other hosts in Caterpillar LX14 that do not have an LMC analog (thin grey lines).}
    \label{fig:cater_sums}
\end{figure*}

As a baseline for the galaxy formation model we use halo catalogs and halo mass accretion histories (MAHs) derived from the Caterpillar suite \citep[][]{Griffen.etal.2016} of zoom-in simulations of MW-sized haloes.\footnote{\url{https://www.caterpillarproject.org}}  The Caterpillar suite consists of simulations of 32 MW-like host haloes\footnote{A total of 35 MW-like haloes were simulated, but a few of them are not used here as they are contaminated by low resolution particles. Refer to \citet{Griffen.etal.2016} for more details on the contamination study.}. Each of these host haloes has been re-simulated at three different resolution levels, namely LX12, LX13 and LX14, with LX14 being the highest resolution of the three \citep[see][for details]{Griffen.etal.2016}. The existence of several resolution levels allows checks for convergence of the results and LX14 resolution is higher than the resolution of ELVIS simulations used in our previous study \citep[][]{Kravtsov.Manwadkar.2022}. We therefore adopt the Caterpillar suite for our modelling of dwarf satellite population. 
 
The haloes were identified using the modified version of the Rockstar halo finder and the Consistent Trees Code \citep{Behroozi.etal.2013}.
As discussed by \citet[][see their Section 2.5]{Griffen.etal.2016}, the modification is related to the misclassification of subhaloes with large initial fraction of unbound particles by the Rockstar, which results in a loss of $\approx 30\%$ of massive subhaloes. 

Figure~\ref{fig:cat_res} shows the average cumulative subhalo mass functions for the Caterpillar suite at the LX13 and LX14 resolution levels. 
The mass resolution limit of a simulation can be approximately deduced as the mass at which the cumulative subhalo mass function begins to flatten. The figure shows that the LX14 suite has a resolution limit of $\mu=\Mpeak/M_{\rm host} \approx 4 \times 10^{-6}$ ($\Mpeak \approx 4 \times 10^{6} \Msun$) even in the innermost regions of the host ($r < 50$ kpc).  Thus, the LX14 suite allows us to model MW satellites with a full range of subhalo masses that are expected to host faintest UDF galaxies probed  by current observations. A detailed convergence test via comparisons of the LX13 and LX14 simulations is presented in the  Appendix~\ref{app:reso_effect}.

We run our model using haloes with $\Mpeak > 10^6 \Msun$ and $t_{\rm start} < 4$ Gyr, where $t_{\rm start}$ is the earliest time when a given halo is detected in the simulation. This cut is used to avoid spurious haloes and ill-defined MAH. Only $\lesssim$1\% of haloes with $\Mpeak > 10^7 \Msun$ are removed with this cut and it will thus will not affect our results significantly.

In our model, we do not track and model for `orphan haloes', that is, dark matter haloes which have disrupted below the detection resolution limit of Rockstar halo finder. We discuss the rationale of this choice detail in Section~\ref{sssec:artificial_disrupt}.

Throughout this paper, we use halo masses defined within the radius enclosing either the ``virial'' density contrast \citep[][]{Bryan.Norman.1998}, $\Mvir$, or the contrast of 200 times the critical density of the universe at the redshift of analysis, $M_{\rm 200c}$. $\Mvir$ "virial" is used when computing subhalo disruption probabilities with the \citet[][]{Nadler.etal.2018} model and in the studies of numerical effects on subhalo population in Section~\ref{sssec:artificial_disrupt}, while we use $M_{\rm 200c}$ elsewhere to compute evolution of halo mass in the galaxy formation model, $M_{\rm peak}$ for subhaloes, etc. because it is much less affected by the pseudo-evolution effect \citep{Diemer.etal.2013}.

\subsection{Subhalo disruption model}
\label{ssec:subhalo_disrupt}

To account for subhalo disruption due to the extra tidal forces exerted by the central baryonic disk, we use the random forest subhalo disruption model from \citet{Nadler.etal.2018}. This model computes a disruption/survival probability of a subhalo given its internal and orbital properties, namely, the proper/physical\footnote{physical/proper distance, i.e., the comoving distance multiplied by the scale factor.} distance and scale factor at the first pericentric passage after accretion, and the virial mass, maximum circular velocity and scale factor at time of accretion. All this information is extracted from the evolution tracks of the simulations described in Section~\ref{ssec:dmo_sims}. 

\subsection{Selecting haloes with Large Magellanic Cloud Analog}
\label{ssec:lmc_select}

The satellite population of the MW has been shown to have significantly impacted by the presence of the Large Magellanic Cloud  \citep[LMC, e.g.,][]{lynden.1976,Lu.etal.2016,Dooley.etal.2017,Nadler.etal.2020}. The LMC appears to be on its first infall into the MW \citep[e.g.,][]{Besla.etal.2007,Kallivayalil.etal.2013} and is thought to have brought with it a significant fraction of the MW satellites \citep[e.g.,][]{Donghia.Lake.2008,Nadler.etal.2020}. It is thus important to account for the presence of the LMC-size object along with the satellites it brought when modelling the population of the MW satellites, especially in the surveys with the sky footprint close to the LMC, such as  the Dark Energy Survey (DES).

Thus, we focus on the two Caterpillar LX14 hosts that have a realistic LMC analog defined as a subhalo with i) mass of $\Mpeak > 1 \times 10^{11} \Msun$ consistent with recent estimates of the LMC mass \citep[e.g.,][]{Erkal.etal.2019,Erkal.Belokurov.2020,Vasiliev.etal.2021,Shipp.etal.2021}, and ii) is on its first infall into the MW and iii) is at a physical host-centric distance close to 50 kpc.

Among all the Caterpillar hosts, we did not find a single host that had an LMC-mass subhalo on its first pericentric passage and at a host-centric distance of $\sim 50$ kpc. However, in  Caterpillar 4 and Caterpillar 53 hosts, an LMC-mass subhalo went through a recent first pericentric passage around at $z \sim 0.05$. For these hosts, we extract all halo track data at a previous epoch ($\zextract$) when LMC was on its first infall and at a host-centric distance closest to 50 kpc. The halo properties, subhalo distribution, and our galaxy model predictions at $\zextract$ are then used. 

Although the subhalo disruption prescription described in Section~\ref{ssec:subhalo_disrupt} was calibrated at $z=0$, in the subsequent analysis we assume that disk properties and halo structure to be similar at $z \sim 0.05$ and that the same disruption model can be used in such hosts. Information for the two hosts with LMC analogs is listed in Table~\ref{tab:lmc_summs}. The MAHs and subhalo mass functions of these two hosts are shown in Figure~\ref{fig:cater_sums}. Caterpillar 4 mass function is pretty close to the median, while Caterpillar 53 mass function is on the high edge of the distribution in amplitude, as could be expected if the LMC counterpart brought in a significant number of subhaloes with it.

We find 4 other hosts, Caterpillar $1$, $6$, $27$, and $40$, that have an LMC mass subhaloes on the first infall. However, in these haloes LMC-sized subhaloes are still at distances  of $201$ kpc, $131$ kpc, $235$ kpc and $225$ kpc, respectively. Many of the faint UFD galaxies brought by these subhaloes would not be detectable at these distances and we therefore do not consider them as having an LMC analog. We do find differences in the observed radial distribution and luminosity function of satellites of such hosts, which we discuss in Section~\ref{ssec:lmc_radials}.

\begin{table*}
\caption{Summary of properties of the MW-sized host haloes in the Caterpillar suite that contain a realistic LMC analog (see Section~\ref{ssec:lmc_select} for details). For each host we list the epoch closest to $z=0$ at which LMC analog is on its first approach to the pericentre and is at a distance closet to 50 kpc. At this epoch, namely $z_{\rm extract}$, we extract the host $\Mcrit$ mass, host mass enclosed within 100 kpc ($\menc$), host virial radius ($\rvir$), host halo concentration ($c_{-2, \rm host}$), redshift at which 50\% of the current host halo mass was assembled ($z_{\rm form, 50\%}$, the LMC analog $\Mpeak$ mass and its distance from the host halo centre. Note that all the distances below ($\rvir$ and $d_{\rm host}$) $d_{\rm host}$ is the proper distance, that is, comoving distance times the scale factor.}
\label{tab:lmc_summs}
\centering
\begin{tabular}{ccccccccc}
\hline
Simulation Name & $z_{\rm extract}$ & $\Mcrit$ & $\menc$ & $r_{\rm vir}$(kpc) & c$_{-2}$ & $z_{\rm form, 50\%}$ & LMC $\Mpeak$ ($\Msun$) & LMC $d_{\rm host}$(kpc)  \\
\hline
Caterpillar 4 & 0.057  & $1.15 \times 10^{12}$   & $6.41 \times 10^{11}$   & 299.6 &  9.97  & 0.95 & $1.3 \times 10^{11}$ & 55.5   \\
Caterpillar 53 & 0.057  & $1.09 \times 10^{12}$   & $6.29 \times 10^{11}$   & 293.7 &  10.91  & 0.88 & $1.16 \times 10^{11}$ & 47.7   \\
\hline
\end{tabular}
\end{table*}

\subsection{Correcting for the difference of host and MW halo masses}
\label{ssec:host_halo_mass}

In addition to accounting for the presence of the LMC, it is important to make sure that simulated host haloes have masses consistent with current estimates of the MW halo mass, because the number of satellites scales roughly linearly with host halo mass \citep[e.g.,][]{Kravtsov.etal.2004}. 

In our analyses we adopt recent constraints  for the MW halo mass enclosed within 100 kpc (denoted as $M(<100)$) by \citet{Vasiliev.etal.2021}, who deduced $M_{\rm MW}(<100) = 5.6 \pm 0.4 \times 10^{11} \Msun$. To approximately correct for the differences in halo mass between MW and simulated haloes, we multiply the number of satellites in each host by the fraction $M_{\rm MW,r}(<100 ) / M_{\rm host}(<100)$, where $M_{\rm MW,r}(<100 )$ is a random number drawn from a Gaussian distribution with the mean of  $5.6 \times 10^{11} \Msun$ and {\it rms} of $0.4 \times 10^{11} \Msun$ and $M_{\rm host}(<100)$ is the mass enclosed within 100 kpc in the simulated hosts. We compute the latter by using the virial mass $M_{\rm vir}$, virial radius $r_{\rm vir}$ and scale radius $r_{\rm s}$ from the halo catalog\footnote{Note that  $r_{\rm vir}$ and  $r_{\rm s}$ should be in units of proper/physical distance. Thus, whenever we analyze an epoch with $\zextract > 0$, we do the appropriate conversion by multiplying by scale factor at that epoch.} and assuming an NFW profile:
\be
    M (< 100\, \text{kpc}) = M_{\rm vir} \frac{f(cx)}{f(x)},
\ee
where $c = r_{\rm vir} / r_{\rm s}$ and $x = 100 / r_{\rm vir}$ and 
\be
    f(x) = \ln(1+x) - \frac{x}{1+x}.
\ee

There is an additional uncertainty in this normalization factor due to the differences in the concentration $c$ of the MW halo and our simulated hosts that we do not account for. In the Appendix~\ref{app:mw_mass_unc} we compare results obtained using mass correction above with the correction based on $\Mcrit$ with MW mass constraints from \citet{Deason.etal.2021} or $M(<100)$ from \citet{Shen.etal.2021} and find that these different choices result in a similar median number of satellites.

\subsection{The \texttt{GRUMPY} Dwarf Galaxy Formation Model}
\label{ssec:grumpy_mod}

\begin{figure}
    \centering
    \includegraphics[width = \columnwidth]{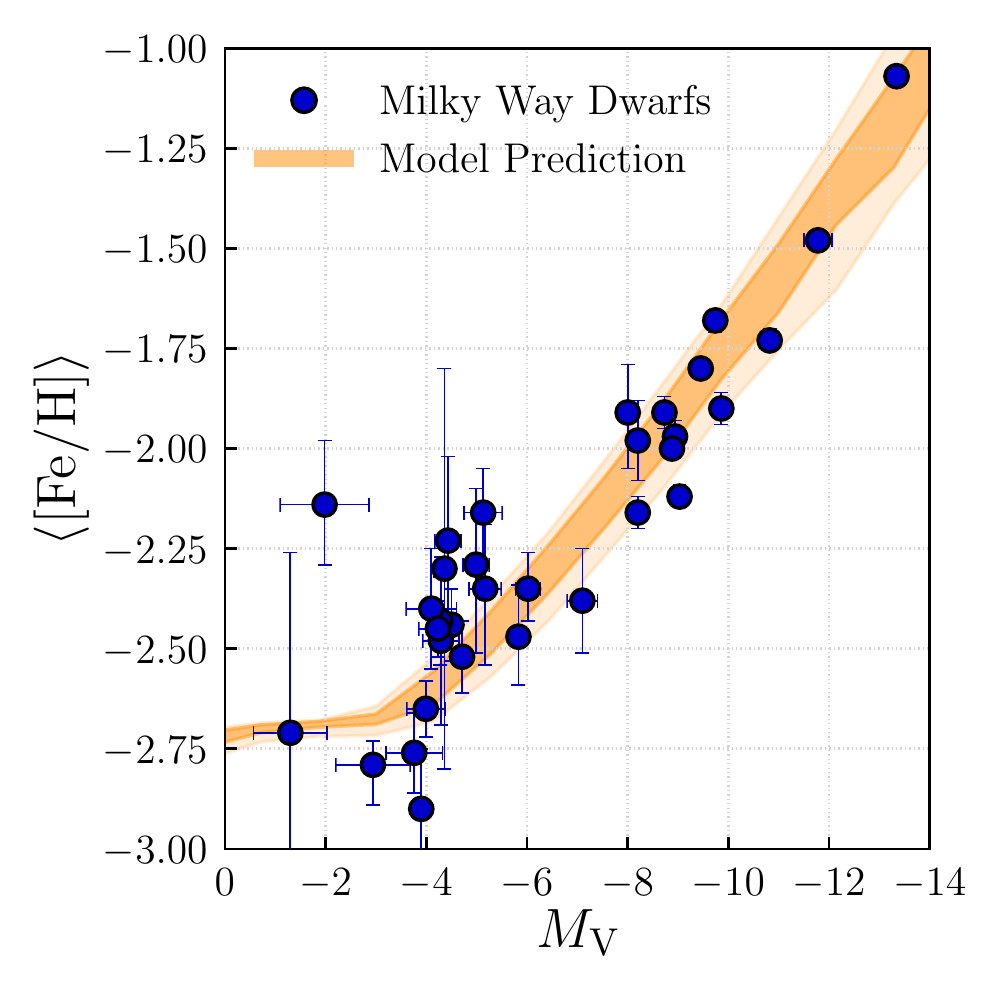}
    \caption{Model prediction for the relation between stellar metallicity $\rm\langle [Fe/H]\rangle$ and absolute V-band magnitude $\Mv$. This is using $\zrei = 8.5$ and the galactic outflow parameters assumed in this analysis. For comparison, we also show observational data of confirmed MW dwarf galaxies taken from the compilation in \citet{Simon.2019} with updated measurements of Antlia II and Crater II from \citet{Ji.etal.2021.Antlia}. Note that this plot is for only demonstration purposes that our model broadly agrees with observed dwarf galaxy stellar metallicities. 
    As seen in Figure 7 of \citet{Kravtsov.Manwadkar.2022}, the form of this relation is sensitive to galactic outflow parameters.}
    \label{fig:dwarf_metals}
\end{figure}

The \texttt{GRUMPY} (Galaxy formation with RegUlator Model in PYthon) is a regulator-type model for dwarf galaxies introduced in \citet{Kravtsov.Manwadkar.2022}. The basic premise of this model is that the mass accretion history of each halo in the simulation, $M_{200\rm c}(t)$, is used to follow evolution of the ISM gas mass, stellar mass, and mass in metals in these two components using a system of ordinary differential equations expressing mass conservation and mass transfer between different components. Using realistic MAHs extracted from high-resolution simulations  allows to account for the intrinsic scatter of observed dwarf galaxy properties arising from the scatter in their MAHs. 

The model includes prescriptions for UV heating after reionization and associated gas accretion suppression onto small mass haloes, galactic outflows, model for gaseous disk and its size, molecular hydrogen mass, star formation, etc. The model includes a prescription for the evolution of half-mass radius of the stellar distribution. This is important for modelling the surface brightnesses of dwarf galaxies and thus their detectability in a particular observational survey. Integrating the galaxy model with the MAHs extracted from simulations, we obtain the time evolution of the baryonic components, including star formation histories and metallicity evolution. 

The galaxy model parameters used in this paper are the fiducial parameters in \citet{Kravtsov.Manwadkar.2022}, except for the galactic outflow parameters $\eta_{\rm norm}$ and $\eta_{\rm p}$ which we assume to be $\eta_{\rm norm} = 1.8$ and $\eta_{\rm p} =-0.45$. We modify these parameters\footnote{The fiducial parameters assumed in \citet{Kravtsov.Manwadkar.2022} were $\eta_{\rm norm} = 3.6$ and $\eta_{\rm p }= -0.35$} to better match the bright end of the MW satellite luminosity function. Such modifications are reasonable as they are within the current scatter/uncertainty of the mass loading factor \citep[e.g.,][]{Mitchell.etal.2020,Pandya.etal.21}.

Our galaxy model produces realistic mean stellar metallicities and metallicity-$M_\star$ relations in good agreement with observations of MW satellites, as shown in Figure~\ref{fig:dwarf_metals}. The normalization and slope of this relation is sensitive to galactic outflow parameters \citep[see][]{Kravtsov.Manwadkar.2022}. We thus choose the outflow model parameters to reproduce the observed relation and a number of other observations of dwarf galaxies as demonstrated in our previous paper. Note that our model galaxies  \textit{simultaneously} match the observed stellar mass--metallicity relation and satellite luminosity functions (as we show in Section~\ref{ssec:lf}) without needing preventative feedback due to preheating as previous studies \citep[e.g.,][]{Lu.etal.2017} have suggested \citep[see Section 4 in][for discussion of why this is the case]{Kravtsov.Manwadkar.2022}.

\subsection{Forward modelling observational satellite samples}
\label{ssec:mock_obs}

We construct mock dwarf galaxy catalog using the galaxy formation model run with the  mass accretion histories of all haloes that satisfy the criteria described in Section~\ref{ssec:dmo_sims}. The star formation histories and stellar metallicity evolution predicted by the model are used along with the Flexible Stellar Population Synthesis \citep[FSPS;][]{Conroy.etal.2009,Conroy.Gunn.2010}\footnote{\href{https://github.com/cconroy20/fsps}{\tt https://github.com/cconroy20/fsps}} model to compute magnitudes of our galaxies in different bands assuming a \citet{Chabrier.2003} initial mass function and standard FSPS parameters. 

Specifically, in this project we compute the absolute V-band magnitude, $\Mv$, and azimuthally averaged effective half-light radius, $r_{1/2}$. In practice, we convert from 3D stellar half-mass radius $r_{\rm 50,3D}$ predicted by the model to project half mass radius, assuming that stellar distribution is approximately spherical: $r_{1/2} \approx r_{\rm 50,3D} / 1.34$ (\citealt{Wolf.etal.2010}). The 1.34 conversion factor is for a spherical galaxy, but is expected to depend on ellipticity \citep[see, e.g., Section 4.2.2 and eq. 10 in][]{Behroozi.etal.2021}: for a galaxy with $c/a=0.5$ the factor is $\approx$1.54 instead of 1.33 -- the $\approx$15\% difference that we ignore in our study. 

We then forward model mock dwarf galaxy populations to compare with observed satellite population in specific surveys. To take into account uncertainty in the observer location, we choose 200 random observer locations at a physical distance of 8 kpc from the host halo centre. As the Caterpillar simulations do not model the central baryonic disk potential, we do not constrain the random observer locations to any plane. For each observer location, we rotate the entire reference frame so that the LMC analog matches with the true sky coordinates of the LMC (RA = 80.89$^{\circ}$, Dec = -69.75$^{\circ}$). In the case of hosts with no LMC analogs, we choose random pointings of survey footprint in the sky. Thus, for each random observer location and each dwarf galaxy in our mock catalog, we have heliocentric distance ($d_{\rm helio}$), absolute V-band magnitude ($\Mv$), effective half-light radius ($r_{1/2}$) and sky coordinates. Similar to \citet{Nadler.etal.2020}, we only consider galaxies with $\rhalf > 10$ pc, which is the threshold used to define observed dwarf galaxy sample to avoid confusion with star clusters \citep[e.g.,][]{DrlicaWagner.etal.2020}. We discuss implications of adopting this threshold and possibility that a significant fraction of observed stellar systems with $\rhalf < 10$ pc may be dwarf galaxies in Section~\ref{ssec:small_dwarf}.

We use the resulting galaxy catalogs and prescriptions for the detection probability for the DES, PS1, SDSS, DELVE-WIDE, HSC-SSP and Vera C. Rubin Observatory/LSST survey to compare with observations and make future predictions. The prescriptions we use in our analysis are detailed in the remainder of this section. 

\subsubsection{The SDSS Selection Function}

To model the SDSS satellite population, we use the selection function described in \citet{Koposov.etal.2009}, which is expressed in terms of threshold magnitude of detection as a function of heliocentric distance:
\be
    \Mv < \frac{1.1 - \log_{10} (d_{\rm helio}/\text{kpc})}{0.228}
\label{eqn:sdss_thresh}
\ee

We assign zero detection probability for satellites with a magnitude fainter than above threshold or with the $V$-band surface brightness, $\muV>30\rm\,mag\, arcsec^{-2}$ where 
\be
    \muV = \Mv + 21.5721 + 2.5 \log_{10} (2\pi r^2_{1/2})
\ee
for $r_{1/2}$ in pc and $\mu_V$ in mag/arcsec$^{2}$.

The SDSS selection function is thus a step function with probability of $1.0$ if conditions are satisfied and zero otherwise. Furthermore, we apply the selection thresholds only to galaxies within the SDSS DR9 imaging footprint of the total area of 14555 deg$^{2}$ (\citealt{Ahn.etal.2012}) 

We use the approach of \citet{Jethwa.etal.2018} to jointly model the classical satellites detected prior to SDSS and SDSS-detected systems. Namely, we assume satellites with $\Mv \leq -11$ are detected over the entire sky, while fainter satellites are modelled using the SDSS selection criterion described above and are restricted to the SDSS footprint. 

\subsubsection{The DES \& PS1 Selection Function}
\label{sec:desps1selfunc}

We use the DES and PS1 survey selection function described in \citet{DrlicaWagner.etal.2020} and implemented in a publicly available code.\footnote{\url{https://github.com/des-science/mw-sats}} The selection functions are based on a machine learning method trained to compute satellite detection probability as a function of absolute magnitude $\Mv$, heliocentric distance $d_{\rm helio}$, azimuthally averaged half-light radius $r_{1/2}$, and sky position (RA, Dec). The information about the local stellar density and survey footprint is contained within the selection function code.

\subsubsection{DELVE-WIDE Selection Function}

The DELVE-WIDE survey\footnote{\url{https://delve-survey.github.io/}} \citep[e.g.,][]{DrlicaWagner.etal.2021.delve} is a survey that will cover the entire high-Galactic-latitude range ($|b| > 10^{\circ}$) in the southern hemisphere using the Dark Energy Camera. DELVE-WIDE will reach a photometric depth that is similar to that of the DES. We thus use the DES selection function discussed above to make predictions for DELVE-WIDE dwarf galaxy searches. 

Note that in the DES selection function, the detection probability calculation is separate from footprint mask value calculation. So we simply modify the footprint mask value calculation by using the DELVE-WIDE footprint instead of the DES footprint. The total area of the DELVE-WIDE footprint is $\sim$17000$\,\rm deg^2$, and includes both the $\sim$2200$\,\rm deg^2$ area around the Magellanic Clouds observed as part of the  DELVE-MC survey and the DES footprint \citep{DrlicaWagner.etal.2021.delve}. 

\begin{figure*}
    \centering
    \includegraphics[width = \textwidth]{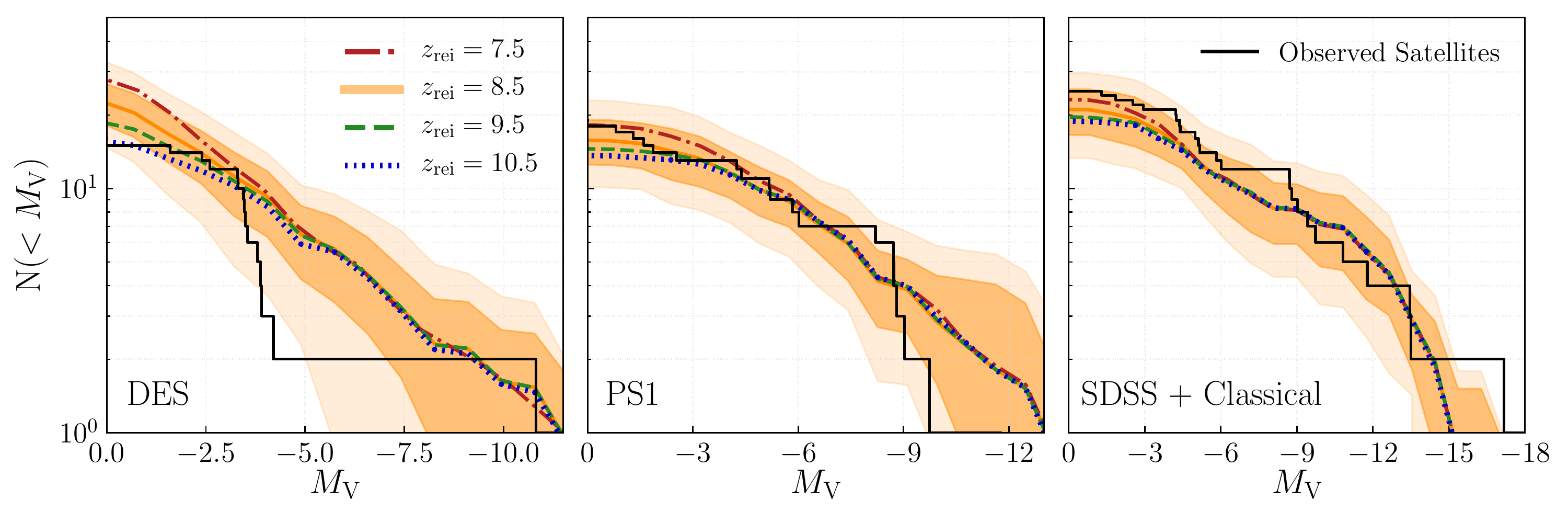}
    \caption{Predicted satellite luminosity function for hosts with LMC analog within $\dhelio < 300$ kpc in the DES (\textit{left}), PS1 (\textit{centre}) and SDSS + Classical (\textit{right}) samples as a function of reionization redshift $\zrei = 7.5,8.5,9.5,10.5$. The dark (light) orange shaded regions for $\zrei = 8.5$ corresponds to the 68\% (95\%) of the model predictions due to host-to-host scatter, bootstrap variations in the disruption model, etc. The colored lines with different line styles denote the median luminosity functions for different $\zrei$ models. The solid black lines shows the observed luminosity function for the respective survey. Note that the `SDSS + Classical' satellite luminosity function is combination of satellites observed in SDSS footprint and all classical satellites ($\Mv \leq -11$) over entire sky as discussed in Section~\ref{ssec:mock_obs}.  }
    \label{fig:fm_all_lfs}
\end{figure*}

\subsubsection{HSC-SSP and LSST Y1 Selection Function}

The selection functions for the ongoing HSC-SSP\footnote{\url{https://hsc.mtk.nao.ac.jp/ssp/}} \citep[e.g.,][]{hsc.survey,Aihara.etal.2018} and upcoming LSST surveys \citep[e.g.,][]{Ivezic.etal.2019} are not yet rigorously defined and the detection efficiency in the presence of high-density of faint background galaxies is not yet fully understood. We therefore adopt approximate selection function using approximations for the detection efficiency through simple surface brightness or magnitude cuts, as was done in  previous recent studies \citep[e.g.,][]{Homma.etal.2018.hsc,Homma.etal.2019.hsc,Nadler.etal.2019, Nadler.etal.2020}. 

Specifically, following \citet{Nadler.etal.2020}, we assume that HSC-SSP detects all satellites with $\Mv < 0$, $\rhalf > 10$ pc, and $\muV < 32$ mag/arcsec$^{2}$ within a heliocentric distance of 300 kpc. LSST Year 1 (Y1) is expected to have a detection efficiency similar to that of the HSC-SSP survey due to similar limiting magnitude. So we assume the same selection function as HSC-SSP for making predictions of dwarf galaxy observations in the LSST Y1 survey. Challenges of dwarf galaxy observations in such deep surveys are discussed further in  Section~\ref{ssec:fut_current_surveys_lf}.

To forward model HSC-SSP observations, we use the HSC-SSP footprint definition described in Table 5 and Figure 3 in \citet{hsc.survey}. At the moment, only 676 deg$^2$ of a total area of $\sim$1400$\,\rm deg^2$ in the HSC-SSP footprint has been searched for satellite galaxies. To forward model the HSC-SSP observations in this current (partially observed) area, we consider a galaxy to be observed if -- i) it passes the HSC-SSP detection threshold, ii) it falls in the HSC-SSP footprint and iii) $\rm U(0,1) < 676/1400 = 0.48$ where $\rm U(0,1)$ denotes a random number drawn from the uniform distribution in the interval $[0,1]$. In other words, we only consider a $676/1400 = 0.48$ fraction of galaxies that pass the HSC-SSP selection function. As the HSC-SSP selection function described above is only an approximation, we do not include HSC-SSP satellite observations in our main analysis. However, we discuss the forward modelled predictions for HSC-SSP in Section~\ref{ssec:fut_current_surveys_lf}.

In the Big Sky approach for the LSST\footnote{\url{https://www.lsst.org/}} survey described in \citet{lsst.footprint.2018}, the survey footprint will cover the area the entire equatorial declination range of $\delta < 30^{\circ}$. However, as central portions of the Galactic plane within this footprint will be removed, we approximate this by considering only the high-Galactic-latitude range ($|b| > 15^{\circ}$) within the LSST footprint. This corresponds to a total footprint area of $\sim$23500$\,\rm deg^2$. 

\section{Model Results and Comparisons to Observations}
\label{sec:comp_to_obs}

Here we present results for the population of satellite galaxies resulting from forward-modelling described in the previous section. When possible, we compare model predictions with observed satellite population in the DES, PS1, and SDSS surveys. The list of observed dwarf galaxies used in these comparisons is taken from the compilation in \citet{DrlicaWagner.etal.2020}. Note that our comparison sample of the MW satellites excludes the recently discovered massive satellites Antlia II (\citealt{Torrealba.etal.2019}) as it is not detected in any of these surveys. Also, even though satellites Kim 2 and Crater I/Laevens 1 have $\rhalf > 10$ pc and have been detected with high significance in the DES and PS1 surveys respectively, we do not include them in our satellite samples because these systems are considered to be likely star clusters (\citealt{Kirby.etal.2015,Weisz.etal.2016}). As a result, these two objects have been listed among probable star clusters in Table~\ref{tab:star_clusters}. The exclusion/inclusion of a single satellite in the DES and PS1 satellite samples respectively does not change our conclusions.   
 
We begin by comparing luminosity function and radial distribution of satellites with respect to halo centre predicted by the forward modelling to observations in Sections~\ref{ssec:lf} and \ref{ssec:rad_dist}, respectively. Unless stated otherwise, in all of the comparisons with observations we use the two hosts with LMC analogs, as discussed in Section~\ref{ssec:lmc_select}. We explore sensitivity of these statistics to reionization redshift of the Milky Way volume, $\zrei$ (Section~\ref{ssec:lf_reion}), and to the presence of the LMC-like object (Section~\ref{sssec:lf_lmc}), and derive the approximate range of $\zrei$ consistent with observed abundance of ultra-faint galaxies (Section~\ref{ssec:mw_reion}). We use this range of $\zrei$ to derive the implied relations between galaxy stellar mass and peak halo mass in our model (Section~\ref{ssec:mw_shmr}) and halo occupation distribution for galaxies with $\Mv \leq 0$ (Section~\ref{ssec:mw_hof}). Furthermore, we discuss the temporal variation of satellite radial distributions, under the influence of LMC infall for instance, in Section~\ref{ssec:time_radials}.

Given the importance of size  (or, equivalently, surface brightnesses) for a galaxy of a given luminosity for its detectability, we also consider distribution of $\rhalf$ sizes of dwarf satellites in Section~\ref{ssec:gal_sizes}.

\subsection{Luminosity Functions}
\label{ssec:lf}

In this section we present comparisons of predicted luminosity functions for different observational surveys and their dependence on the reionization epoch. 

\subsubsection{Luminosity functions and redshift of reionization}
\label{ssec:lf_reion}

\begin{figure*}
\centering
\includegraphics[width = \textwidth]{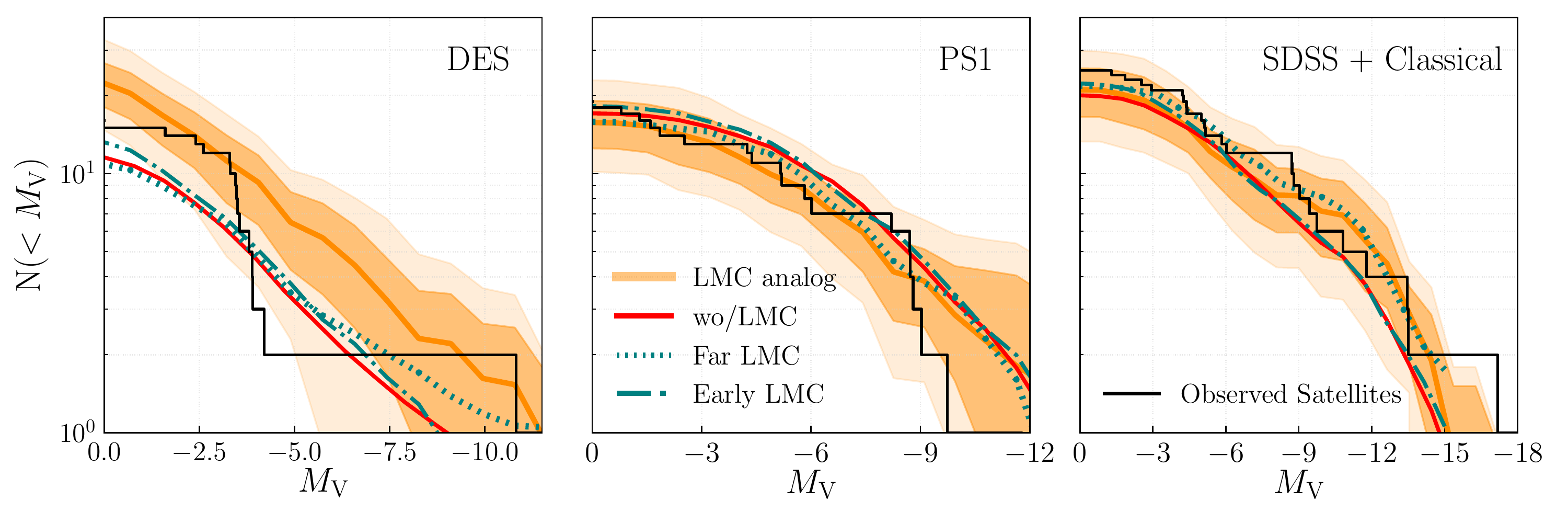}
\caption{Predicted satellite luminosity function for $\zrei = 8.5$ within $\dhelio < 300$ kpc in the DES (\textit{left}), PS1 (\textit{centre}) and SDSS $+$ Classical (\textit{right}) samples for different LMC scenarios. The orange and red solid lines shows the median luminosity function for hosts with `LMC analog' hosts and `without LMC' respectively. The orange contours carry the same meaning as in Figure~\ref{fig:fm_all_lfs} and the red dashed-lines bracket the 68\% confidence intervals of the `without LMC' scenario. The teal colored dotted and dot-dashed lines correspond to the two alternative LMC scenarios considered in Section~\ref{sssec:lf_lmc}. Our forward modelling results prefer a LMC that is on its first infall and near its pericentre (`LMC analog' scenario). }
\label{fig:lmc_nolmc_lf}
\end{figure*}

Figure~\ref{fig:fm_all_lfs} shows the predicted satellite luminosity function for the DES, PS1 and SDSS + Classical satellites in models with reionization redshifts of $\zrei =  7.5, 8.5, 9.5, 10.5$ and compares them with the observed LF in each survey. The colored lines with different line styles show the median luminosity functions for the respective reionization redshift $\zrei$, while the dark (light) shaded regions denote the 68\% (95\%) confidence intervals. 

As was discussed above, reionization is known to directly impact the amplitude of the faint end of the luminosity function. 
This is expected because the UV background after reionization heats the intergalactic medium and prevents gas accretion onto haloes below a certain time dependent characteristic mass scale \citep[e.g.,][]{Gnedin.2000,Okamoto.etal.2008}. This suppresses buildup of stellar mass in low mass haloes and consequently the number of faint satellites. 
This can be seen in Figure~\ref{fig:fm_all_lfs}, which shows that the amplitude of satellite LF at $\Mv \gtrsim -6 $ is increasingly suppressed for higher $\zrei$. The trend with $\zrei$ reflects higher degree of suppression when haloes fall below the critical mass for accretion earlier.  haloes of mass larger than critical mass ($\Mpeak \gtrsim 10^{9} \Msun$) are nearly insensitive to UV heating, and satellite LF is thus insensitive to $\zrei$ for $\Mv \lesssim -6$. The bright end of the luminosity function, however, is sensitive to galactic outflow modelling \citep{Kravtsov.Manwadkar.2022}. 

It is interesting to point out that the luminosity of $\Mv \gtrsim -8$ used to delineate UFDs and classical dwarf galaxies \citep[e.g.,][]{Simon.2019} is very close to the luminosity below which galaxies are significantly affected by UV heating and reionization redshift. This reiterates the importance of UFDs as probes of reionization. 

The faint end LF amplitude  can thus be used to probe the redshift of reionization of the Lagrangian region of the Milky Way. Results shown in Figure~\ref{fig:fm_all_lfs} indicate that $\zrei$ in the considered range are in general agreement with observed LF. The $\zrei\sim 8$ model matches the observed LF most closely and we will consider this model as our fiducial model in this study.  We return to how $\zrei$ for the MW environment could be constrained in more detail in Section~\ref{ssec:mw_reion}.

\subsubsection{Effect of the Large Magellanic Cloud}
\label{sssec:lf_lmc}

A number of UFDs have been discovered around the Magellanic Clouds \citep[e.g.,][]{Koposov.etal.2015,Koposov.etal.2018.lmcsat,DrlicaWagner.etal.2015.lmcsat,DrlicaWagner.etal.2016.lmcsat,Torrealba.etal.2018.lmcsat,Cerny.etal.2021} consistent with the hypothesis that the Magellanic Clouds arrived with their own population of dwarf galaxy satellites  \citep[][]{Donghia.Lake.2008}. Indeed, \citet{Nadler.etal.2020} recently showed that the presence of the LMC has a significant impact on the MW satellite population, especially in the vicinity of the LMC. It is therefore important to account for the presence of LMC when forward modelling satellite populations in surveys with footprint near this object. 

To study the LMC's impact on satellite population, we consider the satellite LF and radial distribution (in Section~\ref{ssec:lmc_radials}) in four different LMC scenarios: i) our fiducial case of hosts with a LMC analog as defined in Section~\ref{ssec:lmc_select} (LMC analog), ii) hosts without any LMC mass object accreted (wo/LMC), iii) hosts that accreted a LMC early on so it has undergone multiple pericentric passages by $z=0$ (Early LMC) and iv) hosts that have an LMC mass object on its first infall, but at distances $> 100$ kpc (`Far LMC'). For the ``Early LMC'' and ``Far LMC'' scenarios we rotate the reference frame such that the LMC halo matches with true sky coordinates of LMC. For the host halos without LMC, we assume a random orientation of the host halo relative to a survey footprint.  

Figure~\ref{fig:lmc_nolmc_lf} shows the predicted satellite luminosity functions at $\zrei = 8.5$ in the DES, PS1 and SDSS$+$Classical samples for the four different LMC scenarios. There is no statistically significant impact of the LMC on the PS1 and SDSS+Classical satellite luminosity function, except the trivial effect that brightest satellites are lacking in the SDSS+Classical model in hosts without LMC mass satellites. However, the figure shows that presence of a LMC analog boosts the amplitude of the faint-end LF predicted for the DES survey. 
This is not surprising as LMC lies close to the DES footprint (see, e.g., Figure 1 in \citealt{DrlicaWagner.etal.2020}).  
Indeed, if we reflect satellite positions in declination in the LMC hosts, effectively moving LMC analog away from the DES footprint, the amplitudes of the DES satellite LFs for hosts with and without LMC (wo/LMC) become similar. This explicitly confirms that LMC has significant impact on the DES satellite population due to its proximity to the DES footprint. 

Furthermore, we find that the `Far LMC' and `Early LMC' have little impact on satellite LF and produce results similar to the host halos without LMC-sized subhalo.  This is because in the `Far LMC' scenario the LMC associated satellite population is still far from the host and hence we observe fewer satellites due to observational survey selection limits. In the 'Early LMC' scenario, the LMC has undergone multiple pericentre passages and so a significant fraction of the associated satellite population has been disrupted or dispersed of into the MW halo \citep[e.g.,][]{Deason.etal.2015}, reducing the over-density of satellites around the LMC. 

\subsubsection{Implications for Reionization of the Local Environment}
\label{ssec:mw_reion}

\begin{figure}
    \centering
    \includegraphics[width = \columnwidth]{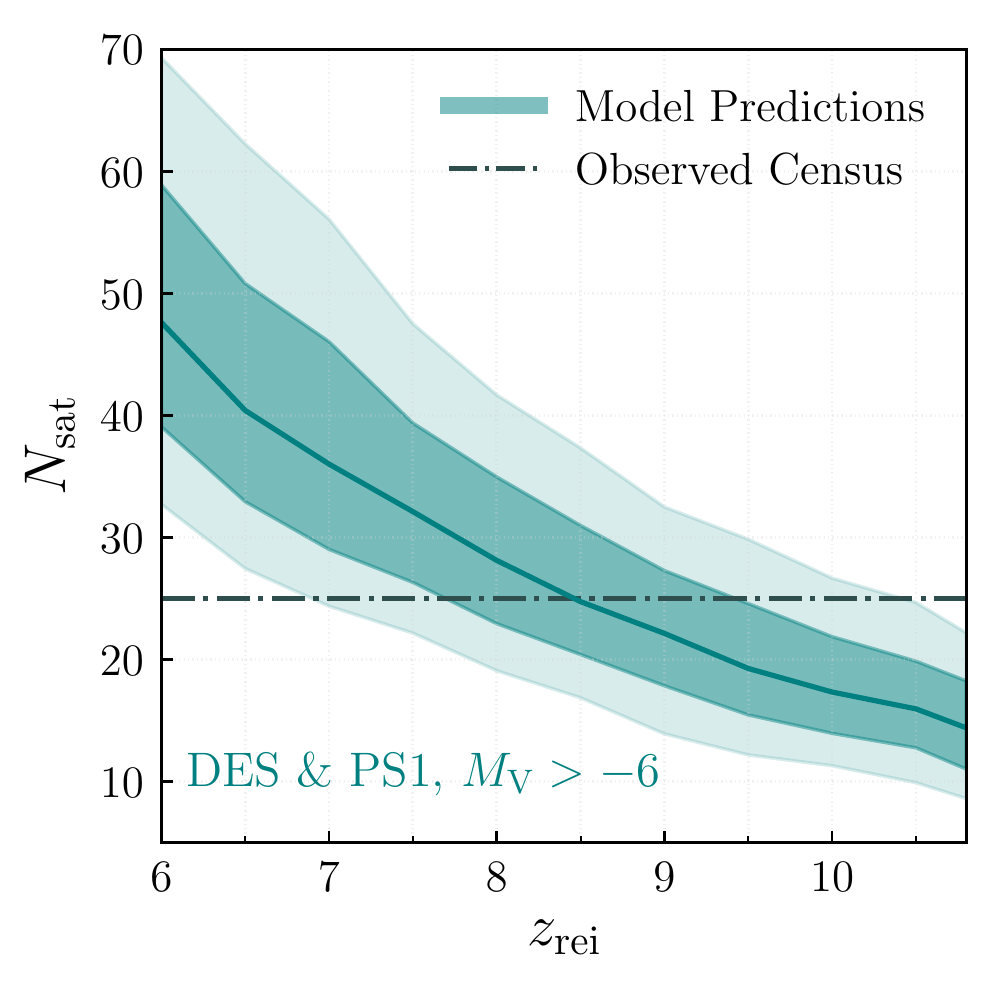}
    \caption{The number of satellites with $\Mv > -6$ and $\dhelio < 300$ kpc predicted to be observed in total in the DES and PS1 surveys as a function of reionization redshift $\zrei$. Note that $\zrei$ here is reionization of the Local Volume, which can be different from the mean reionization redshift of the Universe. The horizontal dot-dash line ($N_{\rm sat} = 25$) shows the number of satellites with $\Mv > -6$ observed in the DES and PS1 surveys above a certain detection threshold as considered in selection function. Even though the $\zrei$ consistent with observed satellite abundance  is quite wide, it is in good agreement with the reionization redshift range of $\zrei = 8.1^{+0.4}_{-0.4}$ (95\% confidence intervals) derived for the MW mass haloes by \citet{Zhu.etal.2019} using simulations of reionization.}
    \label{fig:zrei_nsat}
\end{figure}

Given that the redshift of reionization directly impacts the number of UFD satellites, the observed number of MW UFD satellites can be used to constrain the reionization redshift of the Lagrangian volume of the Milky Way. To this end, we use the total number of UFDs observed in the DES and PS1 surveys and identify the range of $\zrei$ for which predictions of the forward model are consistent with the observed number. 

Figure~\ref{fig:zrei_nsat} shows the number of luminous satellites at $\Mv > -6$ we would expect to observe in the DES and PS1 as a function of $\zrei$. The scatter about the median predicted $N_{\rm sat} - \zrei$ relation is dominated by the host-to-host scatter in the satellite (subhalo) counts, even after accounting for uncertainty in the MW halo mass as described in Section~\ref{ssec:host_halo_mass}. 

Model predictions are consistent with the observed UFD number for reionization redshifts of $\zrei \sim 7 - 10$. The actual uncertainty in $\zrei$ will be slightly larger because we do not account for uncertainties in modelling of galactic outflows, which will impact the number of luminous satellites. Even though our inferred $\zrei$ range is quite wide, it is interesting to see that our inferred $\zrei$ range agrees with the reionization redshift range of $\zrei = 8.1^{+0.4}_{-0.4}$ (95\% confidence intervals) derived for the MW mass haloes by \citet{Zhu.etal.2019} using the CROC simulations of reionization. Our constraint is also in qualitative agreement with results of the CoDa II simulations which show that outer regions of the MW and M31 Lagrangian regions are reionized at $z\gtrsim 7$ \citep[][see their Fig. 11]{Ocvirk.etal.2020} and with the general expectation that the Lagrangian regions around progenitors of the present-day collapsed structures reionize at higher redshifts than the low-density regions and the universe as a whole \citep[e.g.,][]{Trac.etal.21}. 

The sensitivity of the UFD abundance to reionization shows that this is one avenue to probe reionization in the local volume when census of the UFDs will improve with larger and deeper surveys like the DELVE-WIDE with the Dark Energy Camera and LSST survey at the Vera C. Rubin Observatory.

\subsection{Implications for the Galaxy-Halo Connection} 
\label{ssec:mw_ghconn}

\subsubsection{Relations between peak halo mass, stellar mass and $V$-band luminosity}
\label{ssec:mw_shmr}

\begin{figure}
    \centering
    \includegraphics[width = \columnwidth]{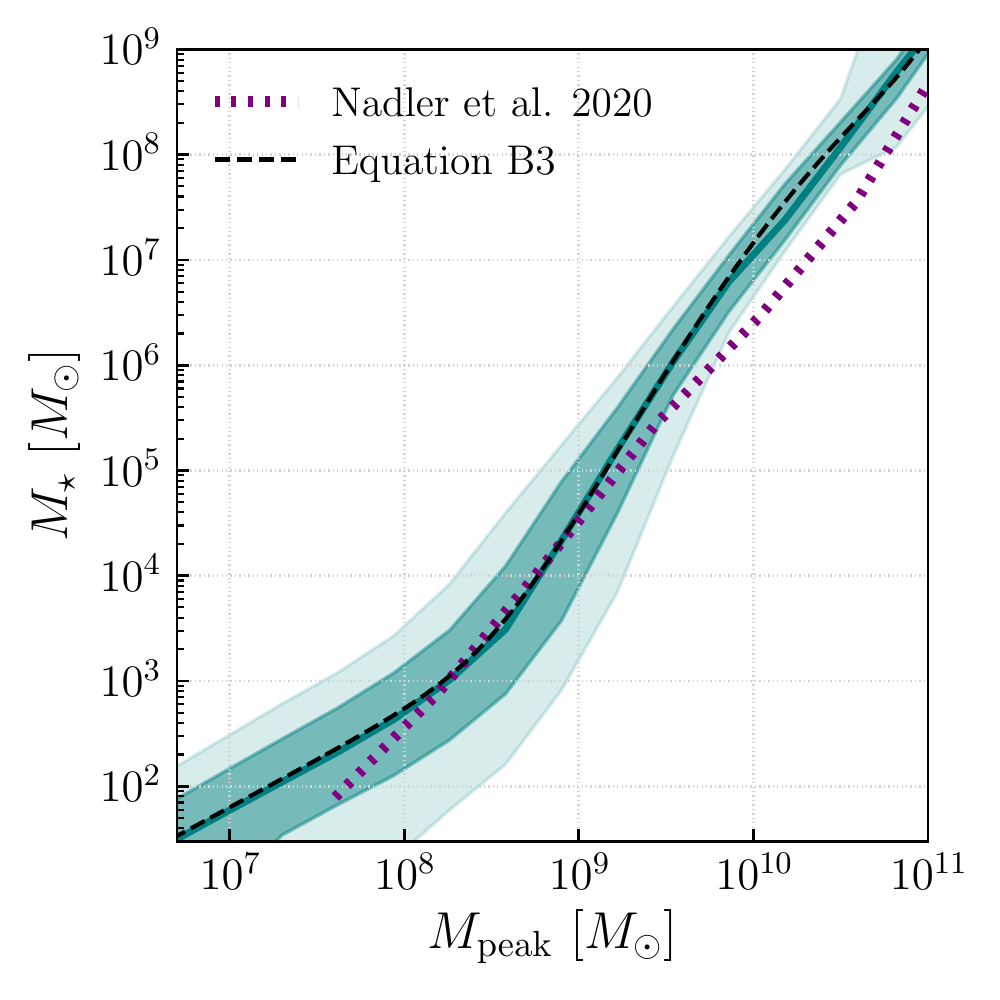}
    \caption{Model prediction for the relation between peak halo mass $\Mpeak$ and stellar mass $M_\star$ for the Milky Way satellite galaxies. The solid colored line denotes the median relation while the dark (light) shaded region correspond to the 68\% (95\%) scatter around the median relation. The purple dotted line shows the best fit relation derived from fitting the MW satellite population by \citet{Nadler.etal.2020}. The black dashed line shows the approximation to the median relation given by Eqs.~\ref{eqn:new_shmr} in Appendix~\ref{app:shmrapprox}. Note that the significant down scatter in this relation at masses $\Mpeak \lesssim 10^9 \Msun$ is due to some fraction of haloes not forming any significant amount of stars. }
    \label{fig:shmr_mw}
\end{figure}

Figure~\ref{fig:shmr_mw} shows relation between peak halo mass $\Mpeak$ and stellar mass $\Ms$. $\Mpeak$ is defined as the maximum mass  that a given halo achieved during its evolution at some redshift $z_{\rm peak}$ and it is measured within the radius enclosing density contrast of $200$ relative to the critical density at $z_{\rm peak}$. The scatter around the median $\Mpeak-\Ms$ relation shown in the figure reflect both halo-to-halo scatter and the range of $\zrei$ consistent with observed abundance of dwarf satellites (Figure~\ref{fig:zrei_nsat}). The figure shows that the $\Mpeak-\Ms$ relation arising in the model {\it is not} a single power law. This is because its slope at $\Mpeak\gtrsim 10^{10}\, M_\odot$  ($\Mv \lesssim -11$) mainly reflects the scaling of the wind mass loading factor with stellar mass, while at smaller masses the slope is determined by suppression of gas accretion due to UV heating after reionization \citep[see][for detailed discussion]{Kravtsov.Manwadkar.2022}. The flattening of the relation at the smallest $\Mpeak$ values reflects the increasing inefficiency of supernovae driven winds in blowing out gas from these galaxies \cite[see also][]{Chen.etal.2022}. 

The dashed line in Fig.~\ref{fig:shmr_mw} shows an accurate analytical approximation to the median $\Mpeak-\Ms$ of the model, which is described in the  Appendix~\ref{app:shmrapprox}. The figure also compares the relation in our model with the relation derived by \citet{Nadler.etal.2020} using observed abundance of satellites in the DES and PS1 surveys. In their analysis, a power law form of the $\Mpeak-\Ms$ was assumed with normalization that was fixed at $\Mpeak\gtrsim 10^{10}\, M_\odot$ to the normalization derived using abundance matching with observed stellar mass function. Overall, the relation in our model is close to the power law relation derived by \citet{Nadler.etal.2020}, but with some differences due to deviations from the power law described above and in amplitude at the largest satellite masses (see Appendix~\ref{app:shmrapprox} for additional comparison). As we show in Figure~\ref{fig:shmr_comp_sim}, our relation is also consistent with recent, state-of-the-art hydrodynamic simulations of dwarf galaxies \citep[e.g.,][]{Applebaum.etal.2021,Orkney.etal.2021,Santistevan.etal.2022,Gutcke.etal.2022,Hopkins.etal.2022} at the halo mass range probed in these simulations.

As our comparisons with observed satellite LF show, the non-power law $\Mpeak-\Ms$ relation in our model is consistent with observed distribution of satellite luminosities. We note that the current data is not sufficient to reliably exclude a single power-law $\Mpeak-\Ms$ relation \citep{Nadler.etal.2020}, although as we show in Section~\ref{ssec:fut_current_surveys_lf} it is easier to explain the abundance of satellite in the HSC-SSP survey with the  flattening of the $\Mpeak-\Ms$ relation in our model. 

\subsubsection{Halo Occupation Distribution of $M_V<0$ Galaxies}
\label{ssec:mw_hof}

\begin{figure}
    \centering
    \includegraphics[width = \columnwidth]{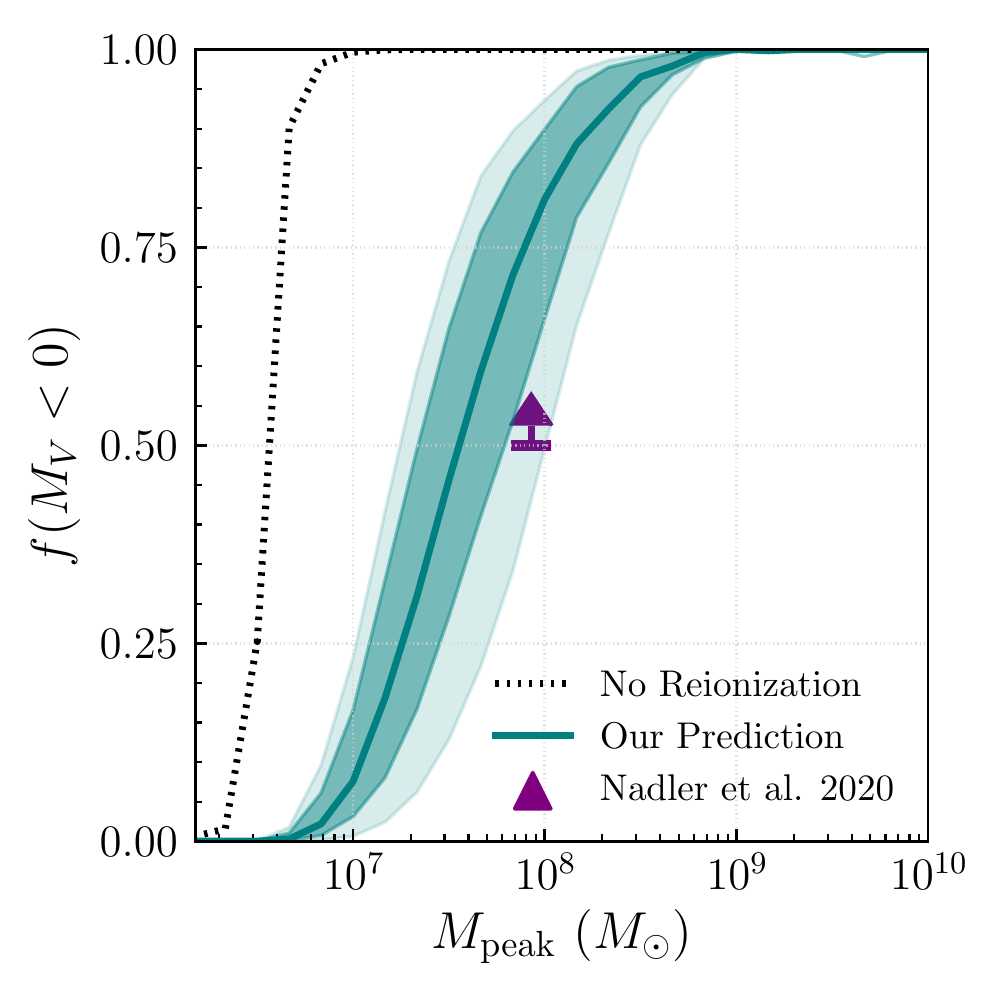}
    \caption{Fraction of subhaloes that host galaxies with absolute V-band magnitudes of $\Mv < 0$ as a function of peak halo mass $\Mpeak$. The dark (light) shaded regions show the 68\% (95\%) confidence intervals on our predictions for the halo occupation fraction for MW satellites. The blue upward arrow shows the 95\% confidence for the peak halo mass that hosts 50\% of detectable satellites from \citet{Nadler.etal.2020}. For comparison sake, we also show the median halo occupation fraction (black dotted line) if reionization was not modeled.}
    \label{fig:halo_occ}
\end{figure}

\begin{figure*}
    \centering
    \includegraphics[width = \textwidth]{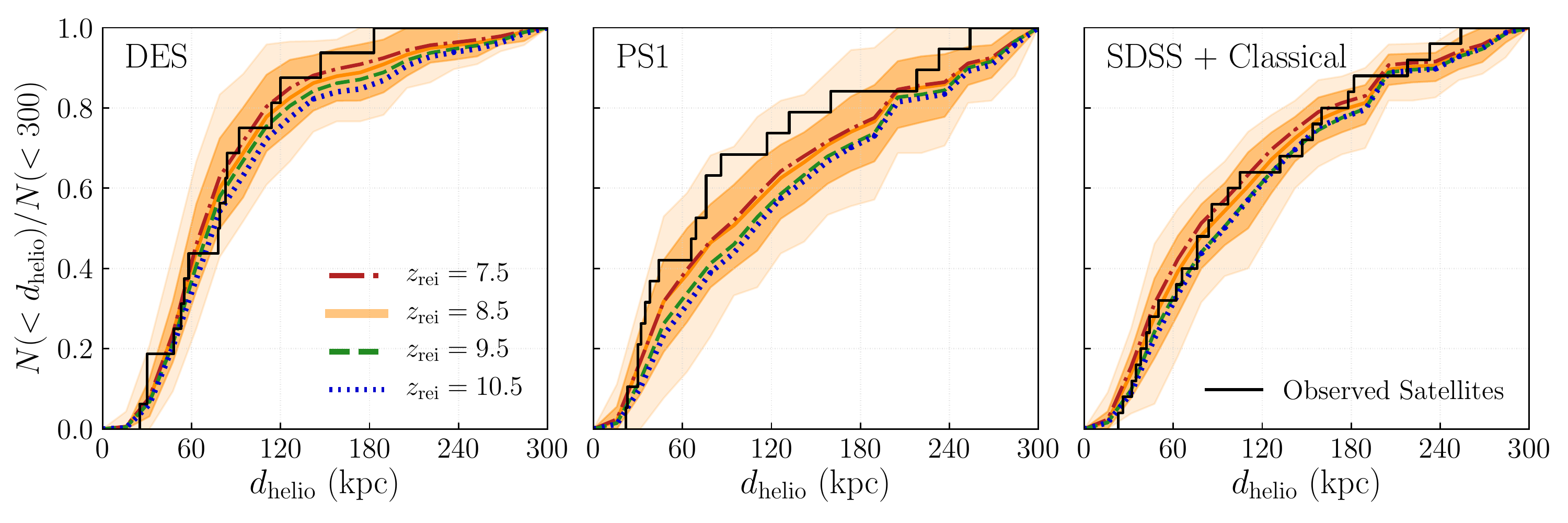}
    \caption{Comparisons of the model and observed distribution of the heliocentric distance of MW satellites in the DES (\textit{left}), PS1 (\textit{centre}) and SDSS + Classical (\textit{right}) as a function of different reionization redshift $\zrei = 7.5,8.5,9.5,10.5$. The distributions are normalized to the total number of satellites within the heliocentric distance of 300 kpc. The solid black line shows the observed satellite radial distribution in each of the surveys, while colored lines show model predictions for different $\zrei$. The dark (light) orange shaded regions for $\zrei=8.5$ correspond to the 68\% (95\%) of the model predictions.}
    \label{fig:fm_radials}
\end{figure*}

Suppression of gas accretion and star formation due to UV heating affects luminosities of galaxies hosted in dark matter subhaloes and thus their occupation distribution. In \citet{Kravtsov.Manwadkar.2022} we showed that the fraction of haloes at a given $\Mpeak$ that host a luminous satellite ($\Mv < 0$) is sensitive to the redshift of reionization for $\Mpeak\lesssim 10^9\,\Msun$. Thus, the range of $\zrei$ favored by the observed number of the DES and PS1 dwarf satellites implies a specific constraint on the halo occupation fraction that we show in Figure~\ref{fig:halo_occ}. The shaded regions around the median shown by the shaded region correspond to the 68\% (95\%) confidence intervals in the halo occupation fraction reflecting the corresponding regions of the $\zrei$ range. In addition, we show 95\% confidence upper bound\footnote{Note that lower $\mathcal{M}_{50}$ corresponds to higher halo occupation and vice-a-versa.} on the halo mass corresponding to $f(M_V<0)=0.5$ of $\mathcal{M}_{50} < 8.5 \times 10^{7}\Msun$ derived by \citet{Nadler.etal.2020}. The model consistent with the observed number of DES and PS1 satellites corresponds to $\mathcal{M}_{50} = 3.6^{+2.6}_{-1.4} \times 10^7 \Msun$  (68\% confidence intervals).

This constraint is qualitatively consistent with results of previous studies \citep[e.g.,][]{Maccio.etal.2010,Jethwa.etal.2018,Nadler.etal.2020}, which concluded that a significant fraction of UFDs must be hosted by subhaloes with $\Mpeak\lesssim 10^8\, M_\odot$. In particular, as can be seen in the figure, the $f(M_V<0)$ of our model is consistent with the $\mathcal{M}_{50}$ of \citet{Nadler.etal.2020}. However, the range of halo masses hosting $M_V<0$ galaxies allowed by the data that was quantified in the latter study is quite broad (reflecting limitations of the current satellite sample). The current model predicts the HOD form and $\mathcal{M}_{50}$ range that are quite narrow. These reflects the specific assumptions made our model. 

The halo occupation fraction implied by observations {\it in the context of our model}, indicates that observed UFD satellites of the MW form in haloes as small as $\sim 10^7\,\Msun$, which is below the putative atomic cooling limit  \citep[e.g.,][]{Haiman.etal.1997,Barkana.etal.1999}. 
We discuss this further in Section~\ref{ssec:total_sat_mw}.

\subsection{Radial Distributions}
\label{ssec:rad_dist}

As discussed in Section~\ref{sec:intro}, radial distribution of satellites with respect to host halo centre is expected to be sensitive to subhalo disruption modelling, details of merger history, observational biases, and possibly even reionization. In this section, we present a comparisons of predicted radial distributions for different observational surveys and their dependence on reionization epoch and presence of the LMC.

\subsubsection{Radial distribution of satellites and redshift of reionization}

Figure~\ref{fig:fm_radials} shows the predicted satellite radial distributions for the DES, PS1 and SDSS+Classical satellites for different values of $\zrei$ and radial distributions of observed satellites in these surveys, both normalized to the number of detected satellites within heliocentric distance of 300 kpc. Overall, we see that the radial distributions are in good agreement with observations and this agreement is achieved without modelling orphan galaxies to account for subhaloes missing due to numerical effects. 

For a more quantitative assessment, Table~\ref{tab:ks_vals} presents the $p$-values of the Kolmogorov-Smirnov (KS) test for the radial distributions predicted by the model with different $\zrei$ compared to observed radial distributions in different surveys. The $p$-values for models with $\zrei\sim 7-10$ are generally high in all survey comparisons for models that use host haloes with LMC analogs, which indicates that the radial distribution of model galaxies is in good agreement with observed distribution of the MW satellites. 

As discussed in Section~\ref{ssec:lf_reion}, earlier reionization (higher $\zrei$) results in an increased suppression in the number of luminous satellites. This can be seen in Figure~\ref{fig:fm_radials} where the amplitude of the radial distribution decreases with increasing $\zrei$. 
However, the overall shape or concentration of the radial distribution does not change with $\zrei$. This is not surprising because the model assumes the same $\zrei$ for all subhaloes regardless of their spatial location or accretion redshift. Although reionization, in general, is expected to proceed inhomogeneously \citep[e.g.,][]{ocvirk.etal.2014} and satellites closer to the MW could be reionized earlier, mixing of satellite orbits within MW will weaken any such trend considerably. Moreover, the difference in $\zrei$ across Lagrangian region of the Milky Way is expected to be only $\Delta \zrei\lesssim 1$ \citep[e.g., see Figure 11 in][]{Ocvirk.etal.2020} which would make any such trend with $\zrei$ small.

\begin{table}
\begin{center}
 \caption{Kolmogorov-Smirnov (KS) test results for model predicted radial distribution compared to observed radial distribution in the DES, PS1 and SDSS + Classical satellite samples.}
 \label{tab:ks_vals}
 \begin{tabular}{lcc}
  \hline
  Model & $p$-value \\
  \hline
  DES, LMC analog, $\zrei = 7.5$ & 0.65 \\
  DES, LMC analog, $\zrei = 8.5$ & 0.73 \\
  DES, LMC analog, $\zrei = 9.5$ & 0.90 \\
  DES, LMC analog, $\zrei = 10.5$ & 0.85 \\
  DES, \textit{without} LMC, $\zrei = 8.5$ & 0.40 \\
  DES, \textit{far} LMC, $\zrei = 8.5$ & 0.01 \\
  DES, \textit{early} LMC, $\zrei = 8.5$ & 0.54 \\
  \hline
  PS1, LMC analog, $\zrei = 7.5$ & 0.46 \\
  PS1, LMC analog, $\zrei = 8.5$ & 0.34 \\
  PS1, LMC analog, $\zrei = 9.5$ & 0.21 \\
  PS1, LMC analog, $\zrei = 10.5$ & 0.13 \\
  PS1, \textit{without} LMC, $\zrei = 8.5$ & 0.03 \\
  PS1, \textit{far} LMC, $\zrei = 8.5$ & 0.02 \\
  PS1, \textit{early} LMC, $\zrei = 8.5$ & 0.08 \\
  \hline
  SDSS + Classical, LMC analog, $\zrei = 7.5$ & 0.95 \\
  SDSS + Classical, LMC analog, $\zrei = 8.5$ & 0.99 \\
  SDSS + Classical, LMC analog, $\zrei = 9.5$ & 0.96 \\
  SDSS + Classical, LMC analog, $\zrei = 10.5$ & 0.93\\
  SDSS + Classical, \textit{without} LMC, $\zrei = 8.5$ & 0.49 \\
  SDSS + Classical, \textit{far} LMC, $\zrei = 8.5$ & 0.07 \\
  SDSS + Classical, \textit{early} LMC, $\zrei = 8.5$ & 0.92 \\
  \hline
 \end{tabular}
 \end{center}
\end{table}

\subsubsection{Effect of the Large Magellanic Cloud}
\label{ssec:lmc_radials}

\begin{figure*}
    \centering
    \includegraphics[width = \textwidth]{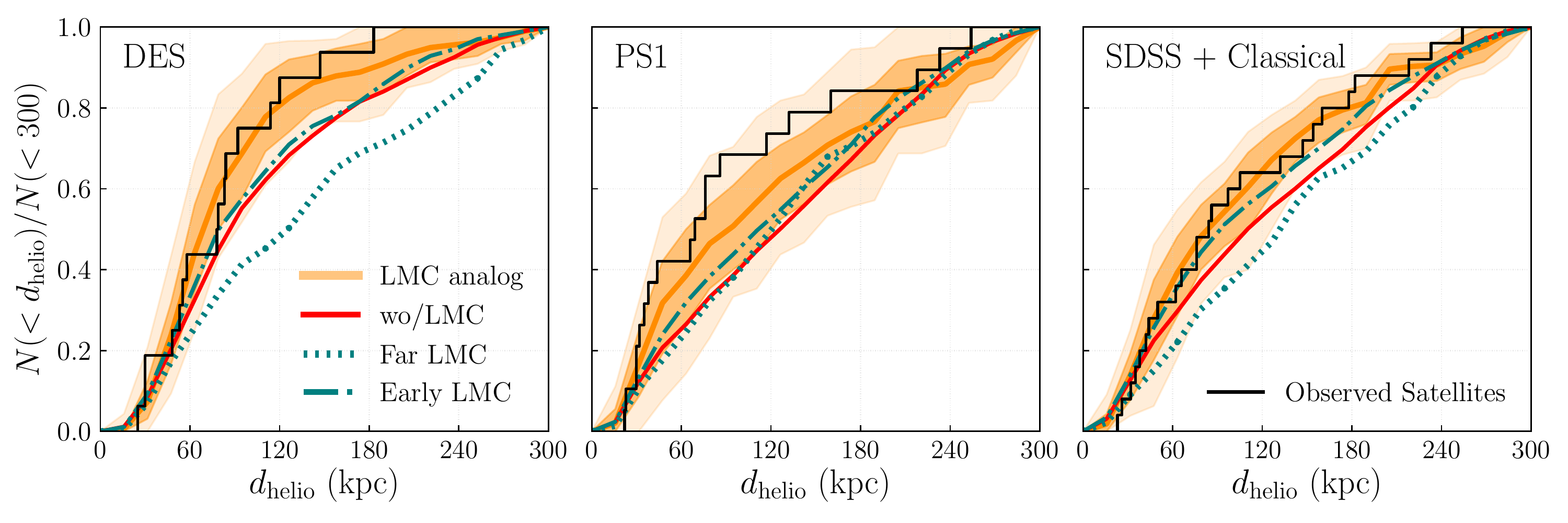}
    \caption{Predictions for the satellite radial distribution in DES (\textit{left}), PS1 (\textit{centre}) and SDSS + Classical (\textit{right}) satellite samples for $\zrei = 8.5$ for different LMC scenarios. Similar to Figure~\ref{fig:lmc_nolmc_lf}, the orange and red lines show the show the radial distributions for hosts with `LMC analog' and `without LMC' respectively. The teal colored dotted and dot-dashed lines correspond to `Far LMC' and `Early LMC' scenario as discussed in Section~\ref{sssec:lf_lmc}. Similar to luminosity functions, our forward modelling results prefer a LMC that is on its first infall and near its pericentre (`LMC analog' scenario).   }
    \label{fig:lmc_rads}
\end{figure*}

\begin{figure*}
    \centering
    \includegraphics[width = \textwidth]{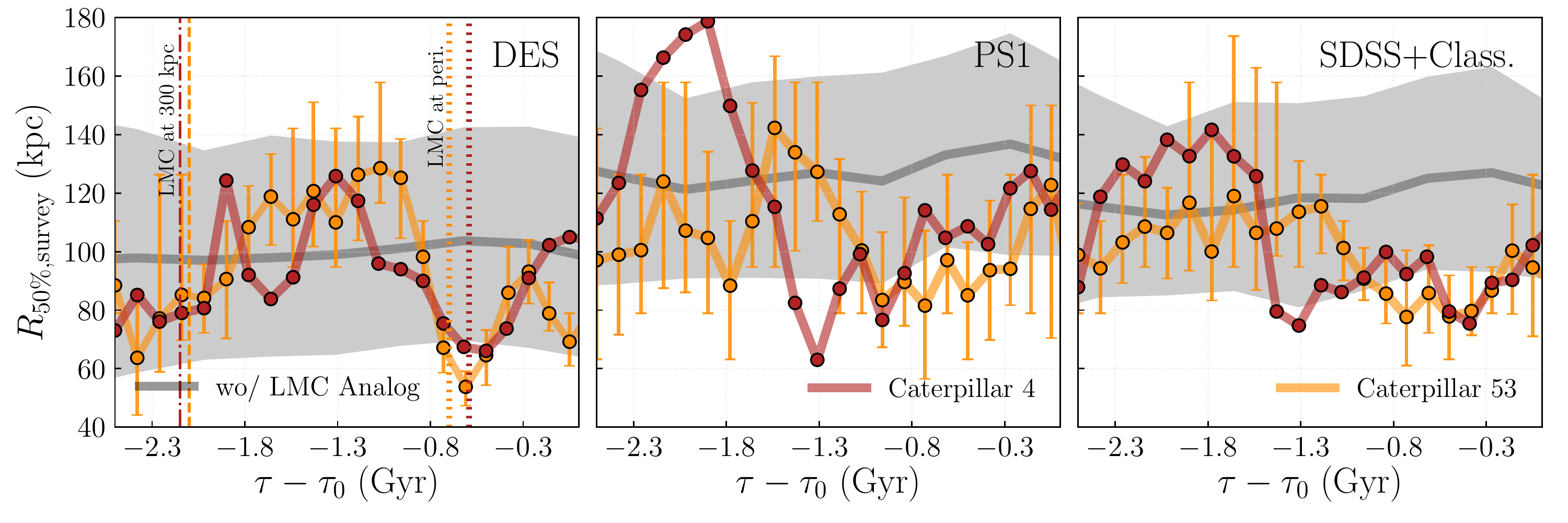}
    \caption{The temporal variation in the concentration of the predicted satellite radial distributions ($R_{\rm 50\%, survey}$) for $\zrei = 8.5$ in the DES (\textit{left}), PS1 (\textit{middle}) and SDSS+Classical (\textit{right}) samples. Note that $\tau_{0}$ is the age of the Universe in Gyr and $R_{\rm 50\%, survey}$ is the heliocentric radius at which half of the detected satellites within $\dhelio < 300$ kpc are enclosed. The two colored lines (red and orange) show the variation for the two hosts with LMC analogs. The 1$\sigma$ uncertainties on $R_{\rm 50\%, survey}$ are only shown for Caterpillar 53, however, the uncertainties for Caterpillar 4 are of similar scale. The solid grey line and shaded grey region show the median and 68\% confidence interval on the variation of concentration in hosts with no LMC analogs. The vertical colored dot-dashed and dotted lines show the epoch when the LMC analog is crossing $300$ kpc and at pericentre respectively. One observes a significant decrease in $R_{\rm 50\%, survey}$ of DES satellites as the LMC approaches pericentre. }
    \label{fig:temporal_rads} 
\end{figure*}

As with the satellite LF, LMC has a significant impact on the predicted satellite radial distribution for the DES, and a much smaller impact on the predicted distributions for the PS1 and SDSS + Classical samples. In addition to the amplitude of the radial distribution, the presence of the LMC affects concentration of the radial distribution of the DES satellites, as can be seen in the comparison of the normalized radial distributions in Figure~\ref{fig:lmc_rads}.

The effect of the LMC on the radial distribution does depend on the location and accretion redshift of the LMC analog. In the `Early LMC' analogs (see Section~\ref{ssec:lmc_select} for details on different LMC scenarios) already experienced multiple pericentre passages. The satellites brought by the LMC analog are likely already dispersed into the MW halo after $\gtrsim 2$ pericentre passages \citep[e.g.,][]{Deason.etal.2015}. In the `far LMC' analogs, the majority of the LMC satellites are located at large distances and thus would not be detectable  \citep[e.g.,][]{Zhang.etal.2019,Bose.etal.2020}, which minimizes their impact. Indeed the `Far LMC' analogs have the least concentrated satellite radial distribution among different LMC analogs. Not surprisingly, the `LMC analog' scenario in which the current distance to the LMC is matched agrees best the observed DES satellite radial distribution. 

The agreement of the radial distributions in hosts with different LMC analogs can be quantitatively assessed using the KS $p$-values  in Table~\ref{tab:ks_vals}. In the DES survey, where the effect of LMC is expected to be most noticeable, the observational data clearly prefers the `LMC analog' scenario over the other LMC scenarios we consider here. In other words, our forward modelling framework along with the DES satellite observations provide strong evidence for a LMC that is on its first infall into the MW and is very close to its pericentre, in good agreement with conclusions of \citet{Nadler.etal.2020}. Interestingly, even though a difference in amplitudes of satellite luminosity functions for PS1 and SDSS + Classical was not seen, the radial distributions for these satellites are sligthly more concentrated in the 'LMC analog' scenario, as can been seen from Figure~\ref{fig:lmc_rads} and $p$-values in Table~\ref{tab:ks_vals}.

In conclusion, predictions of our forward modelling of the MW satellite population are in excellent agreement with the observed satellite radial distributions in different surveys. Additionally, we show that the radial distribution of the observed MW satellites clearly prefers the scenario in which LMC is on its first infall, is close to the pericentre, and has contributed a significant number of UFD satellites to the MW satellite population. 

.
\subsubsection{Temporal variation of satellite radial distributions}
\label{ssec:time_radials}

Due to the relatively small number of observed satellites in a given survey and their respective motions within the MW halo, the observed satellite radial distributions could experience significant fluctuations over relatively short time intervals of $\sim$1 Gyr. When comparing to observations we used model predictions for a single time epoch $z=0$. However, in the case of the LMC-analog hosts, we present results at the epoch ($z=0.057$) when LMC-analog is on its first infall and is at a distance closest to 50 kpc. 

We examine temporal variations of the radial distribution of satellites using  $R_{\rm 50\%, survey}$ -- the heliocentric distance enclosing half of the detected satellites within 300 kpc in a survey. To compute the $R_{\rm 50\%, survey}$ at each epoch, we follow the same forward modelling procedure as described in Section~\ref{ssec:mock_obs} using the $\Mv$ and $\rhalf$ of the subhaloes at $z=0$. We do this because we wish to focus on the satellite motion in MW halo and so decouple the possible systematic of evolution in satellite observable properties. Note that for Caterpillar 4 and Caterpillar 53 (the host haloes with LMC analogs), we shift our reference frame/survey pointings such that the it matches with the true sky co-ordinates of the LMC.

\begin{figure*}
    \centering
    \includegraphics[width = \textwidth]{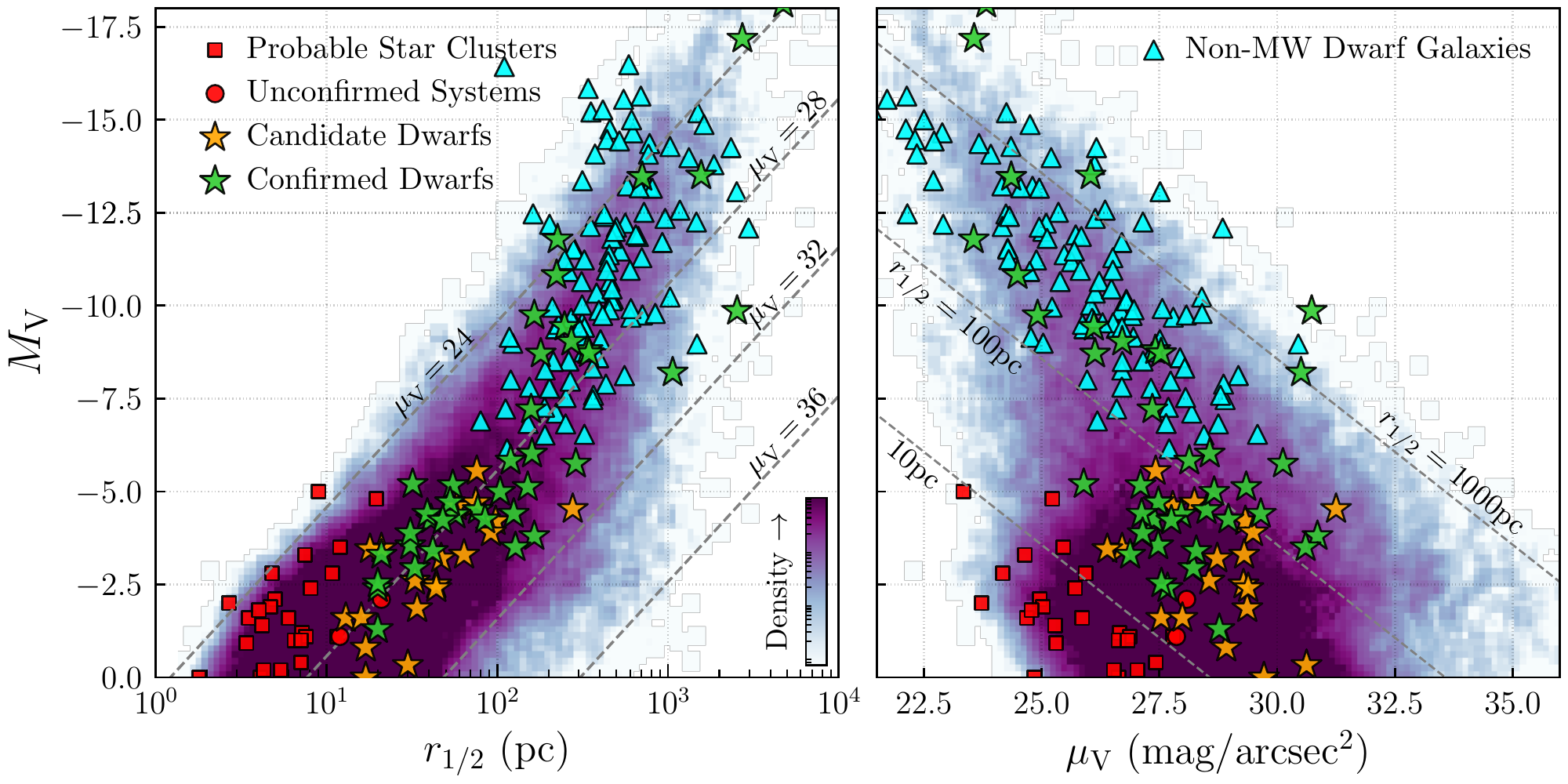}
    \caption{(\textit{Left}) The 2D distribution of all our model dwarf galaxies in azimuthally averaged physical half light radius ($r_{1/2}$) and absolute V-band magnitude ($\Mv$) space (that is size-luminosity plane) for $\zrei = 8.5$. The grey lines show lines of constant surface brightness $\muV$. For comparison, we plot observed MW confirmed/candidate dwarf galaxies (stars; from \citealt{DrlicaWagner.etal.2020}), probable ultra-faint star clusters (red squares; Table~\ref{tab:star_clusters}) and unconfirmed systems (red cirles) as DES J0225+0304 from \citet{Luque.etal.2017.sc} and DELVE 2 from \citet{Cerny.etal.2021}. We also add the recently discovered MW dwarf galaxies Eridanus IV \citep{Cerny.21.eriIV} and Pegasus IV \citep{Cerny.etal.2022.PegIV}. We also plot the data from other Local Volume dwarf galaxies \citep[e.g.,][]{Sharina.etal.2008,McConnachie.2012,Romanowsky.etal.2016,  Carlin.etal.2016.madcash1,Crnojevic.etal.2019,Delgado.etal.2021,MutluPakdil.etal.2021.NGC253,Carlin.etal.2021.madcash2,Sand.etal.2022.TucB}. (\textit{Right}) The 2D distribution of our model dwarf galaxies in absolute V-band magnitude ($\Mv$) and surface brightness ($\muV$) space (that is surface brightness - luminosity plane). The grey lines show lines of constant physical half-light radius $r_{1/2}$. Note that this plot includes all confirmed/candidate MW dwarf galaxies and not just systems used in our analysis (e.g., above plot includes Antlia II). We use the updated measurements for Antlia II from \citet{Ji.etal.2021.Antlia}.}
    \label{fig:mv_r12_muv}
\end{figure*}

Figure~\ref{fig:temporal_rads} shows evolution of $R_{\rm 50\%, survey}$ for each survey over a period of $\sim 2.5$ Gyr. The figure shows that as the LMC approaches its pericentre, $R_{\rm 50\%, survey}$ predicted for the DES survey decreases (i.e. concentration of the satellite radial distribution increases), reaching minimum around the epoch of the LMC pericentre passage. However, no significant variation around LMC pericentre is observed in the predictions for the PS1 and SDSS + Classical samples. 

Figure~\ref{fig:temporal_rads} also shows that $R_{\rm 50\%, survey}$ for DES increases around $\tau - \tau_0 \sim -1.3$ when the LMC is far from its pericentre resulting in an extended satellite radial distribution. Over a very short period of $\sim 0.5$ Gyr, the $R_{\rm 50\%, survey}$ for DES satellite radial distribution is doubled. Such behavior of satellite radial distributions based on position of a massive subhalo (like LMC) within a host halo has been observed in other studies (see, e.g., Figure 13 in \citealt{DSouza.Bell.2021}).  


\subsection{Dwarf Galaxy Sizes}
\label{ssec:gal_sizes}

Size of a galaxy is one of its key properties. Among other things, size, for a given luminosity, defines galaxy surface brightness and detectability, which makes it important for our forward-modelling of UFD population. We therefore present comparisons of the size distributions of our model galaxies to observations in this section.

Detailed description of the size modelling in our framework is presented in \citet{Kravtsov.Manwadkar.2022}. Briefly, the model assumes that the gaseous disk scale length $\Rd$  at any epoch $z$ is proportional to $R_{\rm 200c}(z)$: $\Rd = \chid \Rcrit$. The extent of the distribution of newly formed stars is computed from the radial surface density of H$_2$, which is used to compute radial star formation surface density profile. Finally, the size of the overall stellar distribution is computed from the total stellar mass profile formed during evolution assuming that there is no significant radial re-distribution of stars during evolution.

\begin{figure*}
    \centering
    \includegraphics[width = \textwidth]{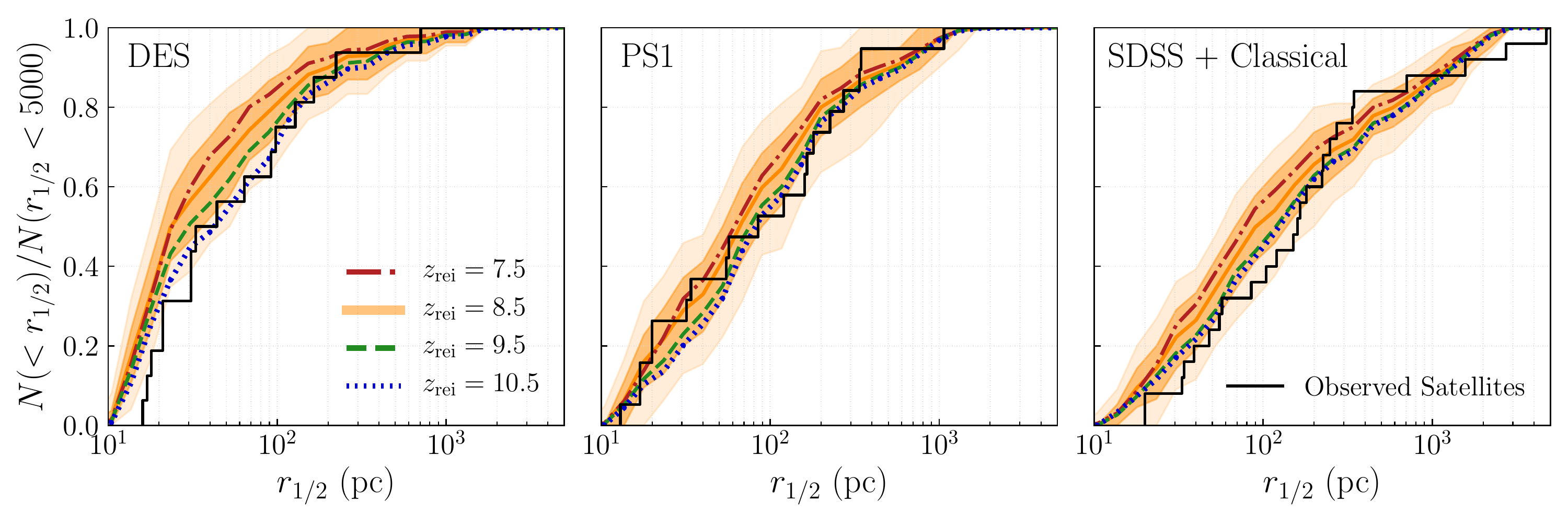}
    \caption{Predicted half-light radius $r_{1/2}$ distribution for satellites within $\dhelio < 300$ kpc in DES (\textit{left}), PS1 (\textit{centre}) and SDSS + Classical (\textit{right}) samples as a function of reionization redshift $\zrei = 7.5,8.5,9.5,10.5$. The distributions are normalized to the total number of galaxies with sizes smaller than 5000 pc. The color scheme and shaded regions carry the same meaning as in Figure~\ref{fig:fm_all_lfs}. Note that we are only including mock and observed galaxies with $\rhalf >10$ pc in the above forward modelling. }
    \label{fig:fm_rsize}
\end{figure*}

The proportionality constant for each satellite is set by $\chid = 0.06 \times 10^{\mathcal{N}(0,0.25)}$, where $\mathcal{N}(0,0.25)$ is normal probability distribution with zero mean and rms of $0.25$. Both the constant of $0.06$ and scatter of $0.25$ dex were chosen to approximately match the normalization of the size-stellar mass relation of dwarf galaxies in the Local Volume \citep{Kravtsov.Manwadkar.2022}.

Figure~\ref{fig:mv_r12_muv} shows the 2D distribution of our model dwarf galaxies in the size--luminosity ($\rhalf - \Mv$) and surface brightness - luminosity ($\muV - \Mv$) planes along with stellar systems (confirmed and candidate dwarf galaxies, star clusters, and unconfirmed systems) detected around Milky Way in recent surveys. For comparison, we also show dwarf galaxies around other Local Group galaxies. The model prediction in Figure~\ref{fig:mv_r12_muv} is shown for $\zrei = 8.5$. However, we find that distribution is not very sensitive to $\zrei$. 

The figure shows that sizes of model galaxies reproduces the wide range of observed sizes of dwarf galaxies. Remarkably, the model also reproduces the scaling of size and surface brightness with $M_V$ and subtle features in the scaling of observed galaxies, although there was no tuning of the model to match these features. 

For example, there is a distinct break in the $\rhalf-\Mv$ relation in observed dwarf galaxies at $\Mv \sim -6$. A similar break at $\Mv \sim -6$ is also seen in the model galaxy distribution. This break occurs at the luminosity below which dwarf galaxies become sensitive to reionization (see Section~\ref{sssec:lf_lmc}). Such a break in the $\rhalf-\Mv$ relation is expected due to reionization in the context of our size modelling because dwarf galaxies with $\Mv \gtrsim -6$ have formed most  of their stellar population before reionization. Their size therefore reflects the small virial radius of their parent haloes at these early epochs. 
This effect is likely enhanced due to the fact that molecular hydrogren fraction is high only at the central regions of these galaxies and this makes stellar distribution even more compact.

Scatter of the model galaxy sizes is due to the scatter in $\chid$ described above,  and scatter of size and luminosity arising from the scatter in the mass assembly history of their haloes in the presence of gas accretion suppression due to UV heating. The overall scatter is $\approx 0.5-0.7$ dex, consistent with observational constraints for MW satellites \citep[][]{Nadler.etal.2020}. The fact that the accretion rate around  $M_{\rm c}(z)$ changes smoothly contributes to the scatter. We find that in the scenario where accretion is shut down immediately once halo mass drops below $M_{\rm c}(z)$ \citep[as assumed, e.g., by][]{Bose.etal.2018}, produces a significantly smaller scatter in the $\rhalf - \Mv$ plane and smaller than the scatter exhibited by observed galaxies in Figure~\ref{fig:mv_r12_muv}. 

Note, however, that our dwarf galaxy model does not include physical processes that potentially can lead to increase of sizes due to feedback or tidal heating, as is likely for galaxies like Antlia II \citep[e.g.,][]{Ji.etal.2021.Antlia}.

\begin{figure*}
    \centering
    \includegraphics[width = \textwidth]{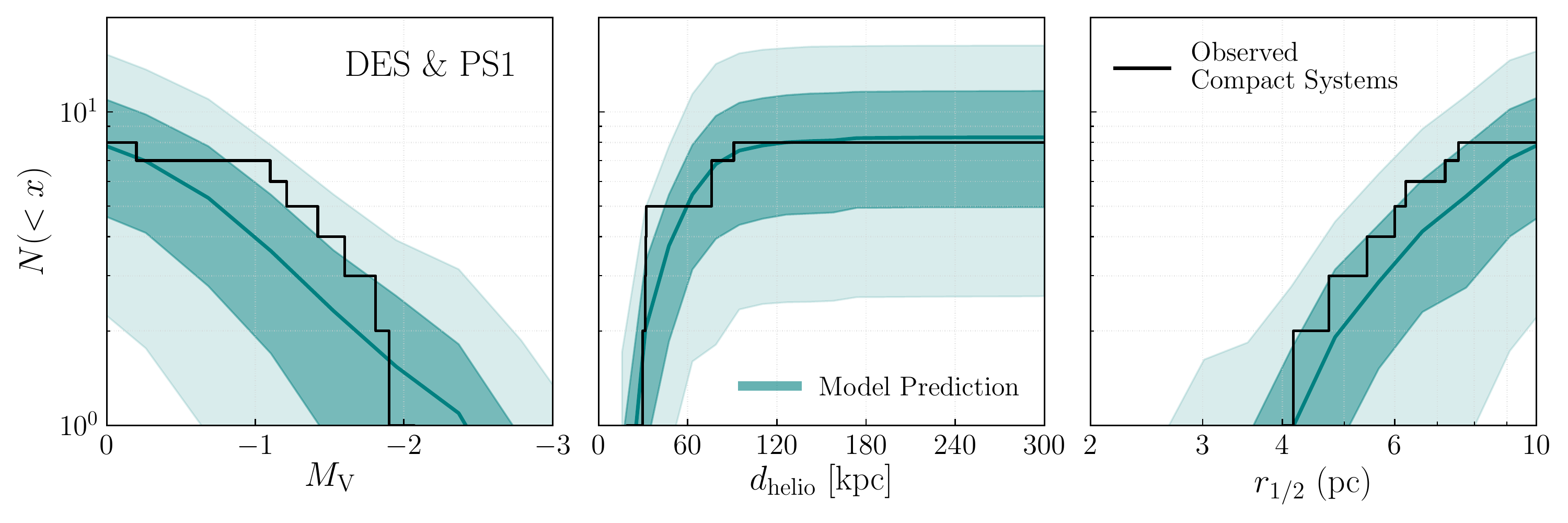}
    \caption{The predicted luminosity function, radial distribution and half-light radius $\rhalf$ distribution for clusters with $2 \text{ pc} < \rhalf < 10$ pc, $\muV > 24$ mag/arcsec$^2$ and $\Mv < 0$ that are detected with high significance in DES and PS1. The cuts of $2 \text{ pc} < \rhalf$ and $\muV > 24$ mag/arcsec$^2$ are applied to only consider observed systems that occupy the same region in the size-luminosity plane as our model dwarf galaxies. The cut of $\Mv < 0$ is applied to ensure that our model predictions are not affected by numerical effects present in low-mass subhaloes. The observed systems considered here are denoted in Table~\ref{tab:star_clusters} with the $\dagger$ symbol. }
    \label{fig:small_dwarf_dist}
\end{figure*}

To make the comparison more quantitative, Figure~\ref{fig:fm_rsize} shows the projected half-light radius $\rhalf$ distribution for satellites predicted by our model for the  DES, PS1, and SDSS+Classical samples for models with different $\zrei$ and compares the with observed $\rhalf$ distribution of galaxies in these samples. As before, here we only include model galaxies with $\rhalf > 10$ pc -- the cut used to define the observational samples we are comparing with.
Figure~\ref{fig:fm_rsize} shows good agreement between our model predictions and observed satellite size distributions. 
This agreement is non-trivial as the size model normalization was calibrated to match the size-stellar mass relation of nearby dwarf galaxies, not the {\it distribution} of sizes of the MW satellites. 

Interestingly, our model predicts that a sizable population of ultra-faint dwarf galaxies with $\Mv \gtrsim -2$  and  $5\text{ pc} < \rhalf < 20$ pc.  A significant fraction of these are in the ultra-compact area of $\rhalf < 10$ pc. This part of the $\rhalf - \Mv$ plane is occupied predominantly by star clusters, but our model implies that some of the currently unconfirmed stellar systems in this regime may be dwarf galaxies. We further discuss potential implications of such population of small-size dwarfs and its plausibility in Section~\ref{ssec:small_dwarf}.  

\section{Discussion}
\label{sec:discuss}

\subsection{The smallest dwarf galaxies}
\label{ssec:small_dwarf}

Galaxies are usually defined as gravitationally bound stellar systems with stellar and/or gas kinematics that cannot be explained
by a combination of baryons and Newton's law of gravity  \citep[e.g.,][]{Willman.etal.2012}. Such definition distinguishes between star clusters and galaxies embedded in a dark matter halo. 
This distinction is important because many of the faintest stellar systems detected around MW lie in the region of the size-luminosity plane ($L_V\lesssim 10^5\, L_\odot$, $r_{1/2} \lesssim 10\text{pc}$) that is occupied both by star clusters and the UFD galaxies (see Figure~\ref{fig:mv_r12_muv}).

Thus, to avoid confusion with star clusters, galaxy samples are often defined using a lower limit of $\rhalf = 10$ pc \citep[e.g.,][]{Conn.etal.2018.sc,DrlicaWagner.etal.2020}. We applied such cut to our model galaxy sample in the comparisons presented in the previous section. Nevertheless,  our model predicts that UFDs have sizes ranging continuously down to $\rhalf \sim 2$ pc (Figure~\ref{fig:mv_r12_muv}). The implication is that a significant fraction of stellar systems with $\Mv \gtrsim -2$ and $\rhalf < 10$ pc can be dwarf galaxies. 

Figure~\ref{fig:small_dwarf_dist} shows the predicted luminosity, distance and half-light radius $\rhalf$ distribution for compact, faint stellar objects with $\rhalf < 10$ pc and $\Mv < 0$ detected with high significance in DES and PS1. We compare our predictions with observations of objects with $2 \text{ pc} < \rhalf < 10$ pc, $\muV > 24$ mag/arcsec$^{2}$ and $\Mv < 0$. The cuts of  $2 \text{ pc} < \rhalf < 10$ pc and $\muV > 24$ mag/arcsec$^{2}$ are applied to only consider objects that occupy the same region in size-luminosity plane as our model dwarf galaxies (refer to Figure~\ref{fig:mv_r12_muv}). The cut of $\Mv < 0$ is applied to ensure that our model predictions are not affected by numerical effects present in low-mass subhaloes. The observed sample used in this comparison is denoted by $\dagger$ symbol in Table~\ref{tab:star_clusters}.  

Figure~\ref{fig:small_dwarf_dist} shows that the predicted abundance and distributions of distances and sizes of such compact systems predicted in our model are in agreement with observations. This agreement is non-trivial as our model predictions are calibrated of dwarf galaxies at $\rhalf > 10$ pc and only on normalization of sizes at a given stellar mass. There is thus no a priori expectation of agreement between observations and theoretical predictions for $\rhalf < 10$ pc regime. Although the number of systems in this comparison is small, the agreement indicates that such compact, faint stellar systems might indeed be dwarf galaxies formed within small-mass halos. 

We note that such compact faint stellar systems do form in recent high-resolution of cosmological simulations of dwarf galaxies  \citep[e.g., see the dashed line in Fig. 5 of][]{Agertz.etal.2020}, which reflects the small mass and virial radius of their host halo at high $z$ where most of the stars form.
Observationally, measurements of velocity and metallicity dispersions of compact low-luminosity stellar systems detected in the recent surveys would be needed to confirm this possibility.

\subsection{The Predicted MW Satellite Luminosity Function within 300 kpc}
\label{ssec:total_sat_mw}

\begin{figure*}
    \centering
    \includegraphics[width = \textwidth]{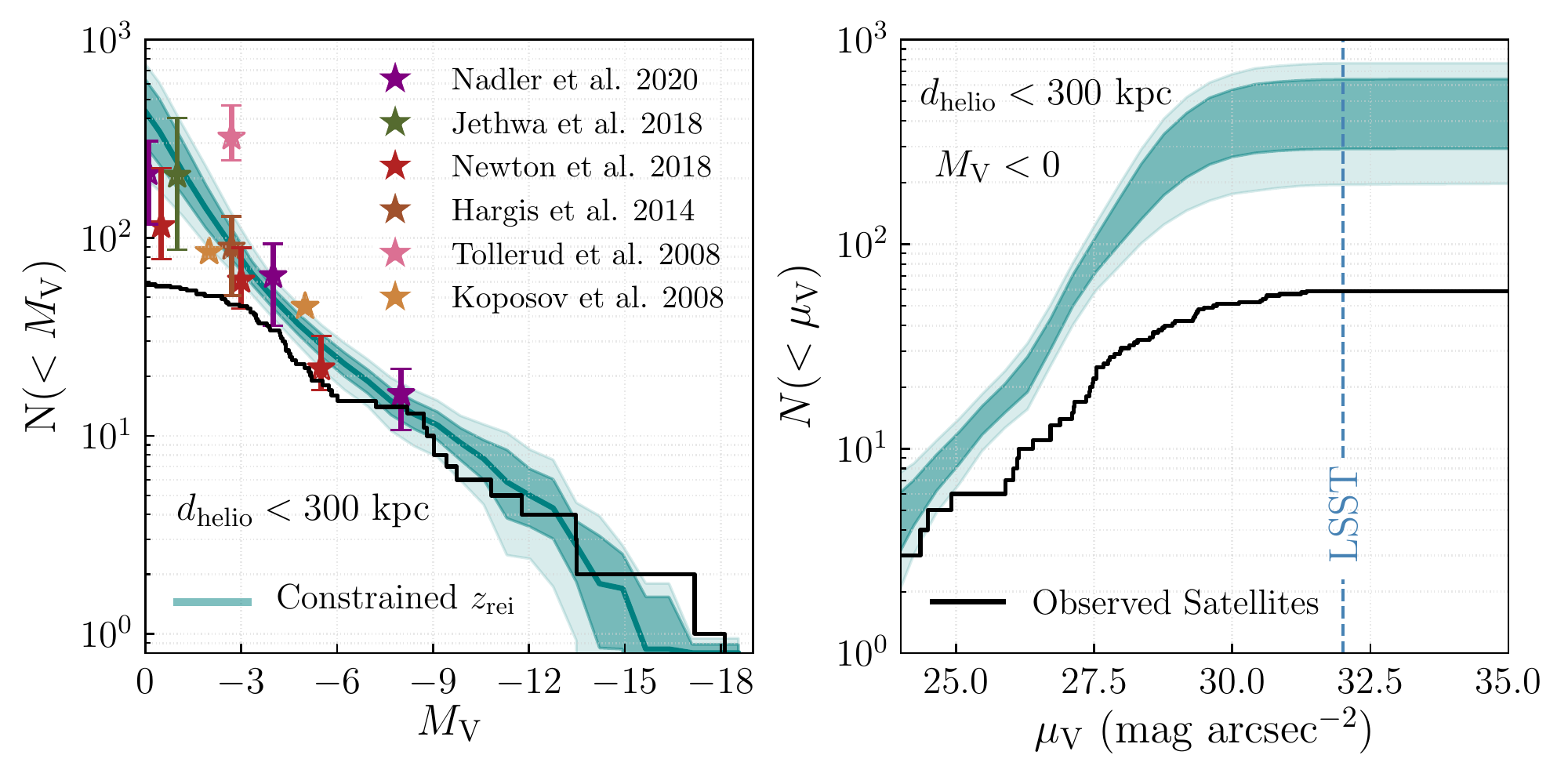}
    \caption{(\textit{Left}) The predicted MW satellite luminosity function within $\dhelio < 300$ kpc for satellites with effective half light radii $r_{1/2} > 10$ pc. For comparison, we show predictions for the total number of luminous MW satellites in other studies, namely \citet{Nadler.etal.2020}, \citet{Jethwa.etal.2018}, \citet{Newton.etal.2018}, \citet{Hargis.etal.2014} and \citet{Tollerud.etal.2008}. The solid black line shows the current observed MW satellite luminosity function. (\textit{Right}) The predicted surface brightness $\muV$ distribution for MW satellites with $\Mv < 0$ and $r_{1/2} > 10$ pc. The vertical dashed line indicates the predicted detection limit for LSST Y1 survey. The solid black line shows the current observed MW satellite surface brightness distribution.}
    \label{fig:total_lf_300}
\end{figure*}


Figure~\ref{fig:total_lf_300} shows predictions of our model for the total MW satellite luminosity function and surface brightness function within a heliocentric distance of 300 kpc and for satellites with effective half light radii $r_{1/2} > 10$ pc. The 68\% and 95\% contours in Figure~\ref{fig:total_lf_300} are due to a combination of host-to-host scatter and the inferred range of $\zrei$ values from current data (see Section~\ref{ssec:mw_reion}). The model predicts $440^{+201}_{-147}$ (440$^{+328}_{-242}$) with 68\% (95\%) range satellites with $\Mv < 0$, i.e. roughly an order of magnitude larger than the currently known number. The right panel of Figure~\ref{fig:total_lf_300} shows that $>$99\% of these satellites have surface brightness $\muV < 32\,\rm mag\,arcsec^{-2}$ and thus should be detectable in the LSST survey with the Vera C. Rubin Observatory (subject to its footprint constraints and other selection effects) if background contamination (e.g., faint galaxies) will not be overwhelming. 

For comparison, the figure shows estimates of the expected satellite abundance within this volume from several previous studies. 
Our estimate for the total number of luminous satellites at the faint end $M_V>-3$ is a factor of $2-3$ higher then recent estimates of \citet{Newton.etal.2018} and \citet{Nadler.etal.2020}, but the difference is within 2 $\sigma$. This difference is mainly due to the flattening of the $M_{\rm h,peak}-M_V$ relation in the UFD regime due to UV heating (see Section~\ref{ssec:mw_shmr}, and is partly due to increased scatter of luminosities at smaller masses, which increases the number of small-mass subhalos that get scattered into observable range. We discuss and evaluate this difference in the context of current estimates of the UFD satellite abundance in the HSC-SSP survey in Section~\ref{ssec:fut_current_surveys_lf}.
Our estimates of the abundance of dwarf satellites with $\Mv \lesssim -3$, on the other hand, are consistent with previous studies.  

The estimate of \citet{Jethwa.etal.2018} shown in Figure~\ref{fig:total_lf_300} are for their model assuming 100\% halo occupancy and fiducial power law SHMR model. The occupation fraction and scatter assumed in their study are quite different from the results of our model, but the overall UFD abundance they estimate is consistent with our model results.

The prediction for the number of satellites with $\Mv<-2.7$ within 300 kpc  from \citet{Tollerud.etal.2008} is the most discrepant from our model and results of other studies. The reasons for this large discrepancy are twofold. First, estimates from \citet{Tollerud.etal.2008} are based on the SDSS DR5 observations of 11 MW dwarf galaxies within a footprint of 8000 deg$^2$, which was rescaled to the entire sky. However, only 3 new satellites were discovered within an additional 6500 deg$^2$ of data in the SDSS DR9.  Second, the SDSS selection function used in \citet{Tollerud.etal.2008} is shallower (less sensitive) compared to other SDSS selection functions used \citep[e.g.,][]{Walsh.etal.2009}, as was also pointed out by \citet{Newton.etal.2018}. This biases satellite census prediction from \citet{Tollerud.etal.2008} to larger values.

In fact, our discussion on halo occupation fraction (Section~\ref{ssec:mw_hof}) and results from \citet{Jethwa.etal.2018} and \citet{Nadler.etal.2020} show that the observed abundance of UFDs in the DES/PS1 surveys cannot be explained if star formation is suppressed in haloes with $\Mpeak \lesssim 3 \times 10^8 \Msun$ \citep[see also][]{Graus.etal.2019}. Recent numerical studies show that haloes with masses as low as $\Mpeak \lesssim 3 \times 10^{7} \Msun$ do form stars prior to reionization  \citep[e.g.,][see also \citealt{Cote.etal.2018} for a compilation of recent galaxy formation results]{Ricotti.Gnedin.2005,Bovill.Ricotti.2009,Bland.Hawthorn.etal.2015}. Suppression of star formation in the haloes with masses below ``the atomic cooling limit'' corresponding to $T_{\rm vir}\sim 10^4$ K of $V_{\rm circ} \lesssim 15$ km/s ($\Mpeak \lesssim 3 \times 10^{8} \Msun$) is thus strongly disfavored both on theoretical and observational grounds.

\subsection{Prospects for current and future surveys}
\label{ssec:fut_current_surveys_lf}

\begin{figure*}
    \centering
    \includegraphics[width=\textwidth]{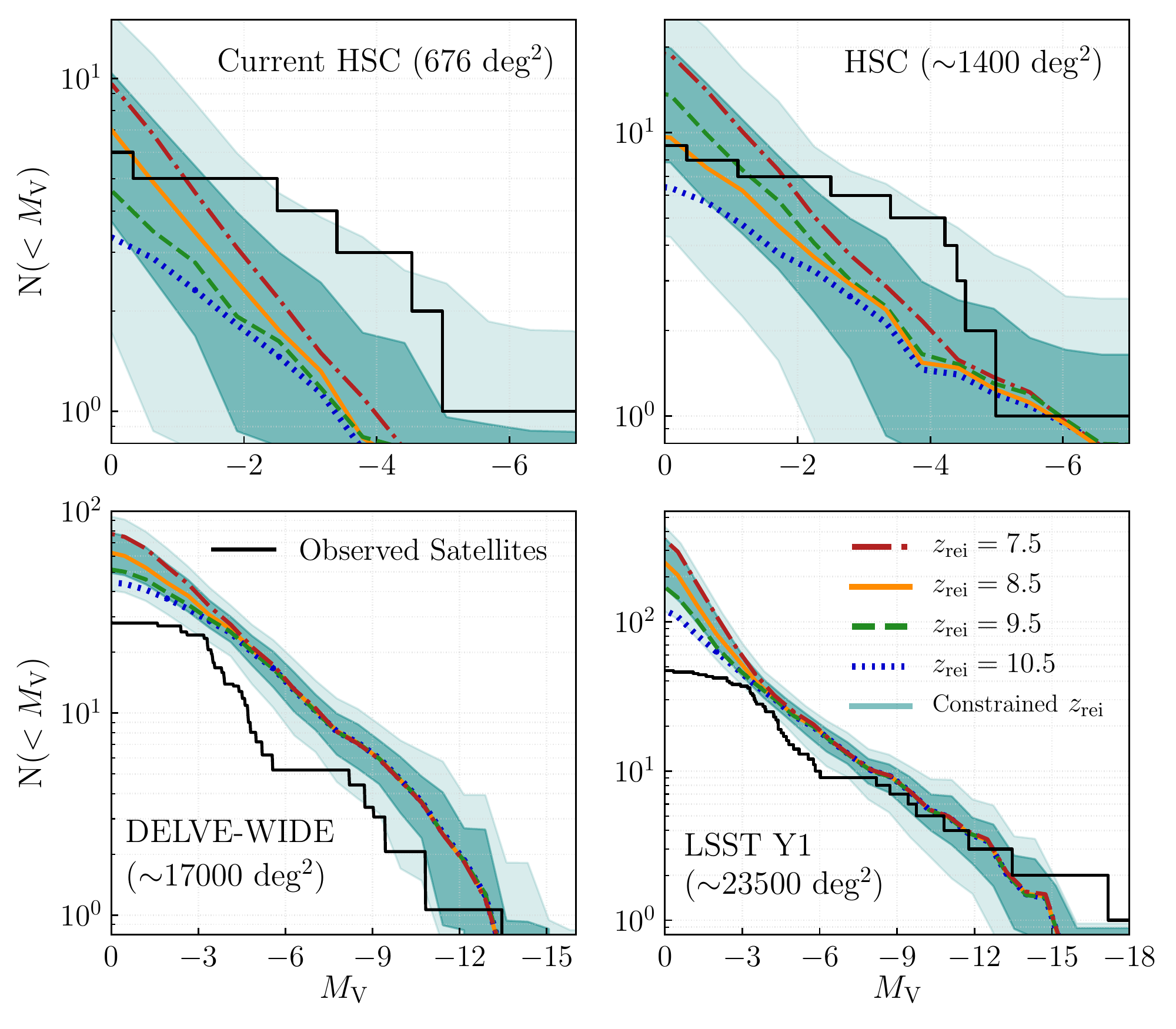}
    \caption{Our forecasts for satellite luminosity functions within $\dhelio < 300$ kpc in current HSC-SSP footprint (\textit{top left}), total HSC-SSP footprint (\textit{top right}), DELVE-WIDE (\textit{bottom left}) and LSST Y1 (\textit{bottom right}). Similar to before, these are satellite galaxies with effective half-light radii $\rhalf > 10$ pc and heliocentric distance $\dhelio < 300$kpc. The different colors denote different reionization histories and carry same meaning as previous figures (Figure~\ref{fig:fm_all_lfs} for instance). For predictions for current HSC-SSP observations (\textit{top left}), we show comparison with currently observed HSC-SSP satellite luminosity function (solid black line) and agreement is observed in the total number of satellites observed. For total HSC-SSP survey (\textit{top right}), we show the satellite luminosity function of satellites currently observed in HSC-SSP and other known satellites that fall in its total footprint. For predictions for the ongoing DELVE-WIDE (bottom left) survey and future LSST Y1 (bottom right) survey, we show the satellite luminosity function of known satellites that fall within respective footprint and can be detected by the respective survey.}
    \label{fig:fut_survey_lf}
\end{figure*}

Given that our model reproduces satellite luminosity functions in the DES, PS1 and SDSS surveys, we can use our model to forecast LFs that will be measured in ongoing surveys, such as HSC-SSP \citep[][]{Homma.etal.2018.hsc,Homma.etal.2019.hsc} and DELVE-WIDE \citep[][]{DrlicaWagner.etal.2021.delve}, and the near-future LSST survey that will use Vera C. Rubin Observatory \citep{Ivezic.etal.2019}. We present these forecasts for different redshifts of reionization $\zrei = 7.5, 8.5, 9.5, 10.5$ and for our inferred range for $\zrei$ (Section~\ref{ssec:mw_reion}) in Figure~\ref{fig:fut_survey_lf}. The black solid lines in Figure~\ref{fig:fut_survey_lf} show the current observed luminosity function in the corresponding survey footprint and accounting for the survey selection function. This gives an indication of how many satellites are currently known and how many more we expect to find in each of these surveys.

The top left panel in Figure~\ref{fig:fut_survey_lf} shows that our forecast for the observed satellite luminosity function in the current HSC-SSP footprint of $\sim$676$\,\rm deg^2$ is consistent to within 1$\sigma$ for the total observed abundance of UFDs, but is somewhat lower than the abundance of galaxies with $M_V\lesssim -2$ (still consistent at the 2$\sigma$ level). \citet{Nadler.etal.2020} does a different version of this calculation where they compute the total number of satellites observable in HSC-SSP, but not observable in DES and/or PS1. While the observed number of satellites for such a calculation is 4, \citet{Nadler.etal.2020} predicts a count of $1.75 \pm 0.6$ \citep[cf. also][]{Homma.etal.2019.hsc}. Performing the same calculation using our model, we predict a count of $6^{+3}_{-2}$. Our higher prediction is reflective of the higher occupation fraction $f(M_V<0)$ in haloes of $\Mpeak \lesssim 10^8 \Msun$ compared to \citet{Nadler.etal.2020}. Assuming Poisson statistics, the uncertainty in the observed satellite count is 2. Thus, both model predictions are consistent with observations at the 1$\sigma$ level. Additional satellite observations in HSC-SSP will be helpful in distinguishing between the different halo occupation fraction scenarios.

The top right panel in Figure~\ref{fig:fut_survey_lf} shows our forecast for the observed satellite luminosity function in the entire HSC-SSP footprint of $\sim$1400$\,\rm deg^2$. The solid black line is the LF for satellites detected in the survey so far and other satellites that fall in the total HSC-SSP footprint, namely, Pisces II, Leo V and DESJ0225+0304. The total number of satellites with $\Mv < 0$ and $\rhalf > 10$ pc predicted by our model for the total HSC-SSP survey is $13^{+7}_{-5}$  at the 68\% level of confidence.

The bottom left panel in Figure~\ref{fig:fut_survey_lf} shows our forecast for observed satellite luminosity function in the ongoing DELVE-WIDE survey and a comparison with the current satellite LF in the DELVE-WIDE footprint denoted by solid black line. Specifically, we Monte Carlo sample the detection probabilities using the DELVE-WIDE selection function to get the median observed LF, which is plotted as the black solid line. The total number of satellites with $\Mv < 0$ and $\rhalf > 10$ pc expected in the final DELVE-WIDE survey according to our model is $64^{+17}_{-13}$ with 68\% confidence compared to 30 currently known in the footprint with detection probability greater than 0.5. Most of the new expected satellites are in the UFD regime ($M_V\gtrsim -4$).

The bottom right panel in Figure~\ref{fig:fut_survey_lf} shows our model forecast for the satellite LF in the LSST Year 1 survey. As discussed in Section~\ref{ssec:mock_obs}, we assume that LSST Y1 selection function will be similar to that of the HSC-SSP, that is, it will detect all satellites with $\Mv < 0$, $\rhalf > 10$ pc, $\dhelio < 300$ kpc and $\muV < 32\,\rm mag\,arcsec^{-2}$ The solid black line is LF for all known satellites satisfying above detection criterion that fall within LSST footprint. The total number of satellites with $\Mv < 0$ and $\rhalf > 10$ pc expected in the LSST Year 1 survey is $273^{+119}_{-92}$ at the 68\% confidence level. We thus predict that LSST Year 1 survey will bring our total observed census to $\approx 60\%$ of the MW satellite population.  

We caution that the above predictions for the HSC-SSP and LSST Year 1 surveys are subject to systematic uncertainty due to uncertain selection function at the extremely faint magnitude limit expected for these surveys. These systematic uncertainties also make it very difficult to forecast the results of the LSST survey based on 10 years of co-added data, which is expected to have a 5-$\sigma$ point source magnitude of $r_{\rm lim} = 27.5$. At this depth, the number of unresolved background galaxies acting as contamination far exceeds the number of stars \citep[e.g.,][]{Willman.2010}, while their colors are similar to those of the main sequence turn off stars in the local dwarf galaxies. This can make detection of stellar over-densities tricky or impossible compared to current surveys of shallower depth. Thus, the detection efficiency of LSST 10 Year co-added survey will require detailed analyses of the survey data and modelling of the corresponding affects. Thus, we do not attempt to make predictions for LSST 10 Year co-added in this work. 

Comparison of the LF predicted for the DELVE-WIDE and LSST surveys shows difference in the LF shape at the faintest luminosities: the cumulative LF in the former flattens, while LF in the LSST survey continues to increase as a fairly steep power law down to $M_V=0$. This difference is due to different surface brightness limits of these surveys.

\begin{figure*}
    \centering
    \includegraphics[width=\textwidth]{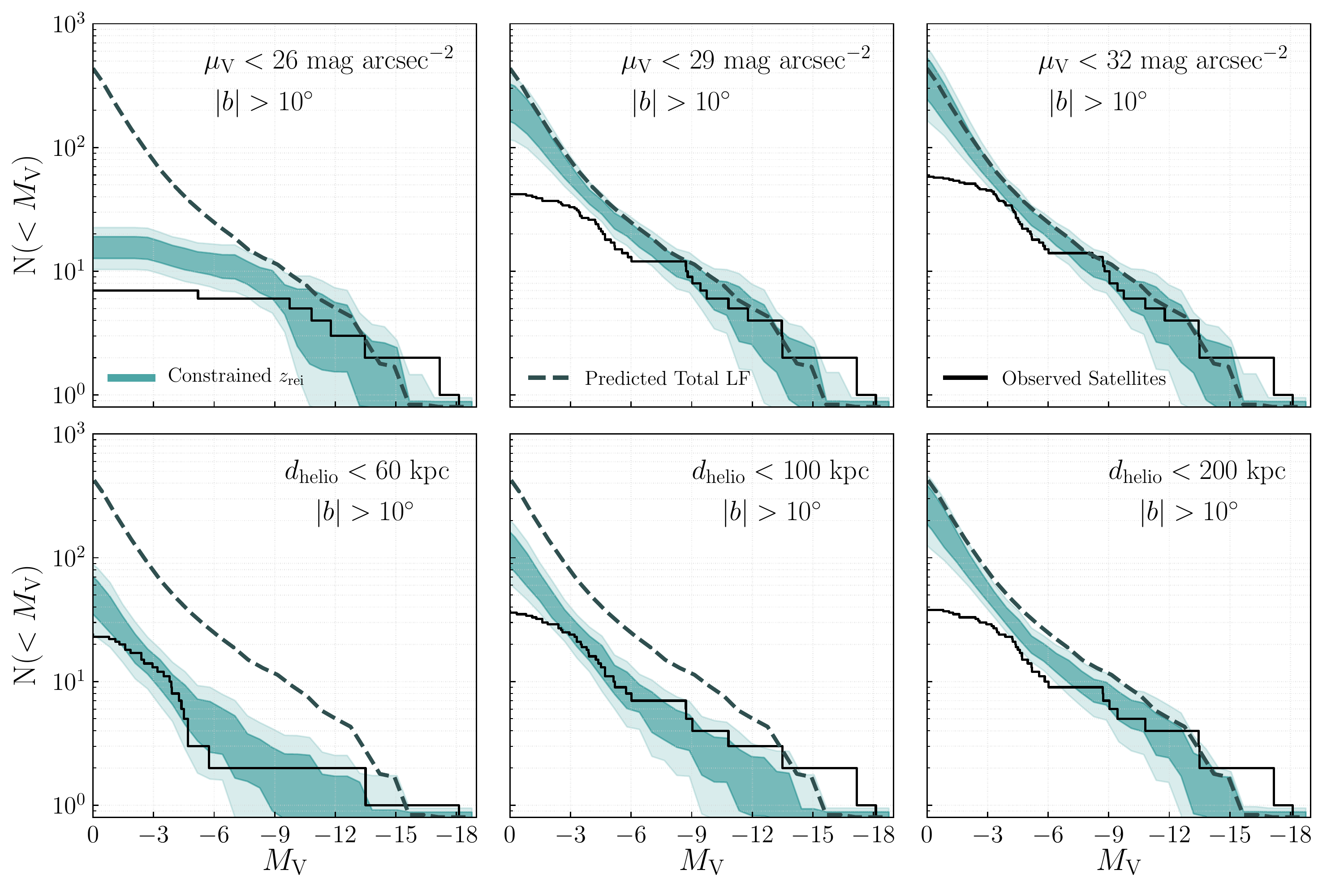}
    \caption{Predicted MW satellite luminosity functions ($\dhelio < 300$ kpc, $\rhalf > 10$ pc) for different cuts on surface brightness $\muV$ (namely $\muV < 26,29,32\ \rm mag\,arcsec^{-2}$) and heliocentric distance $\dhelio$ (namely $\dhelio < 60,100,200$ kpc). For comparison, we show the current observed satellite luminosity functions as solid black lines for the same cuts on surface brightness or heliocentric distance. The dashed line in each panel shows the model predicted median total MW satellite luminosity without any cuts on surface brightness or heliocentric distance. In other words, the dashed line is the same as the solid median line from left panel in Figure~\ref{fig:total_lf_300}.}
    \label{fig:fut_obs_lf}
\end{figure*}

Figure~\ref{fig:fut_obs_lf} show the satellite luminosity functions predicted by our model for three different cuts in surface brightness or heliocentric distance. In all cases, we apply a cut of $|b| > 10^{\circ}$ to mimic the missing coverage around the Galactic plane. The solid black lines in each panel show the current observed satellite luminosity with those same respective cuts. The dashed teal colored line shows the median predicted total MW satellite luminosity function. As one might expect, the satellite luminosity function completeness increases as a function of decreasing heliocentric distance. We appear to be essentially complete at $\dhelio < 60$ kpc except for $\Mv \gtrsim -2$. As shown above, ongoing surveys like DELVE-WIDE and upcoming surveys like LSST will help detect a significant fraction of these satellites. 

The LF is flattened at faintest luminosities for the $\muV < 26\,\rm mag\,arcsec^{-2}$ limit but steepens for the samples with fainter limits. One might expect to be complete at the $\muV < 26\,\rm mag\,arcsec^{-2}$ limit after accounting for galactic plane incompleteness, however, our model predicts a surplus of satellites relative to observations at $-2 \gtrsim \Mv \gtrsim -8$ with $\muV < 26\,\rm mag\,arcsec^{-2}$. For instance, the DES and PS1 surveys are expected to be complete at $\Mv \gtrsim -4$ and $\muV < 26$ at $\dhelio < 300$ kpc. However, the difference could be due to a lack of complete coverage at $|b| > 10^{\circ}$ with depths equivalent to DES/PS1. Uncertainties in size modelling, especially around the break in the $\rhalf - \Mv$ could contribute to this as well.

\subsection{Modelling Caveats}
\label{ssec:model_caveats}

In the preceding sections we demonstrated that our model is able to correctly and non-trivially reproduce a number  of properties of the observed dwarf galaxy population of the Milky Way and their distributions (see also additional comparisons in \citealt{Kravtsov.Manwadkar.2022}). Nevertheless, the model has a number of uncertainties or assumptions that might influence our determination of MW satellite statistics and have to be discussed as caveats, which is the focus of this section. 

\subsubsection{Galactic Outflow Modelling}

Feedback-driven outflows are undoubtedly the key process shaping properties of dwarf galaxies \citep[e.g.,][]{Collins.Read.2022}. 
In the GRUMPY model the outflows are parameterized using the stellar mass dependent mass loading factor \citep[][]{Kravtsov.Manwadkar.2022}.
In this work, we slightly modify the galactic outflow parameters to be $\eta_{\rm norm} = 1.8$ and $\eta_{\rm p} = -0.45$ compared to the values used in \citet{Kravtsov.Manwadkar.2022} of $\eta_{\rm norm} = 3.6$ and $\eta_{\rm p} = -0.35$. As discussed in Section~\ref{ssec:grumpy_mod}, these modifications are sensible and are done to better match the bright end of the MW satellite luminosity function. The galactic outflow parameters affect the bright end of the satellite luminosity function as they affect the high stellar mass end of the stellar mass - halo mass relation as seen in Figure 3 of \citet{Kravtsov.Manwadkar.2022}. Thus, different galactic outflow parameters will result in a different satellite counts as well. Furthermore, Figure 7 in \citet{Kravtsov.Manwadkar.2022} showed that agreement with observed stellar mass - metallicity relations can be brought about with a range of galactic outflow scaling relations or values of the wind metallicity $\eta_{\rm w}$ factor. Thus, to account for effect of scatter/uncertainties in galactic outflows parameters on our model predictions, we will first have to properly constrain these outflow parameters using a likelihood based analysis on the bright end of satellite luminosity function and stellar mass - metallicity relations together. Thus, our uncertainties on predicted number of satellites and $\zrei$ are likely underestimated. We leave this investigation for future work.

\subsubsection{Uncertainties in Galaxy Size Modelling}
\label{sssec:size_unc}

Our dwarf galaxy size model is able to accurately reproduce the observed dwarf galaxy statistics (for instance Figure~\ref{fig:fm_rsize}). The basic assumption of our size model is that the  gaseous disk size is proportional to $\Rcrit$ with some scatter. The linear relation between half-mass radius of stellar distribution and $\Rcrit$ has been shown to exist for galaxies with $\rhalf \gtrsim 100-200$ pc \citep[e.g.,][]{Kravtsov.2013}, and a similar relation for the disk scale length of gaseous disk and $\Rcrit$ for $\sim L_\star$ galaxies. Such a relation between $\Rcrit$ and gaseous disk size is also theoretically motivated \citep[e.g.,][]{Fall.Efstathiou.1980,Ryden.Gunn.1987,Mo.etal.1998} because the angular momentum acquired by haloes during their collapse will influence the collapse of baryons onto the halo, setting the galaxy size. Nevertheless, in our model this assumption is extrapolated into the UFD regime with $\rhalf \lesssim 50$ pc. Furthermore, the scatter assumed by our size model in $\chid$ is 0.25 dex comparable to the scatter expected for the specific angular momentum, while larger scatter might be expected in our size model due to other factors such as tidal heating  \citep[e.g.,][]{Errani.etal.2020}.

Another uncertainty stems from systematic uncertainties in the $\HH$ modelling, as stars in our model form from the molecular gas and its extent within the gas disk thus defines the extent of newly formed stellar population. As shown in Appendix B in \citet{Kravtsov.Manwadkar.2022}, the current calibrations of the molecular gas models have significant uncertainty in the low-metallicity regime. 

Finally, in our model the size of the gas disk is assumed to be proportional to $\Rcrit$ even when gas accretion is suppressed by UV heating. Thus, the size of the gaseous disk continues to grow even when its mass does not grow. This, however, does not result in increase in stellar half mass radius because the extent of the molecular gas would actually be either constant due to increasing metallicity or would shrink due to decreasing gas surface density. We find that only a small subset of model dwarf galaxies undergoes evolution in this regime which results in very faint surface brightness.

\subsubsection{Merger Histories, Subhalo Disruption \& MW's peculiarities}
\label{sssec:mw_merger_hist}

As described in Section~\ref{sec:model}, for comparisons with observations we select host haloes that have realistic LMC analogs and there are only two hosts of this kind in the Caterpillar suite. This does not allow us to quantify the host-to-host scatter and median or mean statistics of the satellite population. We also cannot impose additional selection criteria using other MW properties, such as presence of the SMC, the existence of a massive M31-like neighbor, details of the MW merger history etc. To our current knowledge, the MW underwent 3 signifcant merger events: i) Gaia-Enceladus-Sausage (GES), ii) Sagittarius and iii) LMC. In fact, \citet{Bose.etal.2020} finds the incidence of hosts with a GES-like \textit{and} LMC-like mergers to be $\sim 3\%$. Reproducing such properties of the Milky Way will require a much larger suite of the high-resolution zoom-in simulations. 

Additionally, the two hosts we select have rather similar merger histories, as can be seen in the right panel of Figure~\ref{fig:cater_sums}. It is not clear whether this is a coincidence of this is due to the LMC analog criterion and using two host haloes with similar merger histories in our analysis could possibly bias our results somewhat. 

\begin{figure*}
    \centering
    \includegraphics[width = \textwidth]{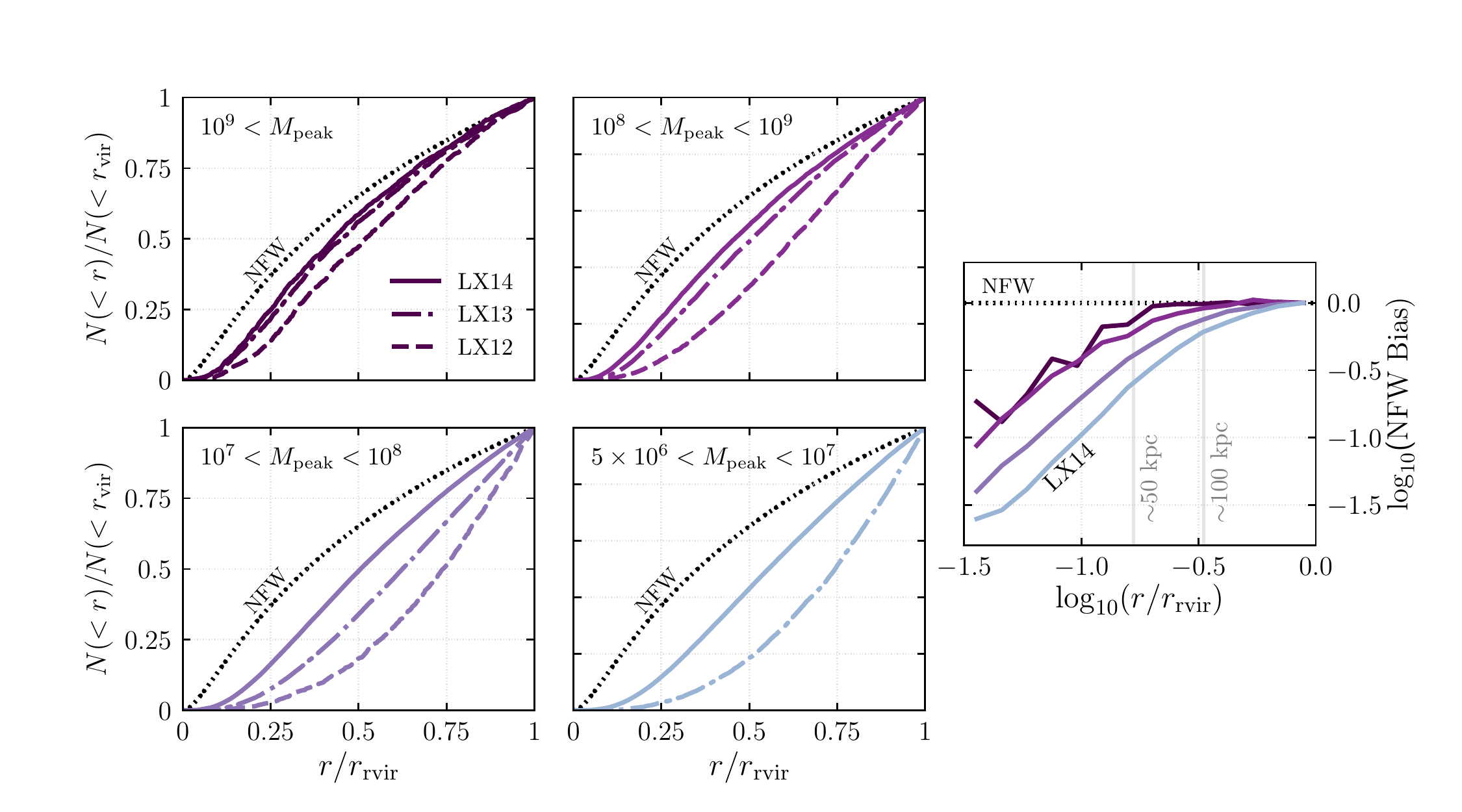}
    \caption{(\textit{Left 4 Panels}): The cumulative subhalo radial distributions in different $\Mpeak$ ranges for the Caterpillar hosts at the different resolution levels of LX12, L13 and LX14. The dashed black lines in each of these correspond to the mean NFW profile for all the hosts. Their is negligible difference in host concentration between different resolution levels so we only show one line for mean NFW profile of host. The different line styles correspond to different resolution suites. Note that the $5 \times 10^6 < \Mpeak < 10^7$ line for Caterpillar LX12 is not shown as it has no subhaloes in that range. (\textit{Right Panel}) The radial bias function between the subhalo radial distribution and host NFW profile shown for different $\Mpeak$ ranges. The shading of the lines carries the same meaning as the shading in the left 4 panels of the figure. The radial bias function (denoted by `NFW Bias') is defined as the ratio of $\rm dN/dx^3|_{\rm subhalo}$ and $\rm dN/dx^3|_{\rm NFW}$ where $\rm dN/dx^3$ is the number of subhaloes per unit shell volume as a function of $x = r/r_{\rm vir}$. The bias function will be equal to one if it is unbiased with respective to the host NFW profile. The above plots show that the effects of withering and artificial disruption are prominent for small mass subhaloes as their radial distribution gets increasingly shallower and biased away from the host NFW profile. }
    \label{fig:radial_mpeak}
\end{figure*}

Details of mass assembly history of the host halo possibly translate into subhalo disruption modelling uncertainties. As we use the Caterpillar simulation suites, which do not model the central galactic disk potential, we apply \citet{Nadler.etal.2018}'s subhalo disruption model to these hosts. The subhalo disruption model from \citet{Nadler.etal.2018} was trained on only two zoom-in simulations with very similar merger histories and so also has systematic uncertainty. Furthermore, when using \citet{Nadler.etal.2018}'s subhalo disruption model, we are implicitly assuming that \textit{all} of the subhalo disruption is due to MW's central galactic disk. However, \citet{Jahn.etal.2019} showed that LMC host halo depletes $\sim 30\%$ of its substructure within $0.2 r_{\rm 200m}$ prior to accretion onto MW-size host. This effect is small compared to the disruption by the MW mass host and disk (\citealt{Jahn.etal.2019}) and thus does not affect our results and conclusions, but such  additional disruption of LMC satellites by the LMC itself may need to be included in the future models if they focus on the predictions in surveys around LMC.  

Finally, recent studies of \citet[][]{Webb.Bovy.2020} and \citet{Green.etal.2022} conclude that effect of disk on subhalo disruption is much smaller than derived from previous simulations due to insufficient resolution of the latter. This effect is thus quite uncertain, which introduces uncertainty in the halo occupation distribution of observable galaxies inferred within the context of our model. An additional uncertainty is due to numerical effects affecting subhalo disruption and their radial distribution and abundance discussed in the next subsection.  

\subsubsection{Withering, Artificial Disruption \& Orphan Galaxies}
\label{sssec:artificial_disrupt}

The predictions for the luminosity function and radial distribution of satellite galaxies are based on the subhalo population down to masses of $\sim 10^7\,\Msun$ in the Caterpillar simulations used in this study and can potentially be subject to numerical effects. For example, \citet{Bose.etal.2020} have recently argued that radial distribution of satellite galaxies cannot be reproduced without accounting for numerical disruption of subhaloes by modelling disrupted subhaloes with an ``orphan galaxy'' model. As shown in Section~\ref{ssec:lf} and Section~\ref{ssec:rad_dist}, our model predicts the luminosity function and radial distribution of satellites in reasonable agreement with observations. The question is thus to what extent numerical disruption affects the subhalo population on which these predictions are based. 

In a recent study, \citet{Green.etal.2021} presented a comprehensive model for the evolution of subhalo populations using the SatGen model of  \citet{Jiang.etal.2021}. This model accounts for the effects of tidal mass loss and potential loss of subhaloes when their bound mass becomes smaller than the resolution limit of simulation and/or halo finder ({\it ``withering''}) and premature disruption of subhaloes due to numerical effects ({\it ``artificial disruption''}). \citet{Green.etal.2021} showed that withering has a large effect on the radial distribution of subhaloes and the amplitude of their mass function, while artificial disruption has a much smaller but still noticeable effect. They also showed that when withering is corrected for, the radial distribution of subhaloes is expected to be close to the radial distribution of dark matter in their host halo.

\begin{table*}
    \caption{The Spearman correlation coefficient $\rho$ of $r/r_{\rm vir}$ with $M_{\rm acc} / M_{\rm host}$ and $\log_{10}(1+z_{\rm acc})$ for subhaloes with different $\Mpeak$ cuts and in the different Caterpillar resolution suites. For comparison, in the rightmost two columns, we show these correlation coefficients from Bolshoi simulations for \textit{all} subhaloes and their predictions from \citet{Green.etal.2021}. `G21(W+D)' denotes predicted value in the presence of W+D and `G21(All)' denotes the same value but in the absence of W+D. Note that we do not quote a correlation coefficient in the [$5 \times 10^6,10^7$] bin for LX12 as no subhaloes are found there. The standard deviation $\sigma$ on all these correlation quantities are negligible ($\sigma <0.01$). The exception is that we have $\sigma \approx 0.02$ for the $[10^9,)$ bin at all resolutions and $[10^7,10^8]$ bin for LX12 due to low subhalo statistics. This standard deviation is computed using the bootstrap method.}
    \begin{tabular}{|c|c|ccc||c|c|}
    \hline
    Correlation Quantities & $M_{\rm peak} (\Msun)$ bins  & LX12 & LX13 & LX14 & &   \\
    \hline
    \hline
    &  $[5 \times 10^6, 10^7]$ & - & -0.07 & +0.00 & Bolshoi & -0.14 \\
    $r/r_{\rm vir} - M_{\rm acc} / M_{\rm host}$  & $[10^7, 10^8]$  &-0.11  & -0.11 & +0.00 & G21(W+D) & -0.15\\
    &  $[10^8,10^9]$ & -0.09 & +0.01 & +0.06 & G21(All) & +0.01 \\
     &  $[10^9,)$ & -0.01 & +0.03 & +0.05 &  & \\
      \hline
      \hline
      &  $[5 \times 10^6, 10^7]$ &  - & -0.27 & -0.49 & Bolshoi & -0.33\\
          $r/r_{\rm vir} - \log_{10}(1+z_{\rm acc})$  & $[10^7, 10^8]$ &  -0.28 & -0.38 & -0.56 & G21(W+D)  & -0.24\\
          & $[10^8,10^9]$  &  -0.35 & -0.55 & -0.62 & G21(All) & -0.46 \\
          & $[10^9,)$  &  -0.53 & -0.60 & -0.62 &  & \\
      \hline
    \end{tabular}
    \label{tab:correlation_coeffs}
\end{table*}


We have examined population of satellites with $\Mv < 0$, which approximately corresponds to subhaloes with $\Mpeak \gtrsim 4 \times 10^6 \Msun$ according to Figure~\ref{fig:halo_occ}\footnote{This $\Mpeak$ limit corresponds to the median halo mass at which $f(\Mv < 0) < 0.01$.}. However, most of the currently observed satellites (and hence our inferred statistics) are at $\Mv \lesssim -1.5$ hosted by subhaloes of $\Mpeak \gtrsim 2 \times 10^7 \Msun$ (see Figure~\ref{fig:shmr_mw}). We will thus examine the effects of withering and artificial disruption on subhaloes of such masses. For purposes of brevity, we will refer to `withering and artificial disruption' together as `W+D'. 

Firstly, note that among the subhaloes that accrete at the same epoch, withering should have  a larger effect on the radial distribution of $5 \times 10^6 \Msun <\Mpeak< 10^7\,\Msun$ subhaloes, which are close to the mass resolution threshold, than on subhaloes with larger $\Mpeak$. Artificial disruption, on the other hand, is expected to affect all subhaloes that lose $\gtrsim 95\%$ of their initial mass independent of that mass \citep{Green.etal.2021} and thus should not affect radial distribution of subhaloes. We can thus probe potential existence of numerical effects by examining radial distribution of subhaloes of different $\Mpeak$. 

Figure~\ref{fig:radial_mpeak} shows the cumulative subhalo radial distributions for different $\Mpeak$ ranges in the Caterpillar LX12, LX13 and LX14 suites. For comparison, we also show the NFW profile that matches the mean dark matter density profile of the Caterpillar hosts. The figure shows that the radial distribution becomes more concentrated with increasing $\Mpeak$ in line with general expectation of the effects of withering outlined above and in agreement with the results of recent study by \citet[][see their Fig. 10]{Grand.etal.2021}. At the same time, the difference in the radial distributions of subhaloes of different mass is very small for $\Mpeak>10^8\,\Msun$. For $10^7<\Mpeak/\Msun<10^7$ subhaloes the difference is larger, but is still $\lesssim 1.6$ for $r\geq 0.1r_{\rm vir}$.

Furthermore, \citet{Green.etal.2021} show that effects of W+D (withering being the dominant factor) are imprinted in correlations of subhalo properties, such as $M_{\rm acc}/M_{\rm host}$ and $\log_{10}(1+z_{\rm acc})$ with their normalized distance to the host centre, $r/\rvir$. This is because W+D effects are expected to be larger for subhaloes orbiting closer to the host halo centre and stronger for haloes that spend more time within the host. In particular, they showed that their SatGen model reproduces the Spearman correlation coefficient for the correlations between $r/\rvir$ and $M_{\rm acc}/M_{\rm host}$ or $\log_{10}(1+z_{\rm acc})$ in the Bolshoi simulation and show what these coefficients are expected to be if W+D were not present. 

In Table~\ref{tab:correlation_coeffs}, we present the correlation coefficients of $r/r_{\rm vir}$ with $M_{\rm acc} / M_{\rm host}$ and $\log_{10}(1+z_{\rm acc})$ for subhaloes with different $\Mpeak$ cuts and in different Caterpillar resolution suites. We compare these numbers with the same correlation coefficients from SatGen model and the Bolshoi simulation. Note that the G21(All) result applies when numerical effects are completely absent, while G21(W+D) is for when they are present at the level of the resolution of the Bolshoi simulation. The numbers marked as ''Bolshoi'' give the actual correlation coefficient measured for the Bolshoi haloes by \citet{Green.etal.2021}. Thus, correlation coefficients close to G21(All) indicate that W+D effects are small, while those close to G21(W+D) or Bolshoi indicate that these effects are significant. 

We find that the $r / r_{\rm vir} - M_{\rm acc} / M_{\rm host}$ correlation for all LX14 subhaloes with $\Mpeak > 10^7 \Msun$ is consistent with a weak positive or zero correlation, similar to the prediction of the SatGen model when there is no W+D (denoted by `G21(All)'). If it were dominant in simulations, it would induce a negative correlation between halo mass ($M_{\rm acc} / M_{\rm host}$) and radial distance ($r/r_{\rm vir}$), as it would remove smaller mass haloes in the inner regions of the host, preferentially leaving them in the outskirts. This is indeed seen in the SatGen prediction with W+D and also for LX12 and LX13 simulations. We find similar results if we consider $M_{\rm peak} / M_{\rm host}$ instead of $M_{\rm acc} / M_{\rm host}$. 

For the $r / r_{\rm vir} - \log_{10}(1+z_{\rm acc})$ correlation, we find a relatively strong negative correlation for subhaloes well above the mass resolution limit, which is similar to prediction of SatGen model with no W+D. When W+D is present, the negative correlation in $r / r_{\rm vir} - \log_{10}(1+z_{\rm acc})$ will be weakened as these effects preferentially remove subhaloes that were accreted earlier and that are orbiting in the inner regions and thus would have lost more mass on average. We indeed observe weakening of the negative correlation for the smallest mass bins in LX 12 and LX 13 suites, where correlation coefficients become similar to the predictions of the SatGen model when W+D is present or the value measured in the Bolshoi simulation. 

Another indication that effect of artificial disruption on the subhalo samples used in our model is small is that there is no difference between subhalo mass functions in the LX13 and LX14 suites in the well-resolved high-mass regime, $\mu = M/M_{\rm host} > 10^{-3}$. Given that artificial disruption does not depend on the initial mass of subhalo, but affects subhaloes that lose $\gtrsim 95\%$ of their initial time, the time it takes for a halo to artificially disrupt will increase with increasing halo mass. However, Figure~\ref{fig:cat_res} shows that this is not the case even in the inner $r < 50$ kpc region of the host halo. Additionally, \citet{Green.etal.2021} found artificial disruption to affect the subhalo mass function amplitude by at most $\sim$20\%. So, even if artificial disruption is present in lower mass subhaloes, it will bias our inferred $\zrei$ by at most $\sim$0.5 (can be inferred by raising contours in Figure~\ref{fig:zrei_nsat} by a factor of 1.2).

The main conclusion that can be drawn from the results discussed in this section is that for a given mass bin, the numerical effects decrease as resolution is increased. Likewise, in a given simulation resolution suite, the numerical effects increase in smaller mass bins. Both Figure~\ref{fig:radial_mpeak} and Table~\ref{tab:correlation_coeffs} indicate that the W+D effects in the smallest mass bin in the LX14 suite we use in our modelling $[10^7,10^8]$ are present, but their effect on the subhalo population is relatively small. In the large mass bins these effects are negligible. 

These results imply that there is no need to include a separate model for orphan galaxies in our analysis of LX14 simulation, as is needed when one uses subhalo samples with masses close to the withering limit of the simulation \citep[e.g.,][]{Bose.etal.2020}. 

\subsection{Comparison with previous models}

Semi-analytic models of dwarf galaxies have been used in many studies over the last three decades \citep[e.g.,][]{Kauffmann.etal.1993,Somerville.2002,Benson.etal.2002a,Benson.etal.2002b,Starkenburg.etal.2013,Lu.etal.2016,Lu.etal.2017,Bose.etal.2016,Bose.etal.2018,Bose.etal.2020}. Such models were used productively to model dwarf satellites observed in the Local Group using mass assembly tracks extracted from the modern generation of high-resolution zoom-in $N$-body simulations \citep[e.g.,][]{Starkenburg.etal.2013,Bose.etal.2018}. 

Our model uses a ``regulator'' like model that strip down semi-analytic modelling down to the minimal number of equations  \citep[e.g.,][]{Finlator.Dave.2008,Krumholz.Dekel.2012,Lilly.etal.2013,Feldmann.2013,Peng.Maiolino.2014}. In particular, we exploit the fact that simulations showed that accretion onto dwarf galaxies is predominantly in the cold-mode to avoid explicit modelling of cold and hot gas phases within dwarf haloes. We also use the fact that due to fairly steep stellar mass--halo mass in the dwarf regime the mergers do not contribute significantly to stellar mass of dwarf galaxies to avoid explicit modelling of mergers and simplify calculations.  

This regulator type of models have been shown to be successful in matching many key observational scaling relations and their evolution \citep[see,e.g.,][for a review]{Tacconi.etal.2020} and they were used for interpretation of halo mass--stellar mass, stellar mass--metallicity correlations, and star formation histories  in the dwarf galaxy regime \citep[][]{Tassis.etal.2012,Ledinauskas.Zubovas.2018,Ledinauskas.Zubovas.2020}. 

At the same time, our model employs refined modelling of several keys processes informed by recent theoretical models and simulations results. 
For example, we model effects of UV heating of gas during and after reionization by modulating gas accretion onto haloes explicitly rather than using a hard cutoff mass. We also explore different reionization redshifts, because the Lagrangian volume of the Local Group is generally expected to reionize at a different time than average volume in the Universe \citep[e.g.,][]{Zhu.etal.2019}. We also use star formation modelling based on a model of molecular hydrogen, which in turn is based on model for gas distribution and galaxy sizes calibrated by observations. Our model also uses outflow prescription calibrated using simulations \citep[see][for details]{Kravtsov.Manwadkar.2022}. 

A study by \citet[][]{Chen.etal.2022} that presented the \texttt{A-SLOTH} model of dwarf galaxies appeared after our paper was submitted. The authors used the model to study $M_\star-M_{\rm h}$ relation and halo occupation distribution of observed dwarf satellites of the Milky Way. Although some aspects of the overall approach
of \texttt{A-SLOTH} model is similar to \texttt{GRUMPY}, details of modelling are rather different. For instance, a rather sophisticated model of both ISM structure and star formation and of stellar feedback is used in the \texttt{A-SLOTH} model with separate modelling of Population III and II stars and explicit modelling of photo-heating by UV radiation of stars and momentum and energy injection by supernovae. Sophistication of the modelling and the correspondingly larger number of differential equations and parameters makes \texttt{A-SLOTH} closer to the classic semi-analytic models. The approach of the \texttt{GRUMPY} model, on the other hand, is to keep the number of equations to a minimum and use simple, simulation calibrated parametrizations of key processes. As we showed in this and previous paper, the model based on this approach can simultaneously reproduce many key observed properties of dwarf galaxies.  

\section{Summary and Conclusions}
\label{sec:summary}

We use a simple regulator-type model for modelling dwarf galaxies presented \citet{Kravtsov.Manwadkar.2022} to study the luminosity function, radial distribution, and size distribution of the dwarf galaxy satellite population of the Milky Way. Although fairly simple, the model reproduces stellar masses, gas mass, gas/stellar-phase metallicities, star formation rate and  diversity in star formation histories of dwarf galaxies.  We forward model observations of the model population accounting for observational biases in surveys through their respective selection functions and current constraints on the MW halo mass and presence of the Large Magellanic Cloud (LMC). 

In particular, we compare model predictions with dwarf galaxy observations in the DES, PS1, and SDSS surveys and  make predictions for ongoing/future surveys, HSC-SSP, DELVE-WIDE and Vera C. Rubin Observatory/LSST. We show that our model is able to reproduce the observed MW satellite luminosity function, radial distributions and the half-light radius $\rhalf$ distribution for reasonable choices of model parameters, such as reionization redshift. We present the halo mass-luminosity relation predicted and halo occupation distribution for the MW satellite population. 

Our main results and conclusions are:
\begin{enumerate}
    \item The amplitude of the ultra-faint end of the MW satellite luminosity function is sensitive to the redshift of reionization $\zrei$ at $\Mv \gtrsim -6$ \citep[in agreement with findings of][]{Bose.etal.2018}, while $\Mv \lesssim -6$ is insensitive to $\zrei$. Our model thus indicates that the empirical boundary between ``classical dwarf galaxies'' and UFDs corresponds to the physical boundary below which galaxies are strongly affected by reionization, as was first suggested by \citet[][]{Ricotti.Gnedin.2005} before the UFDs were discovered.\\

    \item We constrain the connection between stellar mass $\Ms$ and peak halo mass $\Mpeak$ for MW satellites to be Equation~\ref{eqn:new_shmr} and show that this relation has a clear break imprinted by  reionization. This deviation from power-law relation results in a high fraction of luminous galaxies in haloes with $\Mpeak<10^8\,\Msun$. \\

    \item As a consequence, the model predicts a higher halo occupation fraction in low mass haloes ($\Mpeak < 10^8 \Msun$) compared to recent studies. We find that the the peak halo mass at which 50\% of haloes host luminous satellites is $\Mpeak \sim 3.1 \times 10^7\,\Msun$. This is a prediction of our model and the abundance of subhaloes in the Caterpillar simulations used in this study and will need to be tested with future observations. In particular, we find that such high halo occupancy in low mass haloes in our model is helps to reproduce the number of dwarf galaxy observations in the HSC-SSP survey. \\
    
    \item The high occupation fraction in $10^7-10^8\,\Msun$ haloes implies that suppression of star formation in the haloes with masses below ``the atomic cooling limit'' of $V_{\rm circ} \lesssim 10$ km/s ($\Mpeak \lesssim 3 \times 10^{7} \Msun$) is strongly disfavored.\\
    
    \item We demonstrate that our model based on the LX14 Caterpillar suite, which accounts for the disruption of subhaloes due to disk using \citet[][]{Nadler.etal.2018} model, is able to match the observed radial distribution of the MW satellites {\it without} modelling orphan galaxies. Under the assumption of homogeneous reionization in the MW environment, $\zrei$ only affects the normalization of the satellite radial distribution but the concentration of satellite radial distribution is insensitive to $\zrei$.   \\
    
    \item We show that our simple theoretical model can simultaneously reproduce distribution of stellar half-mass radii of observed MW dwarf satellites and scaling of the half-mass radius with stellar mass down to the faintest detected galaxies. Comparison of predicted distribution of model galaxies to observations suggests that some of the observed stellar systems with half mass radii $<10$ pc may be low-luminosity dwarf galaxies, which can be tested with the next generation of extremely large telescopes. \\

    \item We confirm findings of \citet{Nadler.etal.2020} that the presence of the Large Magellanic Cloud (LMC) influences the observed satellite count in the DES footprint due to its proximity to the LMC. Hosts with an LMC analog on its first infall and near pericentre have $\sim$2 times larger number of ultra-faint dwarf galaxies detected in DES footprint, compared to hosts with no LMC analogs, an early accreted LMC or a LMC far from its first pericentric approach. However, the LMC has a negligible impact on amplitude of satellite luminosity function in PS1 and SDSS. \\
    
    \item The presence of the LMC affects the concentration of satellite radial distributions with the effect being most significant for DES satellites. The concentration of the satellite distance distribution increases significantly when LMC analog passes through the pericenter of its orbit. We show that the observed radial distribution concentration in DES strongly prefers a LMC on first infall and near pericentre.
    \\
    
    \item Our fiducial model predicts a total of $440^{+201}_{-147}$ MW satellites (at the 68\% confidence level) with $\rhalf > 10$ pc, $\Mv < 0$, and $\dhelio < 300$. A significant fraction (>99\%) of these satellites have surface brightnesses $\muV < 32$ mag/arcsec$^2$ and should be detectable in upcoming surveys such as the LSST survey.\\
    
    \item We also present predictions for the number of dwarf satellites that will be discovered in specific ongoing and future surveys: $13^{+7}_{-5}$ for the HSC-SSP, $64^{+17}_{-13}$ for the DELVE-WIDE and $273^{+119}_{-92}$ for the LSST Y1 (Big Sky approach). Note that these are numbers for the total count within survey footprint and not for how many more dwarf galaxies will be discovered.
\end{enumerate}

The success of this dwarf galaxy model in matching a wide variety of properties of MW satellites including luminosity functions, radial distributions and galaxy size distributions sets the stage for interpreting the plethora of dwarf galaxy observations by LSST, DELVE-WIDE, HSC-SSP etc. in upcoming years. This study shows the importance of the ultra-faint dwarf galaxy population in not only understanding not only the host galaxy's reionization and merger history, but also constraining the very physical processes that drive their formation. 

Our code for producing all the figures in this paper is available at \url{https://github.com/kibokov/grumpy_mw_forward_model}. The data required for producing the figures is available at \url{https://data.mendeley.com/datasets/zmwh6wxyv3/1}. The \texttt{GRUMPY} model pipeline is available at \url{https://github.com/kibokov/GRUMPY}.

\section*{Acknowledgements}

We would like to thank the referee, Ethan Nadler, for detailed constructive comments that improved presentation of this paper. 
We thank Alexander Ji and the Caterpillar collaboration for providing halo tracks of the Caterpillar simulations used in this study. We are grateful to Alex Drlica-Wagner, William Cerny, Vasily Belokurov and Sidney Mau for useful discussions about observations of dwarf galaxies and star clusters and to Frank van den Bosch and Phil Mansfield for discussions about artificial halo disruption that helped with interpretation of our results. We are also grateful to Alex Drlica-Wagner for a thorough reading of the manuscript draft and detailed comments that improved the paper. This work was supported by the National Science Foundation grants AST-1714658 and AST-1911111 and NASA ATP grant 80NSSC20K0512.

Analyses presented in this paper were greatly aided by \texttt{NumPy} (\citealt{numpy}), \texttt{matplotlib} (\citealt{matplotlib}), \texttt{SciPy} (\citealt{scipy}), \texttt{Jupyter} (\citealt{jupyter}), \texttt{pandas} (\citealt{pandas}), \texttt{Astropy} (\citealt{astropy}), and \texttt{Healpy} (\citealt{healpy}). This research has made extensive use of Astrophysics Data Service (\href{http://adsabs.harvard.edu/abstract_service.html}{\tt ADS}) and \href{https://arxiv.org}{\tt arXiv} preprint repository. Finally, we are grateful to \href{https://www.grumpycats.com/}{Grumpy Cat} for helping us to carry on through the past year. 

\section*{Data Availability}

The data underlying this article will be shared on reasonable request to the authors.



\bibliographystyle{mnras}
\typeout{} 
\bibliography{main} 




\appendix

\section{Effects of resolution on Milky Way Satellite Statistics}
\label{app:reso_effect}

\begin{figure*}
    \centering
    \includegraphics[width = 0.7\textwidth]{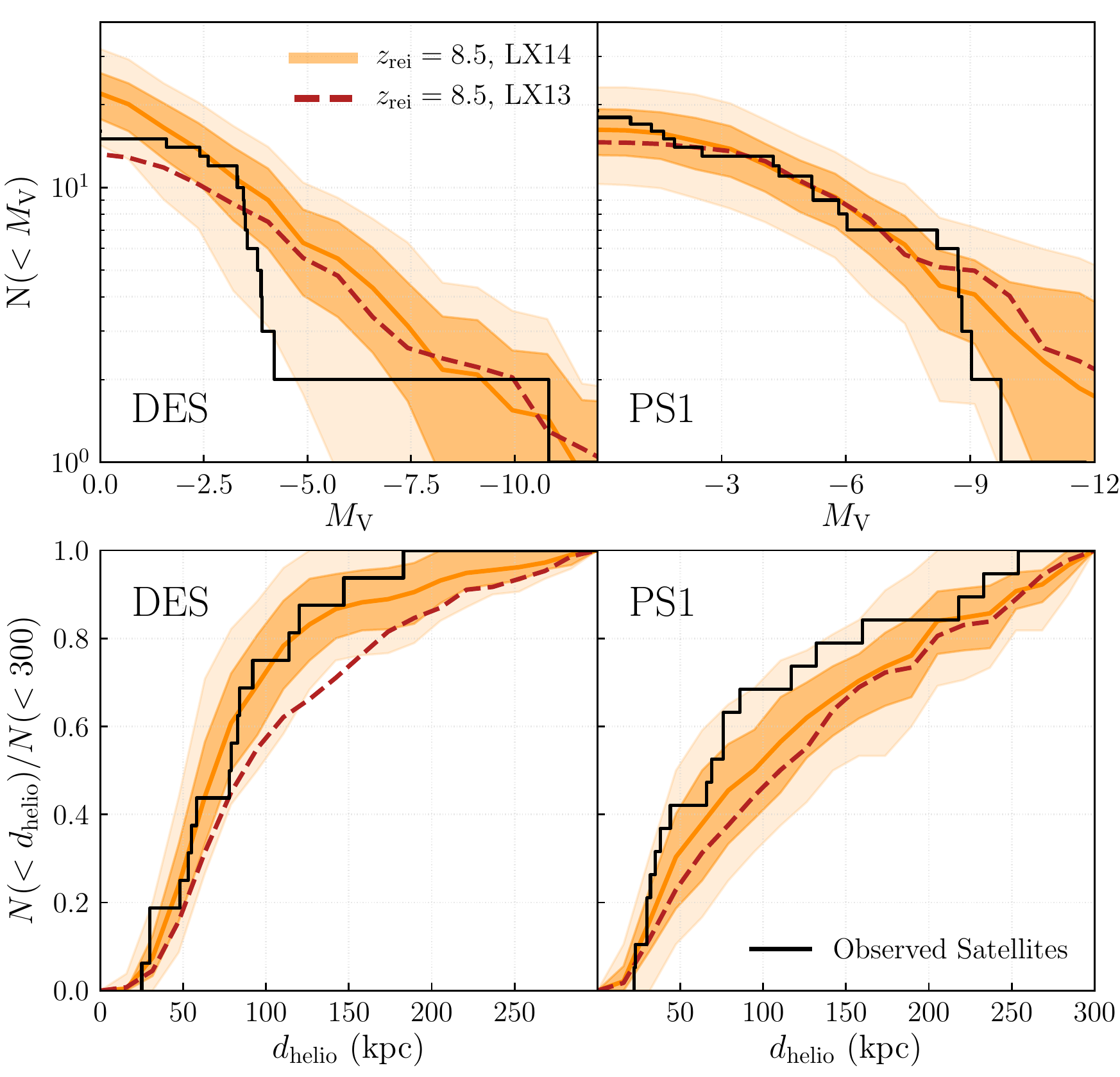}
    \caption{Predicted satellite luminosity function and radial distributions for $\zrei = 8.5$ within $\dhelio < 300$ kpc using the Caterpillar LX14 suite (fiducial) and lower resolution Caterpillar LX13 suite. The red dashed lines denote the median prediction using the Caterpillar LX13 suite. One observes that our predictions are converged for PS1 survey, however, differences in luminosity function at faint end and radial distribution are seen for DES survey. }
    \label{fig:sim_res_lf_rad}
\end{figure*}

To see whether our predicted MW satellite statistics are being influenced by the simulation resolution, we run our galaxy model on the lower resolution Caterpillar LX13 suite and compare the forward modelled survey predictions to the predictions from Caterpillar LX14 (fiducial suite). 

Figure~\ref{fig:sim_res_lf_rad} shows the forward modelled predictions for DES and PS1 surveys for both LX14 and LX13 suite. One observes, that our predictions are fairly converged for the PS1 survey. However, for the DES survey, using the LX13 suite results in a slightly shallower luminosity function at the faint end and less concentrated radial distribution. This is not surprising given the discussion in Section~\ref{sssec:artificial_disrupt} and that the DES survey, being deeper than PS1, is more sensitive to the fainter satellites that reside in low mass subhaloes. Such behavior can also be inferred from looking at the subhalo mass functions in Figure~\ref{fig:cat_res}. LX13 has a shallower mass function relative to LX14 at $10^{-5} < \mu < 10^{-4}$ in the inner regions of host. 

Although we cannot explicitly confirm convergence of our DES predictions by using a higher resolution suite than LX14, the LX14 subhalo mass functions, even in the inner most regions of host, appear converged down to $\mu = 10^{-5}$ where DES is sensitive to. Thus, our satellite predictions in this study are not significantly affected by the simulation resolution.

\section[SHMR Approximation]{Approximations to the $M_{\rm h,peak}-M_\star$ and $M_{\rm h,peak}-M_V$ relations}
\label{app:shmrapprox}

\begin{figure}
    \centering
    \includegraphics[width = 0.5\textwidth]{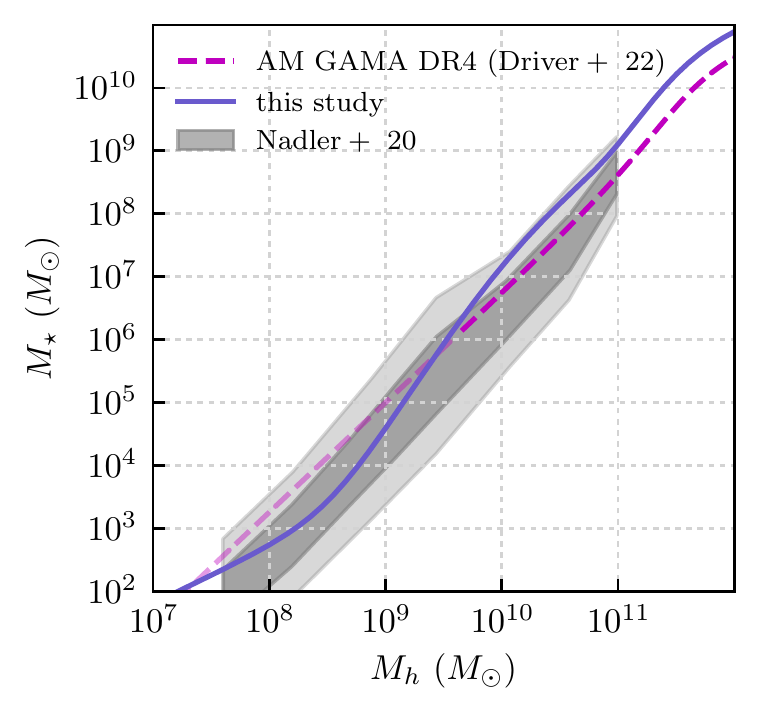}
    \caption{Halo mass--stellar mass relation approximation to the model describing observed population of the Milky Way satellites presented in this paper (solid line; eq.~\ref{eqn:new_shmr}) compared to the constraints obtained by \citet{Nadler.etal.2020} using a power law parametric model for the relation (shaded regions showing $1-$ and $2-\sigma$ constraints) and to the relation obtained by abundance matching of the halo mass function (accounting for subhalos) to the stellar mass function measured in the GAMA DR 4 survey \citet{Driver.etal.2022}. The AM relation is shown with fading color at masses smaller than directly probed by the observed stellar mass function ($\Ms\lesssim 2\times 10^6\,\Msun$).} 
    \label{fig:mhms_comp}
\end{figure}

To parametrize our predicted SHMR, we use the five-parameter parametrization from \citet{Behroozi.etal.2013.shmr} originally given as
\begin{multline}
    \log_{10} G (\Mh | M_1, \alpha, \gamma, \delta, \epsilon)  = \\
    \log_{10} (\epsilon M_{1}) + f\left( \log_{10} \frac{\Mh}{M_1}\right) - f(0)
\label{eqn:b13_shmr}
\end{multline}
where $G$ approximates\footnote{We use the letter G to avoid confusion with notation in Equation~\ref{eqn:full_scatter_shmr}.} $M_\star(\Mh)$ and 
\be
    f(x) = -\log_{10} (10^{\alpha x} + 1) + \delta \frac{\{\log_{10}[1 + \exp(x)]\}^{\gamma}}{1 + \exp(10^{-x})}
\ee
and $M_1 = 10^{11.39} \Msun$, $\alpha = -2$, $\gamma = 0.531$, $\delta = 4.335$, and $\epsilon = 0.021$. The values of these parameters approximate $M_\star(\Mh)$ relation derived by \citet{Kravtsov.etal.2018} at large masses and relation derived by \citet{Nadler.etal.2020} using observations of Milky Way satellites \citep[see][]{Kravtsov.Manwadkar.2022}.   

The modification to above parametrization to fit our model predicted SHMR is given as
\be
    \mu_{\star} (\Mpeak) = \frac{G (\Mpeak| \alpha = -1.6, \epsilon = 0.041)}{1 + (\Mpeak / 10^{10} \Msun)^{-1.3} } + 500\left( \frac{\Mpeak}{10^8 \Msun }\right)^{0.9}   
\label{eqn:new_shmr}
\ee
where we adopt the same values for $M_1,\gamma,\delta$ but set $\alpha = -1.6$ and $\epsilon = 0.041$. The modified SHMR in Equation~\ref{eqn:new_shmr} is shown as the dashed line in Figure~\ref{fig:shmr_mw} and it describes the median SHMR produced by our model quite well. 

Figure~\ref{fig:mhms_comp} compares the approximation to the median to constraints  obtained by \citet{Nadler.etal.2020} using a power law parametric model for the relation (shaded regions showing $1-$ and $2-\sigma$ constraints) and to the relation obtained by abundance matching (AM) of the halo mass function (accounting for subhalos) to the stellar mass function measured in the GAMA DR 4 survey \citet{Driver.etal.2022}. As discussed before, the relation in our model deviates from a power law, but it is in good agreement with power law relations derived by \citet[][]{Nadler.etal.2020} and by extrapolation of the AM-derived relation in the regime of ultra-faint galaxies, $M_{\rm peak}\lesssim 5\times 10^9\, \Msun$. At larger masses relation in our model is somewhat higher than the other relations, but remains within $2-\sigma$ of \citet{Nadler.etal.2020} constraint, which was anchored to the AM result at $M_{\rm peak}>10^{10}\, \Msun$. We note that the AM-derived is also uncertain at least at $\sim 20\%$ level due to uncertainties in the stellar mass function. Moreover, the constraint is derived using the $z=0$ halo mass function, while our model is for the relation of $M_\star$ and peak halo mass. The $M_\star-M_{\rm peak}$ relation of satellites can be somewhat different from the relation of the central galaxies, as indicated by observations for more massive galaxies \citep[][]{Watson.Conroy.2013}. Thus, overall the differences between our model $M_\star-M_{\rm peak}$ and other relation at the masses relevant for modelling MW satellite population ($M_{\rm peak}\lesssim 10^{11}\,\Msun$ is not significant. At the same time, the amplitude of this relation at larger masses can be easily adjusted by changing the wind mass loading scaling slightly. 

We also present  a parametrization of the scatter in our model  SHMR. 
There are two parts to this parametrization: i) quantifying what fraction of haloes at a given $\Mpeak$ that host galaxies (i.e., form stars) and ii) scatter in the stellar masses of haloes {\it that do form stars} at a given $\Mpeak$, $\sigma(\log_{10}M_\star\vert\Mpeak)$. It is expected that haloes with $\Mpeak \lesssim 10^9 \Msun$ cannot all form stars due to significant scatter in MAHs and halo formation time even at a fixed mass. We denote the probability that a halo will form stars (i.e. halo occupation probability) as $\pho(\Mpeak)$. We parametrize $\sigma$ and $\pho$ using our fiducial model as follows (here throughout $\Mpeak$ is in units of $\Msun$):
\begin{equation}
    \pho(\Mpeak) = 1 - 0.17\,\delta^2 \,H(\delta) ;\ \ \ \delta = 8.2 - \log_{10}\Mpeak
\end{equation}
where $H(\delta)$ is the Heaviside step function\footnote{$H(\delta) = 0$ for $\delta \leq 0$ and $H(\delta) = 1$ for $\delta > 0$.} and 
\begin{multline}
    \sigma(\log_{10}M_\star\vert\Mpeak) = \\ 
    \begin{cases}
        0.181\log_{10}\Mpeak - 0.941 & \text{for }\Mpeak < 10^{8.9}, \\
        -0.432\log_{10}\Mpeak + 4.518 & \text{for }\Mpeak \in  [10^{8.9},10^{10}], \\
          0.2 & \text{for }\Mpeak >10^{10}.
    \end{cases}
\label{eqn:sigma_shmr}
\end{multline}
The fixed scatter in SHMR at $\Mpeak > 10^{10} \Msun$ is similar to that observed in recent studies \citep[e.g.,][]{Reddick.etal.2013,Kravtsov.etal.2018}. 

Combining the above, the distribution of $M_\star$ for haloes of mass $\Mpeak$ is given by
\begin{equation}
    \Ms(\Mpeak) =
    \begin{cases} 
          10^{\mathcal{N}\left(\log_{10}(\mu_{\star}),\sigma^2\right)} & \rm U(0,1) \leq p_{\rm HO}(\Mpeak) \\
          0& \rm U(0,1) > p_{\rm HO}(\Mpeak)
       \end{cases}
\label{eqn:full_scatter_shmr}
\end{equation}
where $\rm U(0,1)$ denotes a random number drawn from uniform distribution in interval $[0,1]$, $\mu_{\star}(\Mpeak)$ is the median value of stellar mass formed given by Equation~\ref{eqn:new_shmr} and $\sigma(\log_{10}M_\star\vert\Mpeak)$ is the standard deviation in the stellar masses given by Equation~\ref{eqn:sigma_shmr}. Equation~\ref{eqn:full_scatter_shmr} fully describes the SHMR at $\Mpeak > 10^7 \Msun$ shown in Figure~\ref{fig:shmr_mw} and also similar to the SHMR of \citet{Kravtsov.etal.2018} for galaxies of large masses. 

Using the above parametrization for $\Ms$, we can parametrize our $\Mpeak - \Mv$ relation as follows
\be
    \Mv (\Mh) = 4.8 - 2.5 \log_{10} \left( \frac{M_{\star} (\Mh)}{\Gamma} \right) 
    \label{eq:mvmh}
\ee
where $\Gamma$ is the stellar mass-to-light ratio given by 
\be
    \Gamma = 2.5 - 0.13 \log_{10} (\Ms)
\ee

\begin{figure}
    \centering
    \includegraphics[width = 0.5\textwidth]{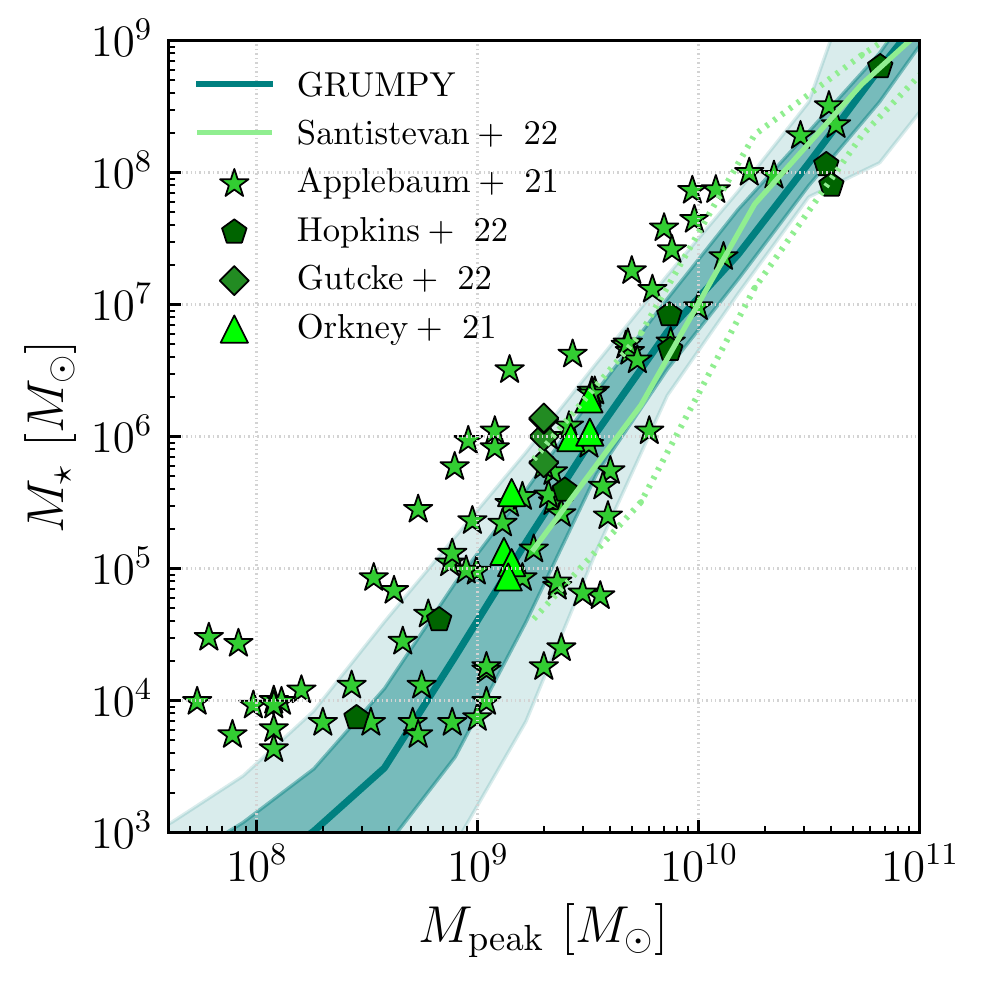}
    \caption{Comparison between the halo mass-stellar mass relation predicted by our model (\texttt{GRUMPY}) and recent hydrodynamic simulations of dwarf galaxies from the EDGE suite \citep{Orkney.etal.2021}, Justice League simulations of MW satellites \citep{Applebaum.etal.2021}, LYRA re-simulations of an isolated dwarf halo \citep {Gutcke.etal.2022}, FIRE-2 simulations of satellites around MW-sized hosts \citep{Santistevan.etal.2022}, and FIRE-3 re-simulations of isolated haloes \citep{Hopkins.etal.2022}. Our model predictions are consistent with simulations at $\Mpeak \gtrsim 10^9 \Msun$.} 
    \label{fig:shmr_comp_sim}
\end{figure}

\section{Effect of Milky Way Mass Constraints}
\label{app:mw_mass_unc}

\begin{figure}
    \centering
    \includegraphics[width = \columnwidth]{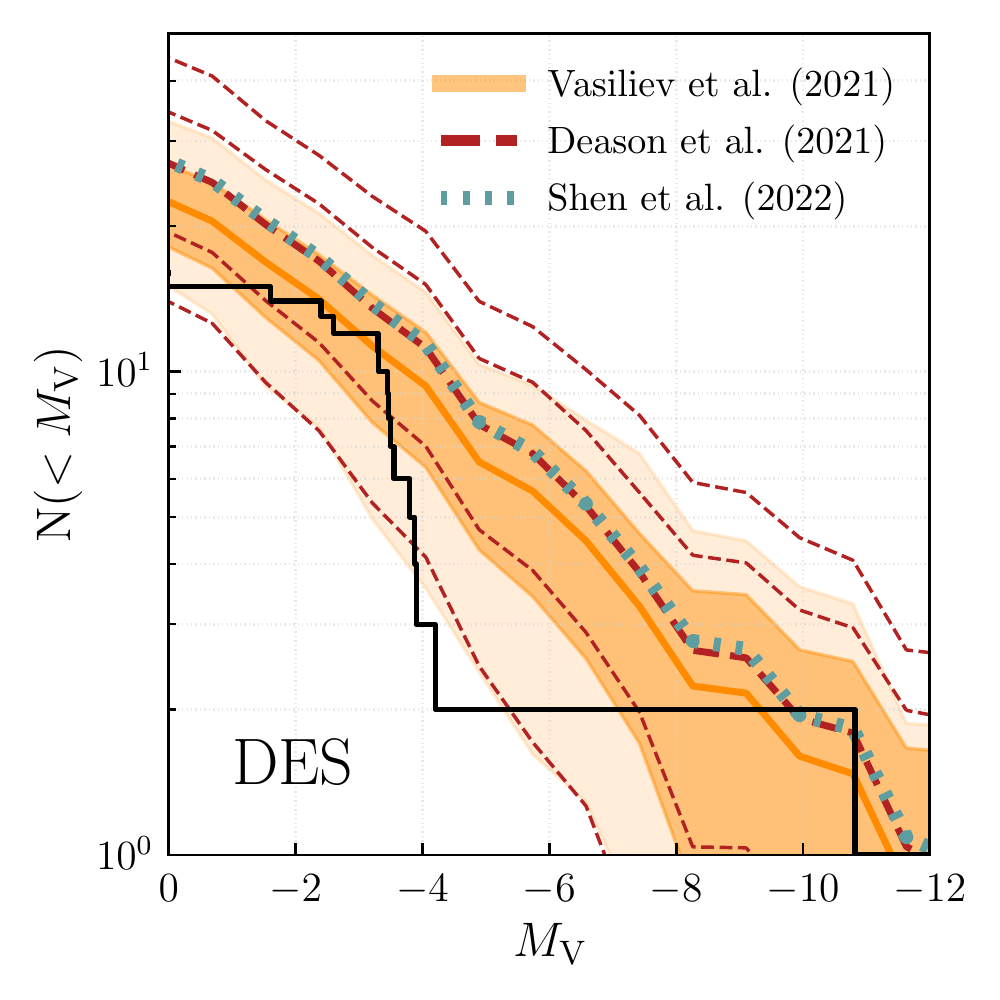}
    \caption{Our model predictions for DES satellite luminosity function for $\zrei = 10$. The orange contours corresponding to prediction using our fiducial MW mass constraints from \citet{Vasiliev.etal.2021} while dashed red lines and dotted blue lines corresponding to predictions using MW mass constraints from \citet{Deason.etal.2021} and \citet{Shen.etal.2021} respectively.}
    \label{fig:mw_mass_appendix}
\end{figure}

To account for the current constraints on the MW halo mass, we use constraints from \citet{Vasiliev.etal.2021} for the mass enclosed within 100 kpc from MW centre as $M_{\rm MW} (<100) = 5.6 \pm 0.4 \times 10^{11}$. There are other constraints on the MW mass using independent methods. For example, the recent constraint from \citet{Deason.etal.2021} on $\Mcrit$ for MW of $1.16 \pm 0.24 \times 10^{12} \Msun$ \citep[see also][]{Correa_Magnus.Vasiliev.2021}. Another constraint on the MW mass of $M(<100) = 6.9^{+0.5}_{-0.4}\times 10^{11} \Msun $ was obtained in \citet{Shen.etal.2021}. The choice of constraint does not significantly affect our final predicted statistics. Figure~\ref{fig:mw_mass_appendix} shows, for example, our model predicted DES satellite luminosity function for $\zrei = 10$. The orange contours show our fiducial predictions using MW mass constraints from \citet{Vasiliev.etal.2021}. The red dashed lines show the median relation and 68\%, 95\% confidence intervals for DES satellite luminosity function using \citet{Deason.etal.2021} MW mass constraints. The light blue colored dotted line shows the median relation using \citet{Shen.etal.2021} MW mass constraints. We do not show the contours for \citet{Shen.etal.2021} as the uncertainty in mass measurement is same as \citet{Vasiliev.etal.2021} (orange contours). All the three constraints have a similar median relations are similar. Note that the confidence intervals are larger for \citet{Deason.etal.2021} due to larger uncertainty in $\Mcrit$ constraint. Thus, our results are not significantly biased by which MW mass constraint we use.

\section{The Milky Way Satellite Population}
\label{app:mw_sats}

The properties of Milky Way satellite dwarf galaxies observed in DES, PS1 and SDSS+Classical are obtained from the tabulation in \citet{DrlicaWagner.etal.2020}. The DES and PS1 satellites we choose in this analysis are objects that pass that detection threshold in the observational selection function in \citet{DrlicaWagner.etal.2020}. Though Eridanus II is detected with high significance in DES, we do not include it in our analysis as we only consider satellites within a heliocentric distance of $300$kpc. Furthermore, Kim II and Crater I/Laevens I also pass the detection threshold and are within heliocentric distance of 300kpc, however, they are considered to be star clusters due to metallicity arguments. As mentioned in Section~\ref{ssec:mock_obs}, we only consider galaxies with $\rhalf > 10$ pc in our main analysis. For instance, DES and PS1 discovered Eridanus III and DELVE I respectively with high significance but they have $\rhalf < 10$ pc. So they were not considered as part of our main analysis. We discuss these objects in Section~\ref{ssec:small_dwarf} though. Table~\ref{tab:hsc_sats} tabulates the properties of Milky Way satellite dwarf galaxies observed in the HSC-SSP survey that we use in this analysis. The properties of newly discovered satellites in HSC-SSP footprint have been taken from \citet{Homma.etal.2016.hsc,Homma.etal.2018.hsc, Homma.etal.2019.hsc}.

When plotting dwarf galaxy statistics not constrained to any surveys (e.g., Figure~\ref{fig:mv_r12_muv}, bottom panel of Figure~\ref{fig:fut_survey_lf}), we again use the MW dwarf galaxy compilation in \citet{DrlicaWagner.etal.2020}. We include the recently discovered MW dwarf galaxy candidate Eridanus IV (\citealt{Cerny.21.eriIV}) as well.

The `probable star clusters' in Figure~\ref{fig:mv_r12_muv} have been tabulated in Table~\ref{tab:star_clusters}. These are systems that have currently been classified as star clusters due to their compact sizes (i.e, $\rhalf \lesssim 10$ pc) and have a brightness $\Mv$ that is fainter than the typical globular cluster. A precise determination of star cluster or galaxy (i.e dark matter dominated) remains ambiguous. We also include systems from the \citet{Harris.1996.clusters} (2010 edition) compilation that satisfy $ 2 \text{ pc} < \rhalf < 10$ pc and $\muV > 24$ mag/arcsec$^2$ as they are used in the comparison in Section~\ref{ssec:small_dwarf}. 

Note that we do not include star clusters like Gaia 4, Gaia 5, Gaia 6, Gaia 7 (\citealt{Torrealba.etal.2019.sc}) or Price-Whelan I (\citealt{Price.Whelan.etal.2019}) as these have been confidently classified as open clusters. 

In all tables, $\rhalf$ denotes the physical azimuthally averaged half-light radius. Thus, where ever appropriate, we convert to $\rhalf$ using the relation $\rhalf = a_{1/2}\sqrt{1 - \epsilon}$ where $\epsilon$ is ellipticity and $a_{1/2}$ is the physical elliptical half-light radius.

\begin{table}
\caption{The properties of HSC-SSP satellites used in our analysis.}
\label{tab:hsc_sats}
\centering
\begin{tabular}{lccccc}
 \hline
 Name  & $M_{\rm V}$& $d_{\rm helio}$ & $r_{1/2}$ & $\muV$\\  
   & (mag) & (kpc)  & (pc) & (mag/arcsec$^2$) \\
 \hline\hline
 Sextans & -8.72 & 86 & 345 & 27.5 \\ 
 Leo IV & -4.99 & 154 &  104 & 28.7 \\
 Pegasus III & -3.4 & 215 & 42 & 28.3\\
 Cetus III & -2.5 & 251 & 44 & 29.3 \\
 Virgo I & -0.33 & 91 & 30 & 30.6 \\
 Bo\"{o}tes IV & -4.53 & 209 & 277 & 31.3 \\
\hline
\end{tabular}
\end{table}

\begin{table}
\caption{A tabulated list of properties of a sample of \textit{objects} who have been classified as ultra faint halo star clusters or classification is yet unknown. The literature references are: (1) \citet{Balbinot.etal.2013.sc}, (2) \citet{Mau.etal.2019.sc}, (3) \citet{Mau.etal.2020}, (4) \citet{Cerny.etal.2021}, (5) \citet{Conn.etal.2018.sc}, (6) \citet{Luque.etal.2018.sc}, (7) \citet{Torrealba.etal.2019.sc}, (8) \citet{Luque.etal.2017.sc}, (9)  \citet{Koposov.etal.2017.sc}, (10) \citet{Homma.etal.2019.hsc}, (11) \citet{Kim.etal.2015.sc}, (12) \citet{Kim.etal.2015.kim2}, (13) \citet{Kim.etal.2016.sc}, (14) \citet{Munoz.etal.2018}, (15) \citet{Laevens.etal.2015.sc}, (16) \citet{Longeard.etal.2019.Laevens3}, (17) \citet{Munoz.etal.2012.sc}, (18) \citet{Laevens.etal.2014.sc}, (19) \citet{Fadely.etal.2011.sc}, (20) \citet{smash1}, (21) \citet{Gatto.etal.2021.YMCA}, (22) \citet{Belokurov.etal.2014.sc}, (23) \citet{Harris.1996.clusters} (2010 edition). The $\dagger$ symbol denotes the objects that are used in the comparison in Section~\ref{ssec:small_dwarf}.}

\label{tab:star_clusters}
\centering
\begin{tabular}{lccccc}
  \hline
 Name  & $M_{\rm V}$& $d_{\rm helio}$ & $r_{1/2}$ & References \\  
   & (mag) & (kpc)  & (pc) &  \\
 \hline\hline
  Balbinot 1 ($\dagger$)  & $-1.2$ & $31.9$ & $7.2$  & (1) \\[0.09cm]
 BLISS 1  & $0.0$ & $23.7$ & $4.1$  & (2) \\[0.09cm]
 DELVE 1 ($\dagger$) & $-0.2$ & $19.0$ & $5.4$ & (3) \\[0.09cm]
 DELVE 2 & $-2.1$ & $71$ & $21$   & (4) \\[0.09cm]
  DES 1 ($\dagger$) & $-1.4$ & $76$ & $4.2$   & (5)  \\[0.09cm]
  DES 3 ($\dagger$) & $-1.6$ & $76.2$ & $6.0$ & (6) \\[0.09cm]
  DES 4 ($\dagger$) & $-1.1$ & 31.3 & 7.6 & (7) \\[0.09cm]
 DES 5  & $0.3$ & $24.8$ & 1.3 & (7) \\[0.09cm]
 DES J0111-1341 & $0.3$ & $26.5$ & $3.9$ &  (8)  \\[0.09cm]
 DES J0225+0304 & $-1.1$ & $23.8$ & $11.6$ &  (8)  \\[0.09cm]
 DES J0222.7-5217 & $-2.4$ & $76.8$ & $8.1$  & (6) \\[0.09cm]
 Eridanus III ($\dagger$) & $-2.1$ & $91$ & $5.0$  & (5) \\[0.09cm]
  Gaia 1  & $-5.0$ & $4.6$ & $9.00$ & (9)  \\[0.09cm]
 Gaia 2 & $-2.0$ & $5.2$ & $2.7$  &  (9) \\[0.09cm]
 Gaia 3  & $-3.3$ & 48.4 & 7.5 & (7) \\[0.09cm]
 HSC 1 & $-0.20$ & 46.0 & 4.3 & (10) \\[0.09cm]
 Kim 1  & $0.3$ & $19.8$ & $5.3$  &  (11,14) \\[0.09cm]
  Kim 2 & $-3.5$ & $100.0$ & $12.0$ & (12,14) \\[0.09cm]
  Kim 3  & $+0.7$ & $15.1$ & $2.1$  &  (13) \\[0.09cm]
   Koposov 1 & $-1.0$ & 48.3 & 6.5 & (14) \\[0.09cm]
 Koposov 2 & $-0.92$ & 34.7 & 3.4 & (14) \\[0.09cm]
  Crater I / Laevens 1 & $-4.80$ & 145.0 & $19.6$  & (14,18,22) \\[0.09cm]
 Laevens 3 & $-2.8$ & $61.4$ & $10.8$ & (15,16) \\[0.09cm]
Munoz 1 & $-0.4$ & $45$ & $7.1$ & (17) \\[0.09cm] 
 PS1 1 ($\dagger$)  & $-1.9$ & 29.6 & 4.7 & (7) \\[0.09cm]
Segue 3  & $0.0$ & 17 & $1.8$  & (19) \\[0.09cm]
 SMASH 1  & $-1.0$ & 57 & $7.1$  & (20) \\[0.09cm]
 To 1  & $-1.6$ & 43.6 & 3.5  & (7) \\[0.09cm]
 YMCA 1 & $-2.8$ & - & 4.8 & (21) \\[0.09cm]
 AM 4 ($\dagger$) & $-1.81$ & 32.2 & 4.03 & (23) \\[0.09cm]
\hline
\end{tabular}
\end{table}

\bsp	
\label{lastpage}
\end{document}